\documentclass[a4paper]{cas-sc}

\usepackage[numbers,sort&compress]{natbib}
\AtBeginEnvironment{thebibliography}{\raggedright}   

\usepackage{mathtools}

\usepackage{float}

\ExplSyntaxOn
\RenewDocumentCommand \printorcid { }
  {
    \seq_if_empty:NF \g_stm_orcid_seq
      {
        \group_begin:
          \tex_let:D \thefootnote \relax
          \footnotetext
            {
              \raggedright
              \textsc{orcid}(s):\c_space_token
              \seq_use:Nn \g_stm_orcid_seq { ;~ }
            }
        \group_end:
      }
  }
\ExplSyntaxOff

\usepackage{amsthm}
\theoremstyle{plain}
\newtheorem{theorem}{Theorem}[section]
\newtheorem{lemma}[theorem]{Lemma}

\newtheorem{corollary}[theorem]{Corollary}

\theoremstyle{definition}
\newtheorem{definition}[theorem]{Definition}
\newtheorem{example}[theorem]{Example}
\newtheorem{remark}[theorem]{Remark}

\let\sep\relax

\usepackage{mathrsfs}
\usepackage{synttree, bussproofs,scrextend, rotating, xcolor, upgreek}
\usepackage[ruled, linesnumbered]{algorithm2e}
\usepackage{doi}

\usepackage{pgf, tikz, color} 
\usetikzlibrary{arrows, automata}
\usetikzlibrary{trees}
\usetikzlibrary{fit}
\usepackage{tikz}
\usetikzlibrary{positioning,arrows,calc,decorations.markings}
\tikzset{
modal/.style={>=stealth',shorten >=1pt,shorten <=1pt,auto,node distance=1.5cm,semithick},
world/.style={circle,draw,minimum size=0.5cm,fill=gray!15},
point/.style={circle,draw,inner sep=0.5mm,fill=black},
reflexive above/.style={->,loop,looseness=7,in=120,out=60},
reflexive below/.style={->,loop,looseness=7,in=240,out=300},
reflexive left/.style={->,loop,looseness=7,in=150,out=210},
reflexive right/.style={->,loop,looseness=7,in=30,out=330}
}

\usepackage[all]{xy}
\usepackage{changepage,proof,relsize,comment} 
\usepackage{multicol}
\usepackage{trfsigns}

\usepackage{wrapfig}
\usepackage{enumitem}

\renewcommand{\phi}{\varphi}

\newtheoremstyle{boldprop}{}{}{\itshape}{}{\bfseries}{.}{ }{}
\theoremstyle{boldprop}

\theoremstyle{plain}

\makeatletter
\newcommand*{\owedge}{%
  \mathbin{%
    \mathpalette\@owedge{}%
  }%
}
\newcommand*{\@owedge}[2]{%
  \sbox0{$#1\oplus\m@th$}%
  \dimen2=.5\dimexpr\wd0-\ht0-\dp0\relax 
  \dimen@=\dimexpr\ht0+\dp0\relax
  \def\lw{.04}
  \def\radius{.5-\lw/2}%
  \kern\dimen2 
  \tikz[
    line width=\lw\dimen@,
    line join=round,
    x=\dimen@,
    y=\dimen@,
    baseline=\dimexpr-.5\dimen@+\dp0\relax,
  ]
  \draw
    (0,0) circle[radius=\radius]
    (225:\radius) -- (0,.5-\lw) -- (-45:\radius)
  ;%
  \kern\dimen2 
}
\makeatother

\definecolor{tim}{RGB}{0, 0, 250}

\newcommand{\ifandonlyif}{\textit{iff} }
\newcommand{\iffi}{\textit{iff} }
\newcommand{\etc}{$\ldots$ }
\newcommand{\dfn}{Definition}
\newcommand{\fig}{Figure}
\newcommand{\lem}{Lemma}
\newcommand{\thm}{Theorem}

\newcommand{\ih}{IH}
\newcommand{\sect}{Section}

\newcommand{\D}{\mathrm{D}_{\charx}}
\newcommand{\T}{\mathrm{T}_{\charx}}
\newcommand{\four}{\mathrm{4}_{\charx}}
\newcommand{\B}{\mathrm{B}_{\charx}}
\newcommand{\five}{\mathrm{5}_{\charx}}

\newcommand{\sar}{\vdash}

\newcommand{\sat}{\vDash} 

\newcommand{\csat}{\vDash_{\!\fcons}}

\newcommand{\negnnf}[1]{\dot{\neg}{#1}}
\newcommand{\amp}{ \ \& \ }

\newcommand{\prf}{\pi} 
            
\newcommand{\km}{\mathsf{K_{m}}}

\newcommand{\cate}{}

\newcommand{\g}[1]{g(#1)}

\newcommand{\glang}{L_{\g{\fcons}}}

\newcommand{\pto}{\longrightarrow}

\newcommand{\empstr}{\varepsilon} 

\newcommand{\charx}{x}
\newcommand{\chary}{y}
\newcommand{\charz}{z}
\newcommand{\chara}{a}
\newcommand{\charb}{b}
\newcommand{\charc}{c}
\newcommand{\stra}{s}
\newcommand{\strb}{t}
\newcommand{\strc}{r}
\newcommand{\conv}[1]{\overline{#1}}

\newcommand{\albetstr}{\albet^{\ast}}

\newcommand{\strabox}{[\stra]}
\newcommand{\stradia}{\langle \stra \rangle}

\newcommand{\osdr}{\pto_{\g{\fcons}}}
\newcommand{\dr}{\pto_{\g{\fcons}}^{*}}

\newcommand{\imp}{\rightarrow}
\newcommand{\ieq}{\leftrightarrow} 
\newcommand{\inot}{\neg}

\newcommand{\xbox}{[\charx]}
\newcommand{\ybox}{[\chary]}
\newcommand{\xdia}{\langle \charx \rangle}

\newcommand{\xboxc}{[\conv{\charx}]}
\newcommand{\xdiac}{\langle \conv{\charx} \rangle}

\newcommand{\abox}{[\chara]}
\newcommand{\adia}{\langle \chara \rangle}
\newcommand{\aboxc}{[\conv{\chara}]}
\newcommand{\adiac}{\langle \conv{\chara} \rangle}

\newcommand{\diap}[1]{\langle #1 \rangle}
\newcommand{\boxp}[1]{[ #1 ]}

\newcommand{\albet}{\Upsigma}
\newcommand{\albetf}{\albet^{+}}
\newcommand{\albetb}{\albet^{-}}
\newcommand{\lang}{\mathscr{L}}

\newcommand{\atm}{\mathsf{Atm}}
\newcommand{\lit}{\mathrm{Lit}}

\newcommand{\sigp}[1]{\mathsf{sig}^{\!+}\!(#1)}
\newcommand{\sign}[1]{\mathsf{sig}^{\!-}\!(#1)}
\newcommand{\sigb}[1]{\mathsf{sig}^{\isym}\!(#1)}

\newcommand{\lab}{\mathrm{Lab}}

\newcommand{\id}{\mathsf{id}}

\newcommand{\xboxr}{\xbox}

\newcommand{\xdiar}{\xdia}

\newcommand{\conr}{\land}

\newcommand{\disr}{\lor}

\newcommand{\ddr}{\mathsf{d}_{\charx}}

\newcommand{\ctr}{\mathsf{c}}

\newcommand{\cut}{\mathsf{cut}}

\newcommand{\inter}{\mathcal{I}}

\newcommand{\fcons}{\mathcal{C}}
\newcommand{\frames}{\mathfrak{F}}
\newcommand{\framesc}{\mathfrak{F}_{\fcons}}
\newcommand{\fm}{\mathsf{fm}}
\newcommand{\F}{\mathcal{F}}
\newcommand{\M}{\mathcal{M}}
\newcommand{\V}{\mathcal{V}}
\newcommand{\R}{\mathcal{R}}
\newcommand{\W}{\mathcal{W}}

\newcommand{\simplepath}[1]{\overset{\mathclap{#1}}{\leadsto}}
\newcommand{\prpath}[1]{\hspace{3pt}\raise-3pt\hbox{$\simplepath{#1}$}\hspace{3pt}}

\newcommand{\prgr}[1]{\mathsf{pg}(#1)}
\newcommand{\prv}{\mathsf{V}}
\newcommand{\pre}{\mathsf{E}}

\newcommand{\set}[1]{\{#1\}}

\newcommand{\xbi}{\xbox\inter_{\!u}^{w}}

\newcommand{\ru}{\mathsf{r}}

\newcommand{\proves}{\Vdash}

\newcommand{\branch}{\mathcal{B}}

\newcommand{\isym}{\circ}

\newcommand{\xmod}{\triangledown_{\!\charx}} 

\newcommand{\tagx}{\mathsf{X}}

\newcommand{\tagy}{\mathsf{Y}}
\newcommand{\tagz}{\mathsf{Z}}

\newcommand{\tagl}{\mathsf{L}}
\newcommand{\tagr}{\mathsf{R}}

\providecommand{\optag}[1]{{#1}^{\mathsf{op}}}

\newcommand{\ilet}{\mathrm{I}}
\newcommand{\sep}{\parallel}
\newcommand{\splt}{ \ {}^{\tagx}|^{\tagy} \ } 
\newcommand{\spltrev}{ \ {}^{\tagy}|^{\tagx} \ }

\newcommand{\splti}{ \ {}^{\tagl}|^{\tagr} \ }
\newcommand{\spltii}{ \ {}^{\tagr}|^{\tagl} \ }
\newcommand{\itp}{\mathcal{I}}

\newcommand{\cstore}[1]{#1 \mid} 
\newcommand{\orth}[1]{\overline{#1}}
\newcommand{\orthru}{\mathsf{oth}}
\newcommand{\topr}{\top}
\newcommand{\idIi}{\id^{\ilet}_{1}}
\newcommand{\idIii}{\id^{\ilet}_{2}}

\newcommand{\disrI}{\lor^{\ilet}}
\newcommand{\conrI}{\land^{\ilet}}

\newcommand{\kmc}{\km(\fcons)}
\newcommand{\sigm}[1]{\mathsf{sig}^{\albet}\!(#1)}

\newcommand{\tkmc}{\mathsf{TK_{m}}(\fcons)} 
\newcommand{\tel}{\mathcal{T}}
\newcommand{\tseq}{\mathcal{G}}
\newcommand{\tseqii}{\mathcal{H}}
\newcommand{\tseqiii}{\mathcal{K}}
\newcommand{\wk}{\mathsf{w}}

\newcommand{\notp}{\neg p}
\newcommand{\rt}{\rho}
\newcommand{\rta}{\rho}

\newcommand{\tkmci}{\mathsf{TK^{\ilet}_{m}}(\fcons)} 
\providecommand{\iseq}{\mathcal{S}}                    
\providecommand{\untag}[1]{\lfloor #1 \rfloor}         
\providecommand{\toprI}{\top^{\ilet}}
\providecommand{\xdiarI}{\xdiar^{\ilet}}
\providecommand{\xboxrIc}{\xboxr^{\ilet}_{c}}
\providecommand{\xboxrIf}{\xboxr^{\ilet}_{p}}
\providecommand{\ddrIc}{\mathsf{d}^{\ilet}_{\charx,c}}
\providecommand{\ddrIf}{\mathsf{d}^{\ilet}_{\charx,p}}
\providecommand{\partalg}{\mathsf{Part}}

\providecommand{\partprf}{\prf^{\mathsf{p}}}

\providecommand{\intprf}{\prf^{\mathsf{i}}}

\newcommand{\desc}{\mathsf{desc}}
\newcommand{\cst}{\mathsf{C}}                         
\newcommand{\pst}[1]{\mathsf{P}_{#1}}                  
\newcommand{\lst}{\pst{\tagl}}                         
\newcommand{\rst}{\pst{\tagr}}                         
\newcommand{\lsets}[3]{#1 \,;\, #2 \,;\, #3}            
\newcommand{\lsxy}{\lsets{\cst}{\pst{\tagx}}{\pst{\tagy}}}
\newcommand{\lsyx}{\lsets{\cst}{\pst{\tagy}}{\pst{\tagx}}}
\newcommand{\rstr}{\mathord{\upharpoonright}}          

\let\lyonsep\sep

\begin{document}
\let\WriteBookmarks\relax
\def\floatpagepagefraction{1}
\def\textpagefraction{.001}

\shorttitle{Strong Lyndon Interpolation for Multi-Modal Logics with Converses}
\shortauthors{T.S. Lyon}

\title [mode = title]{Strong Lyndon Interpolation for Multi-Modal Logics with Converses: A Constructive Approach with Tree Sequents}

\author[1]{Tim S. Lyon}[orcid=0000-0003-3214-0828]
\cormark[1]
\ead{timothy_stephen.lyon@tu-dresden.de}
\ead[url]{https://sites.google.com/view/timlyon}

\affiliation[1]{organization={Technische Universit\"at Dresden},
            city={Dresden},
            country={Germany}}

\cortext[1]{Corresponding author}

\begin{abstract}
Grammar logics with converses are multi-modal logics that generalize tense logics and whose frame conditions are determined by context-free production rules. We study the \emph{strong Lyndon interpolation property (strong LIP)} for these logics, which requires interpolants to respect both the polarities of propositional atoms and the indices of modalities. We show that the strong LIP properly strengthens the Lyndon interpolation property over grammar logics, since it fails for certain grammar logics with bridge axioms, i.e., path axioms relating modalities with distinct indices. We then prove that every \emph{bridge-free} grammar logic, possibly extended with seriality axioms, has the strong LIP. This class includes the multi-modal tense logics obtained by imposing any combination of seriality, reflexivity, transitivity, symmetry, and Euclideanity on each pair of converse modalities. Our proof is constructive and purely syntactic, and is based on tree sequent calculi. Because standard methods for computing interpolants with tree and nested sequents may introduce modalities whose indices are not common to both sides of an implication, we introduce a novel \emph{tripartite partition method}. It sorts the labels of a proof into common labels and labels private to the left or right component, and only lets relational atoms pointing to common labels contribute modalities to interpolants. Given a proof of an implication, our method computes a strong Lyndon interpolant together with proofs of its correctness.
\end{abstract}

\begin{keywords}
Converse modality \sep Grammar logics \sep Lyndon interpolation \sep Multi-modal logic \sep Strong interpolation \sep Tree sequent
\end{keywords}

\maketitle

\let\sep\lyonsep

\section{Introduction}\label{sec:introduction}

\subsection{Grammar Logics}\label{subsec:intro-grammar-logics}

Grammar logics are multi-modal logics introduced by Fari{\~n}as del Cerro and Penttonen~\cite{CerPen88}, who linked the validity problem for certain classes of grammar logics to the word problem for formal grammars. This connection between modal logic and formal language theory has allowed (un)decidability results to be transferred between the two settings~\cite{BalGioMar98,CerPen88,DemNiv05}, and grammar logics have since been studied extensively~\cite{BalGioMar98,Dem01,DemNiv05,GorNgu05,NguSza11,TiuIanGor12}. We consider grammar logics with converses in the style of Demri and de Nivelle~\cite{DemNiv05}, whose frame conditions are determined by context-free production rules (cf.~\cite{Lyo21thesis,TiuIanGor12}). Each logic has a diamond $\xdia$ and a box $\xbox$ for each character $\charx$ of an alphabet $\albet$, where each $\charx$ has a converse $\conv{\charx}$, so that $\xdia$ and $\xbox$ refer to successors, whereas $\xdiac$ and $\xboxc$ refer to predecessors along the same accessibility relation.

The base grammar logic $\km$ is a multi-modal generalization of the minimal tense logic $\mathsf{Kt}$~\cite{Pri57} (see~\cite{GorPosTiu11,Kas94} for cut-free calculi), and we consider its extensions with axioms of two kinds. The first are \emph{path axioms} $\diap{\charx_{1}} \cdots \diap{\charx_{n}} A \imp \xdia A$, each of which corresponds to a production rule $\charx \pto \charx_{1} \cdots \charx_{n}$ of a context-free grammar and expresses that every sequence of $\charx_{1}$-, \ldots, $\charx_{n}$-transitions in a model yields an $\charx$-transition. Path axioms subsume multi-modal variants of the familiar axioms for reflexivity, transitivity, symmetry, and Euclideanity. The second are \emph{seriality axioms} $\xbox A \imp \xdia A$, which are not path axioms but are needed to capture, e.g., the tense variants of all logics of the modal cube, as well as (tense variants of) deontic~\cite{Aqv84} and doxastic~\cite{FagMosHalVar95} logics such as $\mathsf{KD}$ and $\mathsf{KD45}_{n}$.

The class of grammar logics is significant because it covers many well-known logics, such as standard modal logics (i.e., extensions of $\mathsf{K}$ with any combinations of $\mathsf{T}$, $\mathsf{D}$, $\mathsf{B}$, $\mathsf{4}$, and $\mathsf{5}$), epistemic logics~\cite{FagMosHalVar95}, information logics~\cite{Vak86}, and temporal logics~\cite{CerHer95}. Moreover, grammar logics with converses are closely related to description logics (DLs) with inverse roles and complex role inclusion axioms~\cite{HorSat04}, a prominent family of knowledge representation formalisms used to specify ontologies and reason about them.

Grammar logics have also been placed on an intuitionistic footing. In~\cite{Lyo21b}, \emph{intuitionistic grammar logics (IGLs)} were introduced. These generalize the intuitionistic modal logics of Fischer Servi~\cite{Fis84} and Plotkin and Stirling~\cite{PloSti86} (see also Simpson~\cite{Sim94}), as well as Ewald's intuitionistic tense logics~\cite{Ewa86}, to the multi-modal setting with converses. A faithful embedding of classical grammar logics into IGLs was then used to prove that the validity problem for IGLs is undecidable~\cite{Lyo26}.

\subsection{Interpolation and Definability}\label{subsec:intro-interpolation}

The \emph{Craig interpolation property (CIP)}~\cite{Cra57} states that whenever an implication $A \imp B$ is valid, there exists a formula $I$, called an \emph{interpolant}, such that $A \imp I$ and $I \imp B$ are valid and $I$ only contains propositional atoms common to both $A$ and $B$. The \emph{Lyndon interpolation property (LIP)}~\cite{Lyn59} strengthens the CIP by additionally requiring that every atom occurring positively (negatively) in $I$ occurs positively (negatively) in both $A$ and $B$. Interpolation properties have found a variety of applications in logic and computer science; e.g., in computer-aided verification~\cite{McM18}, to establish Beth definability~\cite{KihOno10}, and in knowledge representation to forget or conceal information in ontologies~\cite{LutWol11}.

Interpolation is intimately connected to \emph{Beth definability}~\cite{Bet56}, which states that whenever a symbol is implicitly defined by a theory, it is explicitly definable in terms of the remaining symbols. Indeed, Beth definability is typically derived from interpolation (cf.~\cite{KihOno10}), and constructive proofs of interpolation give rise to algorithms that compute explicit definitions~\cite{LyoKar24}. In DLs, Beth definability ensures that a concept that is implicitly defined by an ontology in terms of a given signature can be explicitly defined by a concept over that signature. The existence and computation of such explicit definitions have been studied for expressive DLs with inverse roles and role hierarchies~\cite{LyoKar24,TenFraSey13}, and explicit definitions and interpolants have been used to optimize reasoning, generate query plans, and compute referring expressions in query answering~\cite{BorTomWed16,HudTomWed15,TenFraSey13}.

Crucially, in DLs, the common language of an interpolant or explicit definition is specified in terms of a \emph{signature}, which consists of both concept names and role names (cf.~\cite{ArtJunMazOzaWol23,LutWol11,TenFraSey13}). That is to say, in the absence of an ontology, an interpolant of a subsumption between two concepts may only use those concept names \emph{and} role names that occur in both concepts; relative to ontologies, each side of the subsumption additionally includes the names of its ontology (see \sect~\ref{sec:conclusion}). In the multi-modal setting, this amounts to requiring that an interpolant only contains modalities whose indices occur in both the antecedent and consequent of the implication. Combining this requirement with the polarity requirement of the LIP yields what we refer to as the \emph{strong Lyndon interpolation property (strong LIP)} (see \dfn~\ref{def:strong-lip}), which is the central concern of this paper. 

Related interpolation properties that treat the indices of modalities as part of the common language have been studied in the modal logic literature, where indices typically represent agents or programs. Van Benthem~\cite{vBe97} introduced the \emph{strong} interpolation property, which requires that every modality occurring in an interpolant occurs in both the antecedent and the consequent, and Marx~\cite{Mar98} showed by means of bisimulation products that it holds for every canonical multi-modal logic whose frames are defined by a universal Horn theory without interaction axioms, i.e., without conditions relating distinct modalities; he also observed that interaction axioms such as $\langle a \rangle p \to \langle b \rangle p$ may destroy it. Such properties have since been studied for multi-agent epistemic logics~\cite{BilFusKuz25,MurSan20} and indexed tense logics~\cite{SanYam26}.

Recently, Sano and Yamasaki~\cite{SanYam26} established the CIP, counting both propositional atoms and indices as common vocabulary, for the indexed tense versions of $\mathsf{K}$, $\mathsf{KT}$, $\mathsf{KD}$, $\mathsf{K4}$, $\mathsf{K4D}$, and $\mathsf{S4}$ (as well as their extensions with converse seriality) by applying Maehara's method~\cite{Mae60} to Gentzen-style sequent calculi. Since Gentzen-style sequents are not expressive enough to yield natural cut-free calculi for logics with converse modalities, their calculi make use of analytic cuts. However, it is not clear whether cuts respect the polarities of propositional atoms, and so, such calculi do not readily yield the LIP; indeed, Sano and Yamasaki leave the generalization of their result to Lyndon interpolation as future work. Tracking polarities therefore calls for cut-free calculi, which, for logics with converse modalities, are available in the form of nested and tree sequent calculi, as discussed below; similarly, B{\'\i}lkov{\'a} et al.~\cite{BilFusKuz25} obtain Lyndon interpolation for $\mathsf{S5}_{n}$, with agents included in the common language, from a cut-free calculus that combines hypersequents with nested sequents.

\subsection{Proof-Theoretic Interpolation}\label{subsec:intro-proof-theory}

A syntactic, proof-theoretic approach to interpolation was initiated by Maehara~\cite{Mae60} for Gentzen-style sequent calculi. Maehara's method takes a cut-free proof of (a sequent encoding) an implication $A \imp B$ as input. First, it partitions each sequent of the proof into two components according to whether its formulae stem from $A$ or $B$. Second, it assigns interpolants to the initial sequents. Finally, it computes an interpolant for the conclusion of each rule from the interpolants of its premises, ultimately yielding an interpolant of $A \imp B$.

A distinguishing feature of this approach is its \emph{constructivity}: rather than merely establishing the existence of interpolants, as is typical of algebraic and model-theoretic approaches (e.g., those based on amalgamation~\cite{KihOno10} or bisimulation products~\cite{Mar98}), Maehara's method provides an explicit procedure for computing interpolants from proofs. The method has since been generalized to nested sequents~\cite{FitKuz15,Lyo21thesis,LyoTiuGorClo20}, labeled sequents~\cite{Kuz16}, hypersequents~\cite{Kuz16b,KuzLel18}, and multicomponent calculi more generally~\cite{Kuz18}, and has been used to compute interpolants and explicit definitions in DLs~\cite{LyoKar24}.

For logics with converse modalities, nested sequents~\cite{Bru09,Kas94,Pog09} (see~\cite{LelPog24} for a survey) and their notational variants, tree sequents~\cite{GorRam12,IshKik07}, have proven to be particularly fruitful, as they yield cut-free calculi for tense logics~\cite{GorPosTiu11,Kas94} and for grammar logics with converses~\cite{Lyo21thesis,TiuIanGor12}. Two approaches to interpolation with such calculi can be found in the literature. The first is the method of Fitting and Kuznets~\cite{FitKuz15}, later generalized by Kuznets~\cite{Kuz16,Kuz18}, in which interpolants are generalized to multiformulae spread over the components of a sequent, and whose correctness is justified semantically. 

The second is the purely syntactic method of~\cite{LyoTiuGorClo20}, in which interpolants are generalized to sets of sequents, an orthogonality operation is used to negate such sets, and correctness is established proof-theoretically, without resorting to semantic arguments or interpreted connectives external to the language of the logic. A notable benefit of this latter method is that it not only computes an interpolant $I$ of an implication $A \imp B$, but also outputs proofs of $A \imp I$ and $I \imp B$. Using this method, the author showed that every context-free grammar logic with converse possesses the \emph{effective} LIP~\cite[\sect~6.2]{Lyo21thesis},\footnote{As context-free grammar logics with converses are undecidable in general~\cite{BalGioMar98}, `effective' means that a Lyndon interpolant can be computed from a given proof of an implication.} thus generalizing the effective CIP for regular grammar logics due to Gor{\'e} and Nguyen~\cite{GorNgu05}. The present paper builds on the syntactic method of~\cite{LyoTiuGorClo20} as well as the tree sequent systems of~\cite{Lyo21thesis}, which we extend with seriality rules (cf.~\cite{Lyo26}). We note that the LIP result of~\cite{Lyo21thesis} does not cover seriality axioms; however, it can be easily extended to cover them by adapting the techniques for seriality presented in this paper.

Although the methods discussed above are well-suited for establishing the CIP and LIP, they are not sensitive to the indices of modalities. In these methods, the labeled formulae of each sequent in a proof are partitioned into a left and a right component, while the relational atoms (i.e., the tree structure of the sequent) are shared by both components. Consequently, whenever a box rule (bottom-up) introduces a relational atom $w R_{\chary} u$, the interpolant computed for its conclusion may acquire a modality indexed with $\chary$, regardless of whether both components actually make use of the relational atom. For example, the standard method computes the interpolant $\boxp{\chary}\top$ for the theorem $\xbox p \imp \xbox p \lor \boxp{\chary}(q \lor \inot q)$, which contains the character $\chary$ even though $\chary$ does not occur in the antecedent (see \sect~\ref{sec:inter-calc}). B{\'\i}lkov{\'a} et al.~\cite[Example~22]{BilFusKuz25} encountered the same phenomenon when applying the multicomponent method~\cite{Kuz18} to their calculus for $\mathsf{S5}_{n}$. In general, the interpolants computed by such methods satisfy the requirements of the LIP, but are not guaranteed to lie in the common modal signature, and so, do not establish the strong LIP.

In this paper, we rectify this issue by introducing a novel \emph{tripartite partition method}. Rather than letting both components use all labels, our method sorts the labels of each sequent into three sets: the labels private to the left component, the labels private to the right component, and the \emph{common labels}, which both components may use. Only relational atoms pointing to common labels are permitted to contribute modalities to interpolants. This suffices to guarantee that interpolants lie in the common modal signature for \emph{bridge-free} grammar logics, i.e., grammar logics in which no path axiom relates modalities with distinct indices (up to converse). Bridge-free grammar logics can be viewed as fusions of tense logics, one for each pair of converse characters (cf.~\cite{Kur07}), and the restriction to such logics is a natural one: as we show, the strong LIP fails for grammar logics with bridge axioms such as $\diap{\charb}\diap{\charc} A \imp \adia A$ (see \sect~\ref{subsec:strong-lip}), even though every context-free grammar logic with converses possesses the LIP~\cite{Lyo21thesis}.

\subsection{Contributions}\label{subsec:intro-contributions}

Our contributions in this paper are discussed below:

(1) We establish that every bridge-free grammar logic possesses the strong LIP. In particular, this holds for the multi-modal tense logics obtained by extending $\km$ with any combination of $\D$, $\T$, $\four$, $\B$, and $\five$ for each pair of converse modalities $\xdia$ and $\xdiac$. This class of logics includes the indexed tense logics of Sano and Yamasaki~\cite{SanYam26}, and so, our result strengthens theirs from Craig to Lyndon interpolation, a generalization they left as future work, and extends it to logics with symmetry, Euclideanity, and further path axioms. 
We also show that the strong LIP fails for some grammar logics with bridge axioms, in line with an earlier observation of Marx~\cite{Mar98} for the strong variant of the CIP, which demonstrates that the strong LIP is a proper strengthening of the LIP over the class of grammar logics.

(2) We introduce the tripartite partition method for computing interpolants with tree sequents, which is the key technical innovation of this paper. The method is realized by means of a \emph{partition algorithm} and an \emph{interpolation calculus}: the partition algorithm determines, for each relational atom of a given proof, whether it is used by one or both components, and accordingly sorts the labeled formulae of the proof into the left and right component and its labels into common labels and labels private to either component. The rules of the interpolation calculus then assign interpolants to the resulting partitioned proof. The rules of our interpolation calculus that introduce relational atoms come in \emph{common} and \emph{private} variants, ensuring that modalities only enter interpolants when it is safe to do so, i.e., when both components make use of the relational atom in question.

(3) Our approach is fully constructive and purely syntactic. Given a proof of $A \imp B$ in our tree sequent calculus, our method computes a strong Lyndon interpolant $I$, together with proofs of $A \imp I$ and $I \imp B$. Moreover, the computed interpolants are guaranteed to lie in the common signature \emph{by construction}, meaning that no post-processing of interpolants is required, and their correctness is verified within the tree sequent calculus itself, in the spirit of~\cite{LyoTiuGorClo20}.

(4) We extend the tree sequent systems for grammar logics with converses~\cite{Lyo21thesis} with seriality rules, and show how seriality can be accommodated within our interpolation method. As relational atoms introduced by seriality do not stem from formulae of either component, they may be used by both components. Nevertheless, we show that they can be treated exactly like relational atoms introduced by box rules: such a relational atom only contributes a modality to an interpolant if both components use it. As mentioned above, the LIP result of~\cite{Lyo21thesis} does not cover seriality axioms, though our techniques for seriality can be easily adapted to extend that result.

Our work is most closely related to the recent work of B{\'\i}lkov{\'a} et al.~\cite{BilFusKuz25}, who establish the LIP for the multi-agent logic $\mathsf{S5}_{n}$ while treating agents (i.e., indices) as part of the common language, which they refer to as the \emph{agent Lyndon interpolation property}. Our approach differs from theirs in several respects. First, B{\'\i}lkov{\'a} et al.\ introduce a new proof formalism, called \emph{crossword sequents}, which is tailored to $\mathsf{S5}_{n}$, whereas we work with tree sequent systems that uniformly capture an infinite class of grammar logics with converses. 

Second, and more significantly, their interpolation method first computes a preliminary Lyndon interpolant (i.e., it may not be strong) by means of the multicomponent method~\cite{Kuz18}, and then corrects those modalities that may fall outside the common language: such modalities are either eliminated by means of equivalences such as $\boxp{\chara}\bot \ieq \bot$ and $\diap{\chara}\top \ieq \top$, which hold in $\mathsf{S5}_{n}$ (and more generally, in serial logics), or shown to be indexed with a common agent by means of a subformula argument specific to their calculus. As B{\'\i}lkov{\'a} et al.\ remark themselves, the proof-theoretic method they employ does not natively support agent interpolation. By contrast, our tripartite partition method builds the sensitivity to indices directly into the computation of interpolants, preventing non-common modalities from entering interpolants in the first place. 

Third, since converse modalities can be eliminated from formulae in the presence of symmetry, the agent Lyndon interpolation property for $\mathsf{S5}_{n}$ follows as a special case of our main result. In particular, $\mathsf{S5}_{n}$ is the converse-free fragment of the bridge-free grammar logic obtained by imposing reflexivity, transitivity, and symmetry on each pair of converse modalities.

\subsection{Paper Organization}\label{subsec:intro-organization}

\sect~\ref{sec:log-prelims} introduces classical grammar logics, the grammar-theoretic notions used to encode their path conditions, and the strong LIP; there, we also show that the strong LIP fails for certain grammar logics with bridge axioms and define the class of bridge-free grammar logics. \sect~\ref{sec:nested-calculi} presents our tree sequent systems for grammar logics, which extend those of~\cite{Lyo21thesis} with seriality rules, and recalls their fundamental properties. \sect~\ref{sec:inter-calc} explains why the standard method of computing interpolants with tree sequents fails to respect the indices of modalities, and introduces interpolation sequents as well as our interpolation calculus, which implements the tripartite partition method. \sect~\ref{sec:part-bias} presents the partition algorithm, which transforms a tree sequent proof into a partitioned proof that has the shape of an interpolation proof. \sect~\ref{sec:main-theorem} establishes the correctness of our interpolation method and proves that every bridge-free grammar logic has the strong LIP. Finally, in \sect~\ref{sec:conclusion}, we conclude and discuss future work.

\section{Preliminaries}\label{sec:log-prelims}

In this section, we introduce classical grammar logics (\sect~\ref{subsec:grammar-logics}), the grammar-theoretic notions used to encode their path conditions (\sect~\ref{subsec:grammar-theory}), and the strong Lyndon interpolation property, which takes into account both the polarities of propositional atoms and the indices of modalities occurring in formulae (\sect~\ref{subsec:strong-lip}).

\subsection{Classical Grammar Logics}\label{subsec:grammar-logics}

The language of each grammar logic is defined relative to an \emph{alphabet} $\albet$, which is a non-empty countable set of \emph{characters}, used to index modalities. As in~\cite{DemNiv05}, we stipulate that each alphabet $\albet$ is partitioned into a \emph{forward part} $\albetf := \{\chara, \charb, \charc, \ldots\}$ and a \emph{backward part} $\albetb := \{\conv{\chara}, \conv{\charb}, \conv{\charc}, \ldots\}$. That is to say, we assume the following holds for an alphabet $\albet$: (1) $\albet = \albetf \cup \albetb$, (2) $\albetf \cap \albetb = \emptyset$, and (3) $\chara \in \albetf$ \iffi $\conv{\chara} \in \albetb$.

We refer to the elements $\chara$, $\charb$, $\charc$, \etc of $\albetf$ as \emph{forward characters} and the elements $\conv{\chara}$, $\conv{\charb}$, $\conv{\charc}$, \etc of $\albetb$ as \emph{backward characters}. A \emph{character} is defined to be either a forward or backward character, and we use $\charx$, $\chary$, $\charz$, \etc to denote them. In what follows, modalities indexed with forward characters will be interpreted as making reference to successor (or future) worlds along an accessibility relation within a relational model, and modalities indexed with backward characters will make reference to predecessor (or past) worlds.

We define the \emph{converse operation} to be a function $\conv{\cdot}$ mapping each forward character $\chara \in \albetf$ to its \emph{converse} $\conv{\chara} \in \albetb$ and vice versa. Hence, the converse operation is its own inverse, i.e., $\conv{\conv{\charx}} = \charx$ for every $\charx \in \albet$.

Unless stated otherwise, we fix an alphabet $\albet$ for the remainder of the text. We let $\atm := \{p, q, r, \ldots\}$ be a denumerable set of \emph{(propositional) atoms} and let $\lit := \atm \cup \{\neg p \mid p \in \atm\}$ be the set of \emph{literals}. We define the language $\lang$ relative to the alphabet $\albet$ to be the set of formulae generated by the following grammar in BNF:
$$
A ::= p \mid \neg p \mid \top \mid \bot \mid A \lor A \mid A \land A \mid \xdia A \mid \xbox A
$$
where $p$ ranges over the set $\atm$ of atoms and $\charx$ ranges over the characters in the alphabet $\albet$. We use $A$, $B$, $C$, \etc to denote formulae in $\lang$ and define the \emph{length} of a formula $A$, denoted $\ell(A)$, to be the number of symbols contained in $A$. 

As in~\cite{Lyo21thesis}, formulae are given in \emph{negation normal form}, meaning that only atomic negations are permitted in the language.\footnote{In contrast to~\cite{Lyo21thesis}, we take $\top$ and $\bot$ to be primitive rather than abbreviations of $p \lor \neg p$ and $p \land \neg p$ for a fixed $p \in \atm$. This ensures that $\top$ and $\bot$ do not contribute any propositional atoms to the signature of a formula (see \sect~\ref{subsec:strong-lip}).} We define the negation of arbitrary formulae as follows.

\begin{definition}[Negation]\label{def:negation}
We define the \emph{negation} $\negnnf{A}$ of a formula $A \in \lang$ recursively:
\begin{align*}
  \negnnf{p}          &:= \neg p                      & \negnnf{\neg p}      &:= p \\
  \negnnf{\top}       &:= \bot                        & \negnnf{\bot}        &:= \top \\
  \negnnf{(B \lor C)} &:= \negnnf{B} \land \negnnf{C} & \negnnf{(B \land C)} &:= \negnnf{B} \lor \negnnf{C} \\
  \negnnf{\xdia B}    &:= \xbox \negnnf{B}            & \negnnf{\xbox B}     &:= \xdia \negnnf{B}
\end{align*}
\end{definition}

It is straightforward to verify by induction on $\ell(A)$ that $\negnnf{\negnnf{A}} = A$ for every formula $A \in \lang$. We define $A \imp B := \negnnf{A} \lor B$ and $A \ieq B := (A \imp B) \land (B \imp A)$ as usual.

Formulae are interpreted over \emph{$\albet$-frames} and \emph{$\albet$-models}~\cite{DemNiv05}, which we refer to simply as \emph{frames} and \emph{models}. These are an extension and slight reformulation of the semantics for grammar logics given by Fari{\~n}as del Cerro and Penttonen~\cite{CerPen88}.

\begin{definition}[Frame]\label{def:frame} A \emph{frame} is a tuple $\F = (\W, \{\R_{\charx} \mid \charx \in \albet\})$ such that:
\begin{itemize}

\item $\W$ is a non-empty set of \emph{worlds} $\{w, u, v, \ldots\}$;

\item For each $\charx \in \albet$, the \emph{accessibility relation} $\R_{\charx} \subseteq \W \times \W$ satisfies: (C) For all $w, u \in \W$, $w \R_{\charx} u$ \iffi $u \R_{\conv{\charx}} w$.

\end{itemize}
\end{definition}

\begin{definition}[Model]\label{def:model} We define a \emph{model} based on a frame $\F$ to be a pair $\M = (\F, \V)$ such that $\V : \atm \to 2^{\W}$ is a \emph{valuation function} mapping each atom to the set of worlds where it holds. We will often write a model $\M$ as $(\W, \{\R_{\charx} \mid \charx \in \albet\}, \V)$.
\end{definition}

We refer to (C) as the \emph{converse condition}, which ensures that the backward modalities are truly converse to the forward modalities. That is to say, if we think of a \emph{forward relation} $\R_{\chara}$ (indexed with a forward character $\chara \in \albetf$) as relating worlds to their \emph{successors}, then (C) ensures that the corresponding \emph{backward relation} $\R_{\conv{\chara}}$ (indexed with the backward character $\conv{\chara} \in \albetb$) relates worlds to their \emph{predecessors}. Hence, the modalities $\adia$ and $\abox$ make reference to successor worlds, whereas the modalities $\adiac$ and $\aboxc$ make reference to predecessor worlds, giving our formulae and models a tense character. Indeed, the class of grammar logics we consider includes the minimal tense logic $\mathsf{Kt}$~\cite{Pri67}.

\begin{definition}[Semantic Clauses]\label{def:semantic-clauses} Let $\M = (\W, \{\R_{\charx} \mid \charx \in \albet\}, \V)$ be a model with $w \in \W$. The \emph{satisfaction relation} $\M,w \sat A$ between the world $w$ of $\M$ and a formula $A \in \lang$ is recursively defined as follows:
\begin{itemize}

\item $\M,w \sat p$ \ifandonlyif $w \in \V(p)$, for $p \in \atm$;

\item $\M,w \sat \neg p$ \ifandonlyif $w \not\in \V(p)$, for $p \in \atm$;

\item $\M,w \sat \top$ and $\M,w \not\sat \bot$;

\item $\M,w \sat A \lor B$ \ifandonlyif $\M,w \sat A$ or $\M,w \sat B$;

\item $\M,w \sat A \land B$ \ifandonlyif $\M,w \sat A$ and $\M,w \sat B$;

\item $\M,w \sat \xdia A$ \ifandonlyif there exists a $u \in \W$ such that $w \R_{\charx} u$ and $\M,u \sat A$;

\item $\M,w \sat \xbox A$ \ifandonlyif for all $u \in \W$, if $w \R_{\charx} u$, then $\M,u \sat A$.

\end{itemize}
We say that $A$ is \emph{true} in $\M$, written $\M \sat A$, \iffi $\M, w \sat A$ for all $w \in \W$. A formula $A \in \lang$ is \emph{valid} w.r.t.\ a class of frames $\frames$ \iffi $\M \sat A$ for all models $\M$ based on a frame $\F \in \frames$.
\end{definition}

\iffalse
The negation operation of \dfn~\ref{def:negation} behaves classically, and so, $\imp$ receives its usual (classical) interpretation, as confirmed by the following lemma, whose proof is standard.

\begin{lemma}\label{lem:negation-semantics}
Let $\M = (\W, \{\R_{\charx} \mid \charx \in \albet\}, \V)$ be a model with $w \in \W$. For all $A, B \in \lang$,
\begin{enumerate}

\item[(1)] $\M,w \sat \negnnf{A}$ \iffi $\M,w \not\sat A$;

\item[(2)] $\M,w \sat A \imp B$ \iffi $\M,w \not\sat A$ or $\M,w \sat B$.

\end{enumerate}
\end{lemma}
\fi

The base grammar logic $\km$ is defined to be the set of formulae valid w.r.t.\ the class of all frames. We also consider logics obtained by imposing \emph{path conditions} and \emph{seriality conditions} on frames, which are of the following forms, respectively:
$$
\forall \vec{w} \, (w_{0} \R_{\charx_{1}} w_{1} \amp \cdots \amp w_{n-1} \R_{\charx_{n}} w_{n} \Rightarrow w_{0} \R_{\charx} w_{n})
\qquad \qquad
\forall w \exists u \, (w \R_{\charx} u)
$$
Here, $n \geq 0$ and $\vec{w} = w_{0}, \ldots, w_{n}$ in the path condition, which is taken to be $\forall w_{0} (w_{0} \R_{\charx} w_{0})$ when $n = 0$. 

\begin{definition}[$\fcons$-Valid]\label{def:fcons-valid} Let $\fcons$ be a set of path and seriality conditions. We define a \emph{$\fcons$-frame} to be a frame satisfying each path and seriality condition in $\fcons$ and let $\framesc$ denote the class of all $\fcons$-frames. A \emph{$\fcons$-model} is a model based on a $\fcons$-frame. A formula $A$ is defined to be \emph{$\fcons$-valid}, written $\csat A$, \iffi $A$ is valid w.r.t.\ $\framesc$. The \emph{grammar logic} $\kmc$ is defined to be the set of all $\fcons$-valid formulae, that is, $A \in \kmc$ \iffi $\csat A$.
\end{definition}

Note that $\km = \km(\emptyset)$. Although we have characterized grammar logics semantically, we note that sound and complete axiomatizations and proof systems of these logics have been provided in the literature~\cite{CerPen88,Lyo21thesis,TiuIanGor12}. As our results do not rely on such axiomatizations, we omit them.

\subsection{Grammar-Theoretic Preliminaries}\label{subsec:grammar-theory}

As implied by their name, grammar logics connect concepts from formal language theory to logical concepts. In particular, the information contained in the path conditions of a set $\fcons$ can be encoded as a formal grammar, and the strings generated by this grammar correspond to sequences of accessibility relations that imply the presence of a single accessibility relation in a $\fcons$-frame. Such grammars have been used to parameterize the propagation rules of tree (and nested) sequent calculi for (classical and intuitionistic) grammar logics~\cite{Lyo21thesis,Lyo26,TiuIanGor12}. The current section introduces the grammar-theoretic notions needed in the sequel.

We let $\varepsilon$ be the \emph{empty string} and define the set $\albet^{*}$ of \emph{strings over $\albet$} to be the smallest set such that (1) $\albet \cup \{\varepsilon\} \subseteq \albet^{*}$, and (2) $\text{if } \stra \in \albet^{*} \text{ and } \charx \in \albet \text{, then } \stra \charx \in \albet^{*}$. We use $\stra$, $\strb$, $\strc$, \etc (potentially annotated) to represent strings in $\albetstr$. Note, for the empty string we have $\stra \cate \empstr = \empstr \cate \stra = \stra$. The converse operation on strings is defined accordingly: (1) $\conv{\varepsilon} := \varepsilon$, and (2) $\text{if } \stra = \charx_{1} \cdots \charx_{n} \text{ with } n \geq 1 \text{, then } \conv{\stra} := \conv{\charx}_{n} \cdots \conv{\charx}_{1}$.

Since accessibility relations in a frame are indexed with characters from $\albet$, sequences of accessibility relations correspond to strings from $\albetstr$. We therefore define generalized relations $\R_{\stra}$ relative to strings (cf.~\cite{DemNiv05}).

\begin{definition}[Generalized Relations]\label{def:generalized-relations} Let $\F = (\W, \{\R_{\charx} \mid \charx \in \albet\})$ be a frame and $\stra \in \albetstr$. We inductively define the \emph{generalized accessibility relation} $\R_{\stra}$ relative to the string $\stra$ as follows:
\begin{itemize}

\item $\R_{\empstr} := \{(w,w) \mid w \in \W\}$;

\item $\R_{\stra \cate \charx} := \{(w,u) \mid \text{there exists a } v \in \W \text{ such that } (w,v) \in \R_{\stra} \text{ and } (v,u) \in \R_{\charx}\}$.

\end{itemize}
\end{definition}

A straightforward induction on the length of $\strb$ shows that, for all $\stra, \strb \in \albetstr$, we have $(w,u) \in \R_{\stra \cate \strb}$ \iffi there exists a $v \in \W$ such that $(w,v) \in \R_{\stra}$ and $(v,u) \in \R_{\strb}$. Combined with the converse condition (C), it follows that $(w,u) \in \R_{\stra}$ \iffi $(u,w) \in \R_{\conv{\stra}}$ for every frame and $\stra \in \albetstr$.

To simplify the presentation of formulae, we define strings in modalities as follows: if $\stra = \charx_{1} \cdots \charx_{n}$, then $[ \stra ] A := [ \charx_{1} ] \cdots [ \charx_{n} ] A$ and $\langle \stra \rangle A := \langle \charx_{1} \rangle \cdots \langle \charx_{n} \rangle A$, with $\strabox A = \stradia A = A$ when $\stra = \empstr$. It is easy to verify by induction on the length of $\stra$ that, for every model $\M$ and world $w$ of $\M$, we have $\M, w \sat \stradia A$ \iffi there exists a $u \in \W$ such that $(w,u) \in \R_{\stra}$ and $\M, u \sat A$, and that $\M, w \sat \strabox A$ \iffi $\M, u \sat A$ for all $u \in \W$ such that $(w,u) \in \R_{\stra}$.

Using generalized relations, we can compactly write each path condition (displayed in \sect~\ref{subsec:grammar-logics}) as $\R_{\stra} \subseteq \R_{\charx}$ with $\stra = \charx_{1} \cdots \charx_{n}$, where $\stra = \empstr$ if $n = 0$. Henceforth, we identify path conditions with such inclusions, and as this subset notation is more succinct, we use it to denote path conditions in the sequel. 

For each $\charx \in \albet$, we write $\D$ for the seriality condition $\forall w \exists u \, (w \R_{\charx} u)$. Note that path conditions include the standard frame conditions of reflexivity, transitivity, symmetry, and Euclideanity for $\R_{\charx}$, which are defined as follows:
$$
\T := (\R_{\empstr} \subseteq \R_{\charx})
\quad
\four := (\R_{\charx \charx} \subseteq \R_{\charx})
\quad
\B := (\R_{\conv{\charx}} \subseteq \R_{\charx})
\quad
\five := (\R_{\conv{\charx} \charx} \subseteq \R_{\charx})
$$
Path and seriality conditions correspond to the path and seriality axioms mentioned in the introduction. Indeed, it is straightforward to verify that a frame satisfies $\R_{\stra} \subseteq \R_{\charx}$ \iffi the \emph{path axiom} $\stradia p \imp \xdia p$ is true in every model based on the frame, and that it satisfies $\D$ \iffi the \emph{seriality axiom} $\xbox p \imp \xdia p$ is true in every model based on the frame (cf.~\cite{DemNiv05,Lyo21thesis}). For instance, $\T$, $\four$, $\B$, and $\five$ correspond to the path axioms $p \imp \xdia p$, $\xdia \xdia p \imp \xdia p$, $\xdiac p \imp \xdia p$, and $\xdiac \xdia p \imp \xdia p$, respectively. The logics we consider therefore include the tense variants of all logics of the `modal cube' (cf.~\cite{Bru09,Kas94,SanYam26}).

We now define \emph{$\fcons$-grammars}, which are types of \emph{semi-Thue systems}~\cite{Pos47} encoding the information contained in the path conditions of a set $\fcons$.

\begin{definition}[$\fcons$-Grammar]\label{def:grammar} Let $\fcons$ be a set of path and seriality conditions. We call expressions of the form $\charx \pto \stra$ with $\charx \in \albet$ and $\stra \in \albetstr$ \emph{production rules}, and define the \emph{$\fcons$-grammar} $\g{\fcons}$ to be the following set of production rules:
$$
\g{\fcons} := \{\charx \pto \stra, \ \conv{\charx} \pto \conv{\stra} \mid (\R_{\stra} \subseteq \R_{\charx}) \in \fcons\}.
$$
\end{definition}

Observe that $\g{\fcons}$ is determined solely by the path conditions in $\fcons$, as seriality conditions do not give rise to production rules. Moreover, $\g{\fcons}$ is closed under converses, i.e., if $\charx \pto \stra \in \g{\fcons}$, then $\conv{\charx} \pto \conv{\stra} \in \g{\fcons}$.

A $\fcons$-grammar $\g{\fcons}$ is a type of string rewriting system. For example, if $\charx \pto \stra \in \g{\fcons}$, we may derive the string $\strb \cate \stra \cate \strc$ from $\strb \cate \charx \cate \strc$ in one step by applying the mentioned production rule. Through repeated applications of production rules to a given string $\stra \in \albetstr$, one derives new strings, the collection of which determines a language. Let us make such notions precise by means of the following definitions.

\begin{definition}[Derivation]\label{def:gram-derivation} Let $\fcons$ be a set of path and seriality conditions. The \emph{one-step derivation relation} $\osdr$ holds between two strings $\stra$ and $\strb$ in $\albetstr$, written $\stra \osdr \strb$, \iffi there exist $\stra', \strb' \in \albetstr$ and $\charx \pto \strc \in \g{\fcons}$ such that $\stra = \stra' \cate \charx \cate \strb'$ and $\strb = \stra' \cate \strc \cate \strb'$. The \emph{derivation relation} $\dr$ is defined to be the reflexive and transitive closure of $\osdr$. For two strings $\stra, \strb \in \albetstr$, we refer to $\stra \dr \strb$ as a \emph{derivation of $\strb$ from $\stra$}, and define its \emph{length} to be equal to the minimal number of one-step derivations needed to derive $\strb$ from $\stra$ in $\g{\fcons}$.
\end{definition}

\begin{definition}[Language]\label{def:gram-language} Let $\fcons$ be a set of path and seriality conditions. For a character $\charx \in \albet$, the \emph{language of $\charx$ relative to $\g{\fcons}$} is defined to be the set $\glang(\charx) := \{\stra \mid \charx \dr \stra \}$. When $\fcons$ is clear from the context, we write $L_{\charx}$ to denote $\glang(\charx)$.
\end{definition}

\begin{example}\label{ex:grammar} Let $\albet := \{\chara, \conv{\chara}\}$ and $\fcons := \{\mathrm{T}_{\chara}, \mathrm{4}_{\chara}, \mathrm{B}_{\chara}\}$, so that $\kmc$ is the tense variant of $\mathsf{S5}$. Then,
$$
\g{\fcons} = \{\chara \pto \empstr, \ \conv{\chara} \pto \empstr, \ \chara \pto \chara\chara, \ \conv{\chara} \pto \conv{\chara}\conv{\chara}, \ \chara \pto \conv{\chara}, \ \conv{\chara} \pto \chara\}.
$$
Since $\chara$ can be duplicated, erased, and rewritten to $\conv{\chara}$, we have that $\glang(\chara) = \{\chara, \conv{\chara}\}^{*}$, i.e., every string over $\albet$ can be derived from $\chara$. This reflects the fact that in a $\fcons$-frame, $\R_{\chara}$ is an equivalence relation, and so, any two worlds connected by a sequence of $\R_{\chara}$ and $\R_{\conv{\chara}}$ relations are related by $\R_{\chara}$.
\end{example}

As noted above, seriality conditions do not affect $\g{\fcons}$. For instance, if $\albet := \{\chara, \conv{\chara}\}$ and $\fcons := \{\mathrm{D}_{\chara}, \mathrm{4}_{\chara}, \mathrm{5}_{\chara}\}$, so that $\kmc$ is the tense variant of $\mathsf{KD45}$, then
$$
\g{\fcons} = \g{\{\mathrm{4}_{\chara}, \mathrm{5}_{\chara}\}} = \{\chara \pto \chara\chara, \ \conv{\chara} \pto \conv{\chara}\conv{\chara}, \ \chara \pto \conv{\chara}\chara, \ \conv{\chara} \pto \conv{\chara}\chara\}.
$$
Accordingly, $\fcons$-grammars only encode the path conditions imposed on $\fcons$-frames, and seriality conditions must be accounted for separately (see \lem~\ref{lem:path-relation} below).

The following lemma will be helpful in the sequel. It follows by induction on the number of one-step derivations, since $\g{\fcons}$ is closed under converses.

\begin{lemma}\label{lem:string-and-inverse}
For all $\charx \in \albet$ and $\stra = \charx_{1} \cdots \charx_{n} \in \albetstr$, we have $\stra \in \glang(\charx)$ \iffi $\conv{\stra} = \conv{\charx}_{n} \cdots \conv{\charx}_{1} \in \glang(\conv{\charx})$.
\end{lemma}

The lemma below shows that the languages generated by a $\fcons$-grammar exactly capture the path conditions imposed on $\fcons$-frames.

\begin{lemma}\label{lem:path-relation}
Let $\fcons$ be a set of path and seriality conditions. A frame $\F$ satisfies each path condition in $\fcons$ \iffi $\R_{\stra} \subseteq \R_{\charx}$ for all $\charx \in \albet$ and $\stra \in L_{\charx}$. Therefore, $\F$ is a $\fcons$-frame \iffi (i) $\R_{\stra} \subseteq \R_{\charx}$ for all $\charx \in \albet$ and $\stra \in L_{\charx}$, and (ii) $\R_{\charx}$ is serial for each $\D \in \fcons$.
\end{lemma}

\begin{proof} We prove the first claim, from which the second claim follows by \dfn~\ref{def:fcons-valid}. The backward direction is immediate: if $(\R_{\stra} \subseteq \R_{\charx}) \in \fcons$, then $\charx \pto \stra \in \g{\fcons}$, so $\stra \in L_{\charx}$, and thus, $\F$ satisfies $\R_{\stra} \subseteq \R_{\charx}$ by assumption. For the forward direction, suppose that $\F$ satisfies each path condition in $\fcons$. We show by induction on $n$ that if $\stra$ can be derived from $\charx$ by means of $n$ one-step derivations, then $\R_{\stra} \subseteq \R_{\charx}$.

\emph{Base case.} If $n = 0$, then $\stra = \charx$, and so, the claim trivially holds.

\emph{Inductive step.} Suppose that $\stra$ can be derived from $\charx$ by means of $n+1$ one-step derivations. By \dfn~\ref{def:gram-derivation}, there exists a string $\strb \cate \chary \cate \strc$ that can be derived from $\charx$ by means of $n$ one-step derivations, and there exists a production rule $\chary \pto \charz_{1} \cdots \charz_{k} \in \g{\fcons}$ such that $\stra = \strb \cate \charz_{1} \cdots \charz_{k} \cate \strc$. Suppose that $(w,u) \in \R_{\stra}$. Then, there exist $v, v' \in \W$ such that $(w,v) \in \R_{\strb}$, $(v,v') \in \R_{\charz_{1} \cdots \charz_{k}}$, and $(v',u) \in \R_{\strc}$. By \dfn~\ref{def:grammar}, either $(\R_{\charz_{1} \cdots \charz_{k}} \subseteq \R_{\chary}) \in \fcons$ or $(\R_{\conv{\charz}_{k} \cdots \conv{\charz}_{1}} \subseteq \R_{\conv{\chary}}) \in \fcons$. In the former case, $(v,v') \in \R_{\chary}$ as $\F$ satisfies $\R_{\charz_{1} \cdots \charz_{k}} \subseteq \R_{\chary}$. In the latter case, $(v',v) \in \R_{\conv{\charz}_{k} \cdots \conv{\charz}_{1}}$ by the converse condition (C) (see the remark after \dfn~\ref{def:generalized-relations}), which implies $(v',v) \in \R_{\conv{\chary}}$ as $\F$ satisfies $\R_{\conv{\charz}_{k} \cdots \conv{\charz}_{1}} \subseteq \R_{\conv{\chary}}$, and so, $(v,v') \in \R_{\chary}$ by (C). In both cases, $(w,u) \in \R_{\strb \cate \chary \cate \strc}$, from which $(w,u) \in \R_{\charx}$ follows by \ih.
\end{proof}


\subsection{Strong Lyndon Interpolation}\label{subsec:strong-lip}

The \emph{Craig interpolation property (CIP)}~\cite{Cra57} for a logic states that whenever an implication $A \imp B$ belongs to the logic, there exists a formula $I$, called a \emph{Craig interpolant}, such that $A \imp I$ and $I \imp B$ belong to the logic, and $I$ only contains atoms common to both $A$ and $B$. The \emph{Lyndon interpolation property (LIP)}~\cite{Lyn59} strengthens the CIP by additionally requiring that each propositional atom occurring in $I$ occurs with the same polarity in $A$ and $B$.

In the multi-modal setting, it is natural to further strengthen the LIP by requiring that an interpolant only contains modalities whose indices occur in both the antecedent and consequent of the implication. This is in line with signature-based notions of interpolation in description logics, where signatures consist of both concept and role names (cf.~\cite{ArtJunMazOzaWol23,TenFraSey13}). We make this precise by means of the following notion of signature, which tracks the polarities of propositional atoms as well as the indices of modalities in a formula.

\begin{definition}[Signature]\label{def:signature}
For a formula $A \in \lang$, the set $\sigp{A}$ of propositional atoms \emph{occurring positively} in $A$, the set $\sign{A}$ of propositional atoms \emph{occurring negatively} in $A$, and the set $\sigm{A}$ of characters \emph{indexing modalities} in $A$ are defined recursively as follows:
\begin{itemize}

\item $\sigp{p} = \{p\}$, $\sign{p} = \emptyset$, $\sigp{\neg p} = \emptyset$, and $\sign{\neg p} = \{p\}$ for $p \in \atm$;

\item $\sigm{p} = \sigm{\neg p} = \emptyset$ for $p \in \atm$;

\item $\sigb{\top} = \sigb{\bot} = \emptyset$ for $\isym \in \{+, -, \albet\}$;

\item $\sigb{A \oplus B} = \sigb{A} \cup \sigb{B}$ for $\isym \in \{+, -, \albet\}$ and $\oplus \in \{\lor, \land\}$;

\item $\sigb{\xmod A} = \sigb{A}$ for $\isym \in \{+, -\}$ and $\xmod \in \{\xdia, \xbox\}$;

\item $\sigm{\xmod A} = \{\charx, \conv{\charx}\} \cup \sigm{A}$ for $\xmod \in \{\xdia, \xbox\}$.

\end{itemize}
We refer to $\sigp{A}$, $\sign{A}$, and $\sigm{A}$ as the \emph{positive}, \emph{negative}, and \emph{modal signature} of $A$, respectively.
\end{definition}

Observe that $p \in \sigp{A}$ (resp.\ $p \in \sign{A}$) \iffi the literal $p$ (resp.\ $\neg p$) occurs in $A$. As the lemma below shows, negation swaps the positive and negative signature of a formula, and so, our notion of polarity agrees with the usual one for languages in which implication is primitive: a propositional atom occurs positively in $A \imp B$ \iffi it occurs negatively in $A$ or positively in $B$.

\begin{lemma}\label{lem:signature-negation}
For every $A \in \lang$, $\sigp{\negnnf{A}} = \sign{A}$, $\sign{\negnnf{A}} = \sigp{A}$, and $\sigm{\negnnf{A}} = \sigm{A}$. Consequently, for all $A, B \in \lang$,
\begin{itemize}

\item $\sigp{A \imp B} = \sign{A} \cup \sigp{B}$;

\item $\sign{A \imp B} = \sigp{A} \cup \sign{B}$;

\item $\sigm{A \imp B} = \sigm{A} \cup \sigm{B}$.

\end{itemize}
\end{lemma}


\begin{definition}[Strong Lyndon Interpolation Property]\label{def:strong-lip}
Let $\fcons$ be a set of path and seriality conditions. We say that $\kmc$ has the \emph{strong Lyndon interpolation property (strong LIP)} \iffi whenever $A \imp B \in \kmc$, there exists a formula $I \in \lang$, called a \emph{strong Lyndon interpolant}, such that:
\begin{enumerate}

\item[(1)] $\sigb{I} \subseteq \sigb{A} \cap \sigb{B}$ for each $\isym \in \{+, -\}$;

\item[(2)] $\sigm{I} \subseteq \sigm{A} \cap \sigm{B}$;

\item[(3)] $A \imp I \in \kmc$ and $I \imp B \in \kmc$.

\end{enumerate}
\end{definition}

\begin{remark}\label{rem:lip-cip}
The LIP is obtained from \dfn~\ref{def:strong-lip} by dropping condition~(2), and coincides with the LIP for grammar logics given in~\cite[Definition~96]{Lyo21thesis}, which is formulated in terms of the literals occurring in formulae. The CIP is obtained by additionally weakening condition~(1) to $\sigp{I} \cup \sign{I} \subseteq (\sigp{A} \cup \sign{A}) \cap (\sigp{B} \cup \sign{B})$, i.e., every propositional atom in $I$ occurs in both $A$ and $B$, without regard to polarity. Hence, the strong LIP implies the LIP, which in turn implies the CIP. 


Analogously, imposing condition (2) in addition to the requirements of the CIP yields a strong variant of the CIP, introduced by van Benthem~\cite{vBe97} as the \emph{strong interpolation property}. Marx~\cite{Mar98} established this property for canonical multi-modal logics whose frames are defined by universal Horn theories without interaction axioms. This result concerns Craig rather than Lyndon interpolation and covers neither converse modalities, whose defining condition (C) relates two distinct modalities, nor seriality, which is not a universal Horn condition. Sano and Yamasaki~\cite{SanYam26} established the strong variant of the CIP for several indexed tense logics.
\end{remark}

Two features of \dfn~\ref{def:signature} are worth highlighting. First, since $\top$ and $\bot$ are primitive and have empty signatures, they may serve as interpolants in degenerate cases; e.g., $\top$ is a strong Lyndon interpolant of $\abox p \imp \boxp{\charb}(q \lor \neg q)$. Second, the modal signature of a formula contains each character together with its converse. This reflects the fact that $\xdia$ and $\xdiac$ are interpreted relative to mutually converse relations by condition (C), mirroring the treatment of inverse roles in description logics, where an inverse role $r^{-}$ is built from the role name $r$. Distinguishing $\charx$ from $\conv{\charx}$ in the modal signature would yield a property that fails already for symmetric logics: if $\mathrm{B}_{\chara} \in \fcons$, then $\adiac p \imp \adia p \in \kmc$, yet $\adiac p$ and $\adia p$ would share no index, and one can verify, as in Remark~\ref{rem:strong-lip-failure} below, that no formula without modalities interpolates this implication.

\begin{remark}\label{rem:strong-lip-failure}
In contrast to the LIP, which is possessed by every grammar logic $\kmc$ such that $\fcons$ only contains path conditions~\cite{Lyo21thesis}, the strong LIP is not possessed by every grammar logic. Let $\chara, \charb, \charc \in \albetf \subseteq \albet$ be pairwise distinct and let $\fcons := \{\R_{\charb\charc} \subseteq \R_{\chara}\}$. Then, $\diap{\charb}\diap{\charc} p \imp \adia p \in \kmc$, since $w \R_{\charb} u$ and $u \R_{\charc} v$ imply $w \R_{\chara} v$ in every $\fcons$-frame, though
$$
\sigm{\diap{\charb}\diap{\charc} p} \cap \sigm{\adia p} = \{\charb, \conv{\charb}, \charc, \conv{\charc}\} \cap \{\chara, \conv{\chara}\} = \emptyset.
$$
Hence, a strong Lyndon interpolant $I$ of this implication would contain no modalities and would satisfy $\sign{I} = \emptyset$ and $\sigp{I} \subseteq \{p\}$, that is, $I$ would be built from $p$, $\top$, and $\bot$ using $\lor$ and $\land$, and would therefore be equivalent to $\top$, $\bot$, or $p$. However, one can verify that none of these serves as a suitable interpolant. 
Thus, $\kmc$ does not have the strong LIP, and so, the strong LIP is a proper strengthening of the LIP over the class of grammar logics. A similar observation was made by Marx~\cite{Mar98} for the strong variant of the CIP in a converse-free setting: the logic with the single bridge condition $\R_{a} \subseteq \R_{b}$ has the CIP, but $\langle a \rangle \top \to \langle b \rangle \top$ has no interpolant whose modalities occur on both sides.
\end{remark}

Given this counter-example, we identify a class of grammar logics that \emph{do} possess the strong LIP, which we refer to as \emph{bridge-free grammar logics}. 

\begin{definition}[Bridge-Free Grammar Logics]\label{def:bridge-free} We define a grammar logic $\kmc$ to be \emph{bridge-free} \iffi $\glang(\charx) \subseteq \set{\charx, \conv{\charx}}^{*}$ for each $\charx \in \albet$.
\end{definition}

Note that this condition only restricts the path conditions in $\fcons$, since seriality conditions do not contribute production rules to $\g{\fcons}$. The following lemma confirms that bridge-freeness can be read off directly from the path conditions in $\fcons$.

\begin{lemma}\label{lem:bridge-free-char}
A grammar logic $\kmc$ is bridge-free \iffi $\stra \in \set{\charx, \conv{\charx}}^{*}$ for each path condition $(\R_{\stra} \subseteq \R_{\charx}) \in \fcons$.
\end{lemma}

\begin{proof} The forward direction is immediate: if $(\R_{\stra} \subseteq \R_{\charx}) \in \fcons$, then $\charx \pto \stra \in \g{\fcons}$, and so, $\stra \in \glang(\charx) \subseteq \set{\charx, \conv{\charx}}^{*}$. For the backward direction, the assumption implies that $\strb \in \set{\chary, \conv{\chary}}^{*}$ for every production rule $\chary \pto \strb \in \g{\fcons}$; for the converse production rules $\conv{\charz} \pto \conv{\strb}$ of \dfn~\ref{def:grammar}, this holds since $\strb \in \set{\charz, \conv{\charz}}^{*}$ implies $\conv{\strb} \in \set{\conv{\charz}, \charz}^{*}$. Hence, as $\set{\chary, \conv{\chary}} = \set{\charx, \conv{\charx}}$ for every $\chary \in \set{\charx, \conv{\charx}}$, applying a production rule to a string in $\set{\charx, \conv{\charx}}^{*}$ yields a string in $\set{\charx, \conv{\charx}}^{*}$, and so, $\glang(\charx) \subseteq \set{\charx, \conv{\charx}}^{*}$ follows by induction on the length of derivations.
\end{proof}

We use the term \emph{bridge-free} since, by \lem~\ref{lem:bridge-free-char}, a grammar logic $\kmc$ is bridge-free \iffi $\fcons$ contains no \emph{bridge conditions}, i.e., path conditions $\R_{\stra} \subseteq \R_{\charx}$ such that $\stra$ contains a character outside of $\set{\charx, \conv{\charx}}$. Such conditions relate modalities with distinct indices (up to converse); examples include $\R_{\chara} \subseteq \R_{\charb}$, which corresponds to the bridge axiom $\adia p \imp \diap{\charb} p$, and the path condition $\R_{\charb\charc} \subseteq \R_{\chara}$ of Remark~\ref{rem:strong-lip-failure}.

By contrast, each of the conditions $\D$, $\T$, $\four$, $\B$, and $\five$ only involves the characters $\charx$ and $\conv{\charx}$, and so, $\kmc$ is bridge-free whenever each condition in $\fcons$ is of one of these forms (see \lem~\ref{lem:bridge-free-char}). In general, if $\kmc$ is bridge-free, then each condition in $\fcons$ only constrains the accessibility relations $\R_{\chara}$ and $\R_{\conv{\chara}}$ for a single $\chara \in \albetf$. A bridge-free grammar logic can therefore be viewed as a fusion of tense logics, one for each pair of converse characters $\chara$ and $\conv{\chara}$ (cf.~\cite{Kur07}). As we show in the sequel (see \thm~\ref{thm:strong-lip}), every bridge-free grammar logic has the strong LIP.

The class of bridge-free grammar logics subsumes the indexed tense logics studied by Sano and Yamasaki~\cite{SanYam26}, namely, the indexed tense versions of $\mathsf{K}$, $\mathsf{KT}$, $\mathsf{KD}$, $\mathsf{K4}$, $\mathsf{K4D}$, and $\mathsf{S4}$, as well as their extensions with converse seriality~\cite[Section~7.1]{SanYam26}. 
Consequently, our results extend those of Sano and Yamasaki in three respects. First, we strengthen their result from Craig to Lyndon interpolation, a generalization they leave open~\cite[Section~7.2]{SanYam26}; as in their work, indices are counted as part of the common vocabulary (cf.~Remark~\ref{rem:lip-cip}). Second, we permit distinct conditions to be imposed on distinct accessibility relations, whereas Sano and Yamasaki impose the same conditions on all of them. Third, we cover further path conditions, such as $\B$ and $\five$, and more generally, arbitrary bridge-free sets of path and seriality conditions; extensions with symmetry, for instance, are mentioned by Sano and Yamasaki as a direction for future work~\cite[Section~7.2]{SanYam26}.

\section{Tree Sequent Systems}\label{sec:nested-calculi}

In this section, we recall the tree sequent systems for the grammar logics introduced in~\cite{Lyo21thesis}, which are notational variants of the nested sequent systems for grammar logics given by Tiu et al.~\cite{TiuIanGor12}, and extend them with seriality rules. We opt for the labeled sequent syntax (cf.~\cite{Sim94,Vig00}) rather than the nested sequent syntax as it simplifies the formulation of our interpolation algorithm later on (cf.~\cite{LyoTiuGorClo20}). It was already observed by Gor{\'e} and Ramanayake~\cite{GorRam12} that restricting labeled sequents to be trees, rather than more general binary graphs (which may be disconnected or include cycles), yields tree sequents (cf.~\cite{IshKik07}), which are a notational variant of nested sequents.

We let $\lab := \{w, u, v, \ldots\}$ be a countably infinite set of \emph{labels}, define a \emph{relational atom} to be an expression of the form $w R_{\charx} u$ with $w, u \in \lab$ and $\charx \in \albet$, and define a \emph{labeled formula} to be an expression of the form $w : A$ such that $w \in \lab$ and $A \in \lang$. (NB. We use $w$, $u$, $v$, \etc to denote labels since $\charx$, $\chary$, $\charz$, \etc are reserved for characters.) For a relational atom $w R_{\charx} u$, we call $w$ the \emph{source} and $u$ the \emph{target} of the relational atom.

We use $\tel$ to denote finite sets of relational atoms and upper-case Greek letters $\Gamma$, $\Delta$, $\Sigma$, \etc to denote finite multisets of labeled formulae. For a set $\tel$ of relational atoms and a multiset $\Gamma$ of labeled formulae, we let $\lab(\tel)$, $\lab(\Gamma)$, and $\lab(\tel,\Gamma)$ be the sets of all labels occurring therein. For a multiset $\Gamma$ of labeled formulae and $w \in \lab$, we define $\Gamma(w) := \{A \mid w : A \in \Gamma\}$, and for multisets $\Gamma$ and $\Delta$ of labeled formulae, we let $\Gamma, \Delta$ denote the union of the two.

A set $\tel$ of relational atoms is called a \emph{tree} \iffi either $\tel = \emptyset$ or the edge-labeled graph $G(\tel) := (\lab(\tel), \{(w,\charx,u) \mid w R_{\charx} u \in \tel\})$ forms a directed tree, that is, there exists a label $w \in \lab(\tel)$, called the \emph{root}, with no incoming edges such that every other label in $\lab(\tel)$ has exactly one incoming edge and is reachable from $w$ by a directed path of edges.

\begin{definition}[Tree Sequent]\label{def:tree-sequent} A \emph{tree sequent} is an expression of the form $\tel \sar \Gamma$ such that (1) $\tel$ is a tree, (2) if $\tel \neq \emptyset$, then $\lab(\Gamma) \subseteq \lab(\tel)$, and (3) if $\tel = \emptyset$, then $|\lab(\Gamma)| \leq 1$, i.e., all labeled formulae in $\Gamma$ share the same label.
\end{definition}

We note that conditions (1)--(3) ensure that each tree sequent forms a connected graph that is indeed of a tree shape. We use $\tseq$, $\tseqii$, $\tseqiii$, $\ldots$ to denote tree sequents, and let $\lab(\tseq)$ be the set of labels occurring in $\tseq$. The \emph{root} of a tree sequent $\tel \sar \Gamma$ is the root of $\tel$ if $\tel \neq \emptyset$, and is the single label occurring in $\Gamma$ otherwise. We adopt standard tree terminology when discussing tree sequents (e.g., parent, child, leaf, and subtree). Intuitively, a tree sequent $\tel \sar \Gamma$ states that if all relational atoms in $\tel$ hold, then some labeled formula in $\Gamma$ holds (see \dfn~\ref{def:fm-interp} below).\footnote{Since we consider classical logics with formulae in negation normal form, it suffices to use one-sided sequents~\cite{Lyo21thesis,LyoTiuGorClo20,TiuIanGor12}.} We define a \emph{flat sequent} to be a tree sequent of the form $\sar \Gamma$, that is, a flat sequent is a tree sequent without relational atoms, and where every labeled formula in $\Gamma$ shares the same label. In particular, the \emph{empty sequent} $\sar \emptyset$ is a flat sequent.

Every tree sequent encodes a tree whose nodes are labeled with sets of formulae and whose edges are labeled with characters. Let $\tseq = \tel, w R_{\charx_{1}} u_{1}, \ldots, w R_{\charx_{n}} u_{n} \sar \Gamma$ be a tree sequent such that $w$ is the root and $u_{1}, \ldots, u_{n}$ are all children of $w$. The tree $\mathsf{tr}_{w}(\tseq)$ is graphically depicted below, where $\mathsf{tr}_{u_{i}}(\tseq)$ is the tree encoded by the subtree rooted at $u_{i}$ for $i \in [n]$:\footnote{We define $[n] = \set{1, \ldots, n}$ and therefore use $i \in [n]$ as a shorthand for $1 \leq i \leq n$.}
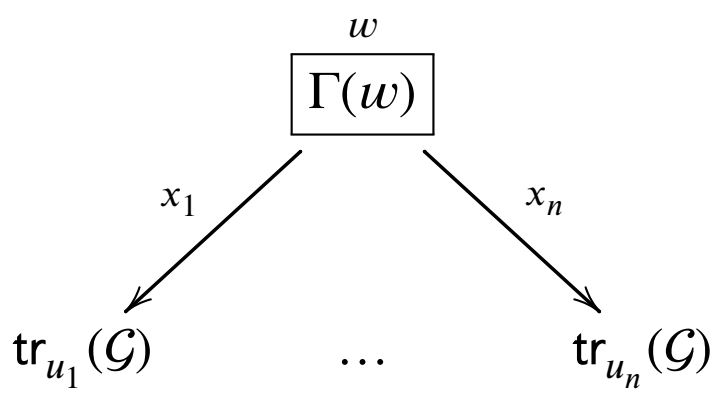
\begin{center}
\begin{tabular}{c}
\xymatrix@C=2em@R=2.5em{
 & \overset{w}{\boxed{\Gamma(w)}}\ar@{->}[dl]_-{\charx_{1}}\ar@{->}[dr]^-{\charx_{n}} & \\
 \mathsf{tr}_{u_{1}}(\tseq) & \hdots & \mathsf{tr}_{u_{n}}(\tseq)
}
\end{tabular}
\end{center}
This tree structure gives rise to a natural formula interpretation of tree sequents.

\begin{definition}[Formula Interpretation]\label{def:fm-interp} Let $\tseq = \tel \sar \Gamma$ be a tree sequent and $u \in \lab(\tseq)$. We define the \emph{formula interpretation} $\fm_{u}(\tseq)$ of $\tseq$ at $u$ recursively on the subtree of $\tel$ rooted at $u$:
$$
\fm_{u}(\tseq) := \bigvee \big( \Gamma(u) \cup \{ \xbox \fm_{v}(\tseq) \mid u R_{\charx} v \in \tel \} \big)
$$
where the empty disjunction is taken to be $\bot$. If $w$ is the root of $\tseq$, then we define $\fm(\tseq) := \fm_{w}(\tseq)$, and we define $\fm(\sar \emptyset) := \bot$. We say that a tree sequent is \emph{$\fcons$-valid} \iffi $\fm(\tseq)$ is $\fcons$-valid.
\end{definition}

\begin{figure}[pos=t]
\noindent

\begin{center}
\begin{tabular}{c c}
\AxiomC{$\phantom{\tseq}$}
\RightLabel{$\id$}
\UnaryInfC{$\tel \sar w : p, w : \inot p, \Gamma$}
\DisplayProof

&

\AxiomC{$\phantom{\tseq}$}
\RightLabel{$\topr$}
\UnaryInfC{$\tel \sar w : \top, \Gamma$}
\DisplayProof
\end{tabular}
\end{center}

\begin{center}
\begin{tabular}{c c}
\AxiomC{$\tel \sar w : A, \Gamma$}
\AxiomC{$\tel \sar w : B, \Gamma$}
\RightLabel{$\conr$}
\BinaryInfC{$\tel \sar w : A \land B, \Gamma$}
\DisplayProof

&

\AxiomC{$\tel \sar w : A, w : B, \Gamma$}
\RightLabel{$\disr$}
\UnaryInfC{$\tel \sar w : A \lor B, \Gamma$}
\DisplayProof
\end{tabular}
\end{center}

\begin{center}
\begin{tabular}{c c c}
\AxiomC{$\tel, w R_{\charx} u \sar u : A, \Gamma$}
\RightLabel{$\xbox^{\dag_{1}}$}
\UnaryInfC{$\tel \sar w : \xbox A, \Gamma$}
\DisplayProof

&

\AxiomC{$\tel \sar w : \xdia A, u : A, \Gamma$}
\RightLabel{$\xdia^{\dag_{2}(\fcons)}$}
\UnaryInfC{$\tel \sar w : \xdia A, \Gamma$}
\DisplayProof

& 

\AxiomC{$\tel, w R_{\charx} u \sar \Gamma$}
\RightLabel{$\ddr^{\dag_{3}}$}
\UnaryInfC{$\tel \sar \Gamma$}
\DisplayProof
\end{tabular}
\end{center}

\begin{flushleft}
\textbf{Side Conditions:}\\
$\dag_{1} :=$ $u$ is fresh.\\
$\dag_{2}(\fcons):= \prgr{\tel \sar w : \xdia A, \Gamma} \models w \prpath{L_{\charx}} u$ with $L_{\charx} = \glang(\charx)$.\\
$\dag_{3} :=$ $u$ is fresh and $w \in \lab(\tel,\Gamma)$.
\end{flushleft}

\caption{The tree sequent calculus $\tkmc$ defined relative to a set $\fcons$ of path and seriality conditions. We have an $\xbox$ rule and an $\xdia$ rule for each $\charx \in \albet$, and a $\ddr$ rule for each $\charx \in \albet$ such that $\D \in \fcons$.\label{fig:tree-calculus}}
\end{figure}

A uniform presentation of our tree sequent systems is provided in \fig~\ref{fig:tree-calculus}. Each tree sequent system $\tkmc$ takes a set $\fcons$ of path and seriality conditions as a parameter and is sound and complete for the logic $\kmc$ (see Theorem~\ref{thm:tkma-sound-compl} below). We note that $\mathsf{TK_{m}} = \mathsf{TK_{m}}(\emptyset)$ is the calculus for the base logic $\km$. Each system contains the \emph{initial rules} $\id$ and $\top$, and we call the conclusion of an initial rule an \emph{initial sequent}. With the exception of the \emph{seriality rules} $\ddr$, which are \emph{structural rules} that (bottom-up) introduce a relational atom without introducing any formula, the remaining rules are \emph{logical rules}, which introduce complex formulae into the conclusion. 

The $\xbox$ rule is subject to the side condition $\dag_{1}$, which requires that the label $u$ be \emph{fresh} in any application of the rule, i.e., the label $u$ does not occur in the conclusion. We remark that the freshness condition on $u$ preserves the tree structure of sequents when applied bottom-up: since $u$ does not occur in the conclusion, adjoining the relational atom $w R_{\charx} u$ extends the tree by a fresh leaf rather than introducing a cycle or disconnected region. Likewise, the $\ddr$ rule is subject to the side condition $\dag_{3}$, which requires that $u$ be fresh and that $w$ occur in the conclusion, so that (bottom-up) applications of $\ddr$ also extend the tree by a fresh leaf $u$, which is a child of $w$.

As in the setting of tree (or nested) sequents~\cite{Lyo21thesis,Lyo26,TiuIanGor12}, the parameter $\fcons$ determines what logic is captured by the system $\tkmc$. This is achieved by (1) including the seriality rule $\ddr$ in $\tkmc$ \iffi $\D \in \fcons$, and (2) changing the functionality of the propagation rules $\xdia$ according to the path conditions in $\fcons$. Propagation rules view tree sequents as labeled transition systems, and enable formulae to be (bottom-up) propagated along certain paths (corresponding to strings generated by a $\fcons$-grammar) within the tree of a tree sequent. To make the functionality of such rules precise, we define \emph{propagation graphs} and \emph{propagation paths} (cf.~\cite{CiaLyoRamTiu21,GorPosTiu11,Lyo21thesis}).

\begin{definition}[Propagation Graph]\label{def:propagation-graph} Let $\tseq = \tel \sar \Gamma$ be a tree sequent. We define the \emph{propagation graph} $\prgr{\tseq} = (\prv,\pre)$ such that
\begin{itemize}

\item $\prv := \lab(\tseq)$;

\item $(w, \charx, u), (u, \conv{\charx}, w) \in \pre$ \iffi $w R_{\charx} u \in \tel$.

\end{itemize}
We may write $w \in \prgr{\tseq}$ and $(w,\charx,u) \in \prgr{\tseq}$ as shorthand for $w \in \prv$ and $(w,\charx,u) \in \pre$, respectively.
\end{definition}

\begin{definition}[Propagation Path]\label{def:propagation-path} Let $\prgr{\tseq} = (\prv,\pre)$ be a propagation graph with $u,w \in \prv$ and $\charx \in \albet$. We write $\prgr{\tseq} \models u \prpath{\charx} w$ \iffi $(u,\charx,w) \in \pre$. Given a string $\charx\stra \in \albetstr$, we define $\prgr{\tseq} \models u \prpath{\charx\stra} w$ as `there exists a $v \in \prv$ such that $\prgr{\tseq} \models u\prpath{\charx} v$ and $\prgr{\tseq} \models v \prpath{\stra} w$', and we take $\prgr{\tseq} \models u \prpath{\empstr} w$ to mean that $u = w$. Additionally, when $\prgr{\tseq}$ is clear from the context we may simply write $u \prpath{\stra} w$ to express $\prgr{\tseq} \models u \prpath{\stra} w$. Finally, given a language $L_{\charx} := \glang(\charx)$ of some $\fcons$-grammar $\g{\fcons}$ and $\charx \in \albet$, we write $u \prpath{L_{\charx}} w$ \iffi there is a string $\strb \in \glang(\charx)$ with $u \prpath{\strb} w$.
\end{definition}

Observe that the premise and conclusion of a $\xdia$ rule have the same propagation graph, as they contain the same relational atoms and labels. Hence, propagation rules can be just as easily applied bottom-up as top-down. The following is a consequence of \lem~\ref{lem:string-and-inverse}.

\begin{lemma}\label{lem:reversing-paths}
Let $\tseq$ be a tree sequent, $L_{\charx} = \glang(\charx)$, and $L_{\conv{\charx}} = \glang(\conv{\charx})$. Then, $\prgr{\tseq} \models u \prpath{L_{\charx}} w$ \iffi $\prgr{\tseq} \models w \prpath{L_{\conv{\charx}}} u$.
\end{lemma}

With the above notions, one can formally specify each propagation rule $\xdia$. Applications of such rules are controlled by the side condition $\dag_{2}(\fcons)$ shown in \fig~\ref{fig:tree-calculus}. Since $\charx \in L_{\charx}$, the side condition is satisfied whenever $w R_{\charx} u$ occurs in the conclusion, meaning that each $\xdia$ rule subsumes the usual diamond rule of labeled and nested calculi for $\km$ (cf.~\cite[Lemma~25]{Lyo21thesis}). Moreover, if $\empstr \in L_{\charx}$ (e.g., if $\T \in \fcons$), then the side condition holds for $u = w$, allowing $A$ to be propagated to the same label where $\xdia A$ occurs. We provide an example to clarify applications of these rules.

\begin{example}\label{ex:propagation-graph-path}
Let $\tseq = w R_{\charx} u, u R_{\conv{\chary}} v \sar w : \inot p, v : \diap{\charz} p$. A graphical depiction of $\tseq$'s propagation graph $\prgr{\tseq}$ is given below:
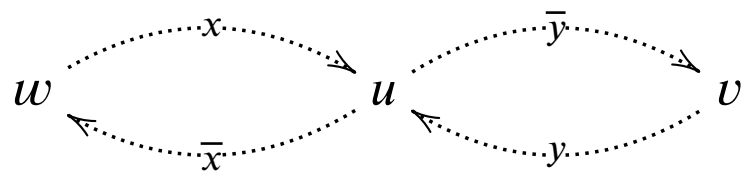
\begin{center}
\begin{minipage}[t]{.5\textwidth}
\xymatrix{
  w \ar@/^1pc/@{.>}[rr]|-{\charx} & &
  u \ar@/^1pc/@{.>}[rr]|-{\conv{\chary}}
    \ar@/^1pc/@{.>}[ll]|-{\conv{\charx}} & &
  v \ar@/^1pc/@{.>}[ll]|-{\chary}
}
\end{minipage}
\end{center}
Let $\fcons := \{\R_{\chary\conv{\charx}} \subseteq \R_{\charz}\}$, so that the corresponding $\fcons$-grammar is $\g{\fcons} = \{\charz \pto \chary \conv{\charx}, \conv{\charz} \pto \charx \conv{\chary} \}$. Then, $\prgr{\tseq} \models v \prpath{\chary\conv{\charx}} w$ with $\chary \conv{\charx} \in \glang(\charz)$ due to the first production rule of $\g{\fcons}$. Therefore, the side condition of the propagation rule $\diap{\charz}$ is satisfied, and so, we are permitted to (bottom-up) apply the rule to $\tseq$ to propagate the formula $p$ from $v$ to $w$. This yields the initial sequent $w R_{\charx} u, u R_{\conv{\chary}} v \sar w : \inot p, w : p, v : \diap{\charz} p$, showing that $\tseq$ is provable. 
\end{example}

As usual, we refer to the labeled formulae and relational atoms explicitly presented in the premises of a rule as \emph{auxiliary}, and to the labeled formulae explicitly presented in its conclusion as \emph{principal}. 
For example, in the $\xbox$ rule, $w R_{\charx} u$ and $u : A$ are auxiliary and $w : \xbox A$ is principal, whereas in the $\ddr$ rule, $w R_{\charx} u$ is auxiliary and there is no principal formula. A \emph{derivation} in $\tkmc$ of a tree sequent $\tseq$ is a (potentially infinite) tree whose nodes are labeled with tree sequents such that: (1) The root is labeled with $\tseq$, and (2) Every parent node is the conclusion of a rule of $\tkmc$ with its children the premises.

We define a \emph{branch} $\branch = \tseq_{0}, \tseq_{1}, \ldots, \tseq_{n}, \ldots$ to be a maximal path of tree sequents in a derivation such that $\tseq_{0}$ is the conclusion of the derivation and each tree sequent $\tseq_{i+1}$ (if it exists) is a child of $\tseq_{i}$. A \emph{proof} is a finite derivation where all leaves are initial sequents. We use $\prf$ and annotated versions thereof to denote both derivations and proofs with the context differentiating the usage. If a proof of a tree sequent $\tseq$ exists in $\tkmc$, then we write $\tkmc \proves \tseq$ to indicate this. Often, when discussing a proof transformation, we will use the following notation to denote that $\prf$ is a proof of $\tseq$:
\begin{center}
\AxiomC{$\prf$} 
\noLine
\UnaryInfC{$\tseq$}
\DisplayProof
\end{center}
We define $\lab(\prf)$ to be the set of all labels occurring in a derivation $\prf$. Last, the \emph{height} of a derivation is defined in the usual way as the maximal length of a branch in the derivation, which may be infinite if the derivation is infinite.

\begin{example}\label{ex:tree-proofs} We provide an example proof of a path axiom. Let $(\R_{\charx_{1} \cdots \charx_{n}} \subseteq \R_{\charx}) \in \fcons$ with $n \geq 1$, meaning $\charx \pto \charx_{1}\cdots\charx_{n} \in \g{\fcons}$ by definition. Recall that $\negnnf{(\diap{\charx_{1}} \cdots \diap{\charx_{n}} p)} = \boxp{\charx_{1}} \cdots \boxp{\charx_{n}} \inot p$, so the end sequent below is the path axiom $\diap{\charx_{1}} \cdots \diap{\charx_{n}} p \imp \diap{\charx} p$. Observe that a propagation path tracing the string $\charx_{1}\cdots\charx_{n}$ exists from $w$ to $u_{n}$, showing that the side condition holds for $\xdia$ as $\charx_{1}\cdots\charx_{n} \in \glang(\charx)$.
\begin{center}
\AxiomC{}
\RightLabel{$\id$}
\UnaryInfC{$w R_{\charx_{1}} u_{1}, \ldots, u_{n-1} R_{\charx_{n}} u_{n} \sar w : \xdia p, u_{n} : \inot p, u_{n} : p$}
\RightLabel{$\xdia$}
\UnaryInfC{$w R_{\charx_{1}} u_{1}, \ldots, u_{n-1} R_{\charx_{n}} u_{n} \sar w : \xdia p, u_{n} : \inot p$}
\RightLabel{$\boxp{\charx_{n}}, \ldots, \boxp{\charx_{1}}$}
\UnaryInfC{$\sar w : \boxp{\charx_{1}} \cdots \boxp{\charx_{n}} \inot p, w : \xdia p$}
\RightLabel{$\disr$}
\UnaryInfC{$\sar w : \boxp{\charx_{1}} \cdots \boxp{\charx_{n}} \inot p \lor \xdia p$}
\DisplayProof
\end{center}
When $n = 0$, i.e., $\T \in \fcons$, the corresponding path axiom $p \imp \xdia p$ is proven by applying $\lor$ and then $\xdia$ with $u = w$, which yields the initial sequent $\sar w : \inot p, w : \xdia p, w : p$.
\end{example}

The soundness and completeness of $\tkmc$ were established in~\cite{Lyo21thesis} for sets $\fcons$ of path conditions (cf.~\lem~\ref{lem:path-relation}), and extends to seriality conditions by standard techniques. 

\begin{theorem}[Soundness and Completeness, cf.~\cite{Lyo21thesis,Lyo26}]\label{thm:tkma-sound-compl} For any tree sequent $\tseq$, $\tkmc \proves \tseq$ \iffi $\tseq$ is $\fcons$-valid, i.e., $\fm(\tseq) \in \kmc$.
\end{theorem}

\begin{figure}[pos=t]
\noindent

\begin{center}
\begin{tabular}{c c c}
\AxiomC{$\tel \sar \Gamma$}
\RightLabel{$\wk$}
\UnaryInfC{$\tel, \tel' \sar \Gamma, \Gamma'$}
\DisplayProof

&

\AxiomC{$\tel \sar w : A, w : A, \Gamma$}
\RightLabel{$\ctr$}
\UnaryInfC{$\tel \sar w : A, \Gamma$}
\DisplayProof

&

\AxiomC{$\tel \sar w : A, \Gamma$}
\AxiomC{$\tel \sar w : \negnnf{A}, \Gamma$}
\RightLabel{$\cut$}
\BinaryInfC{$\tel \sar \Gamma$}
\DisplayProof
\end{tabular}
\end{center}

\caption{Admissible rules in $\tkmc$. The conclusion of $\wk$ is required to be a tree sequent.\label{fig:tree-admissible}}
\end{figure}

The interpolation algorithm presented later on relies on several fundamental properties of our tree sequent systems, which we collect in this section. We define a rule to be \emph{admissible} (\emph{height-preserving admissible}) \iffi the following holds: if the premises of the rule have proofs (of height $h_{1}, \ldots, h_{n}$), then the conclusion of the rule has a proof (of height $h \leq \max\{h_{1}, \ldots, h_{n}\}$). Given an $n$-ary rule $\ru$, we define its \emph{$i$-inverse}, denoted $\ru^{-1}_{i}$, to be the rule whose sole premise is the conclusion of $\ru$ and whose conclusion is the $i^{th}$ premise of $\ru$. We say that $\ru$ is \emph{(height-preserving) invertible} \iffi $\ru^{-1}_{i}$ is (height-preserving) admissible for each $i \in [n]$.

The weakening rule $\wk$, the contraction rule $\ctr$, and the cut rule $\cut$ are displayed in \fig~\ref{fig:tree-admissible}. We note that in the setting of tree sequents, the $\wk$ rule is assumed to preserve the tree shape of sequents when applied, i.e., its conclusion must be a tree sequent. Since tree sequents contain multisets of labeled formulae, we also consider the $\ctr$ rule, which merges two copies of a labeled formula into one; no such rule is needed for relational atoms, as these form a set. The following theorem is a consequence of the work in~\cite{Lyo21thesis}, whose arguments extend to the $\ddr$ rule by standard techniques (cf.~\cite{Lyo26}).

\begin{theorem}[Admissibility and Invertibility, cf.~\cite{Lyo21thesis,Lyo26}]\label{thm:tree-properties} The tree sequent calculus $\tkmc$ satisfies the following:
\begin{itemize}

\item[(1)] Each tree sequent of the form $\tel \sar w : A, w : \negnnf{A}, \Gamma$ is provable in $\tkmc$;

\item[(2)] The $\wk$ and $\ctr$ rules are height-preserving admissible in $\tkmc$;

\item[(3)] All non-initial rules are height-preserving invertible in $\tkmc$;

\item[(4)] The $\cut$ rule is admissible in $\tkmc$.

\end{itemize}
\end{theorem}

\section{Interpolation Calculus}\label{sec:inter-calc}

Over the next three sections, we leverage our tree sequent calculi to establish the strong LIP for bridge-free grammar logics. Our methodology generalizes that of~\cite{FitKuz15,LyoTiuGorClo20,Lyo21thesis,LyoKar24} for computing interpolants with nested and tree sequents, which is itself a generalization of Maehara's method~\cite{Mae60} and can be summarized as a four-step process:
\begin{itemize}

\item[$(1)$] Take a proof of a tree sequent $\sar w : \negnnf{A}, w : B$ as input. (NB. As $A \imp B = \negnnf{A} \lor B$, such a proof exists \iffi $\sar w : A \imp B$ is provable, by the $\disr$ rule and its invertibility; see Theorem~\ref{thm:tree-properties}(3).)

\item[$(2)$] Partition the conclusion as $\sar w : \negnnf{A} \mid w : B$ so that $w : \negnnf{A}$ occurs in the left and $w : B$ occurs in the right of the partition with `$\mid$' separating the partitions. 

\item[$(3)$] Partition the premises of each rule based on the partition of the conclusion to obtain a `partitioned proof' whereby the left partition derives $w : \negnnf{A}$ and the right partition derives $w : B$. 

\item[$(4)$] Assign interpolants to the initial sequents of the proof and specify how the interpolants of the premises of a rule determine an interpolant of the conclusion.

\end{itemize}
This culminates in the construction of an interpolant $w : I$ of the conclusion, yielding a proof of $\sar w : \negnnf{A}, w : I$ (i.e., of $w : A \imp I$) and a proof of $\sar w : \negnnf{I}, w : B$ (i.e., of $w : I \imp B$).

One of the features required for this method to work with tree (or nested) sequents is the generalization of interpolants from formulae to sets of sequents~\cite{LyoTiuGorClo20}. In our setting, the interpolants assigned to sequents in step (4) are sets of sets of labeled formulae (see \dfn~\ref{def:interpolant}), which can be interpreted as a single formula at the conclusion. 

To make the above process concrete, let us illustrate the Lyndon interpolation method of~\cite{Lyo21thesis} by means of an example, which also exposes its shortcoming in obtaining strong interpolants that account for the indices of modalities. Suppose we are given a proof of $\sar w : \xdia \notp, w : \xbox p \lor \ybox(q \lor \inot q)$ (i.e., of $w : \xbox p \imp \xbox p \lor \ybox(q \lor \inot q)$) as shown below left. We partition the conclusion as shown in the proof below right. The entire proof is then bottom-up partitioned by placing the auxiliary labeled formulae of an inference in the same partition as the principal formula. Note that in this method relational atoms are \emph{independent} of the partition and are shared by both partitions.
\begin{center}
\small
\begin{tabular}{@{}c@{\hskip 1.5em}c@{}}
\AxiomC{}
\RightLabel{$\id$}
\UnaryInfC{$w R_{\chary} u \sar w : \xdia \notp, w : \xbox p, u : q, u : \inot q$}
\RightLabel{$\disr$}
\UnaryInfC{$w R_{\chary} u \sar w : \xdia \notp, w : \xbox p, u : q \lor \inot q$}
\RightLabel{$\ybox$}
\UnaryInfC{$\sar w : \xdia \notp, w : \xbox p, w : \ybox(q \lor \inot q)$}
\RightLabel{$\disr$}
\UnaryInfC{$\sar w : \xdia \notp, w : \xbox p \lor \ybox(q \lor \inot q)$}
\DisplayProof

&

\AxiomC{}
\RightLabel{$\id$}
\UnaryInfC{$w R_{\chary} u \sar w : \xdia \notp \mid w : \xbox p, u : q, u : \inot q$}
\RightLabel{$\disr$}
\UnaryInfC{$w R_{\chary} u \sar w : \xdia \notp \mid w : \xbox p, u : q \lor \inot q$}
\RightLabel{$\ybox$}
\UnaryInfC{$\sar w : \xdia \notp \mid w : \xbox p, w : \ybox(q \lor \inot q)$}
\RightLabel{$\disr$}
\UnaryInfC{$\sar w : \xdia \notp \mid w : \xbox p \lor \ybox(q \lor \inot q)$}
\DisplayProof
\end{tabular}
\end{center}

As shown below, we can assign a suitable interpolant to the $\id$ rule, viz., $\{(u : \top)\}$, and an interpolant of the conclusion of a rule can be defined relative to the interpolants of its premises. We note that one can consult~\cite[Section~6.2]{Lyo21thesis} for the explicit description of how Lyndon interpolants for grammar logics are computed; this example is simply intended to give a high-level idea of how proofs are interpolated. What is important to recognize is that $\xbox p \imp \ybox \top$ and $\ybox \top \imp \xbox p \lor \ybox(q \lor \inot q)$ are valid, showing that $\ybox \top$ `interpolates' the input implication.
\begin{center}
\AxiomC{}
\RightLabel{$\id$}
\UnaryInfC{$w R_{\chary} u \sar w : \xdia \notp \mid w : \xbox p, u : q, u : \inot q \sep \{(u : \top)\}$}
\RightLabel{$\disr$}
\UnaryInfC{$w R_{\chary} u \sar w : \xdia \notp \mid w : \xbox p, u : q \lor \inot q \sep \{(u : \top)\}$}
\RightLabel{$\ybox$}
\UnaryInfC{$\sar w : \xdia \notp \mid w : \xbox p, w : \ybox(q \lor \inot q) \sep \{(w : \ybox \top)\}$}
\RightLabel{$\disr$}
\UnaryInfC{$\sar w : \xdia \notp \mid w : \xbox p \lor \ybox(q \lor \inot q) \sep \{(w : \ybox \top)\}$}
\DisplayProof
\end{center}

However, the interpolant produced above does not satisfy condition (2) of \dfn~\ref{def:strong-lip}, i.e., the above methodology cannot guarantee that the interpolant lies in the common modal signature. Indeed, as also recognized by B{\'\i}lkov{\'a} et al.~\cite[Example~22]{BilFusKuz25}, the usual interpolation method on nested and tree sequents (cf.~\cite{FitKuz15,LyoTiuGorClo20,Lyo21thesis,LyoKar24}) only guarantees that the interpolant is built from \emph{common} (polarized) propositional atoms. In the above example, the interpolant contains the character $\chary$, introduced by the $\ybox$ rule, even though $\sigm{\ybox \top} = \{\chary, \conv{\chary}\} \not\subseteq \sigm{\xbox p} = \{\charx, \conv{\charx}\}$.

This issue arises in step (3) of the above method, that is, the usual `bipartite' partition method is not sensitive enough to which relational atoms should and should not contribute modalities to the interpolant. We overcome this issue by means of a novel partitioning technique, which, besides partitioning the labeled formulae, also sorts the labels of each sequent into three sets: the labels private to the left partition, the labels private to the right partition, and the \emph{common labels}, which both partitions may use. (NB. We refer to the left and right partitions as the \emph{left component} and \emph{right component} below.) 

A relational atom pointing to a common label may be used to propagate formulae in \emph{both} partitions; as we will see, in bridge-free grammar logics, the index of such a relational atom is guaranteed to occur in the modal signature of both the antecedent and consequent of the input implication whenever both partitions actually use it. Hence, only relational atoms pointing to common labels are permitted to contribute modalities to interpolants, and only when used by both partitions. We therefore refer to our new partitioning method as a \emph{tripartite partition method} (detailed in \sect~\ref{sec:part-bias}), which is the key technique for establishing the strong LIP for bridge-free grammar logics.

\subsection{Interpolation Sequents}\label{subsec:interp-sequents}

Our interpolation method uses \emph{interpolation sequents}, which are generalizations of tree sequents that distribute their labeled formulae over a left and a right component, and their labels over three sets: the common labels, and the labels private to the left or to the right component. We let $\set{\tagl, \tagr}$ be the set of \emph{tags}, which name the left and right components, and use $\tagx$, $\tagy$, $\tagz$ to range over tags.

We define $\sigb{w : A} := \sigb{A}$ for each $\isym \in \set{+,-,\albet}$, and for a multiset $\Gamma$ of labeled formulae, we let $\sigb{\Gamma}$ be the union of the signatures of its elements.

\begin{definition}[Interpolant]\label{def:interpolant} An \emph{interpolant} $\itp$ is a finite set $\set{ \Sigma_{1}, \ldots, \Sigma_{n}}$, where $\Sigma_{i}$ is a finite set of labeled formulae for each $i \in [n]$. We let $\lab(\itp)$ be the set of all labels occurring in $\itp$, and for $\isym \in \set{+,-,\albet}$, we define $\sigb{\itp} := \bigcup_{i \in [n]} \sigb{\Sigma_{i}}$.
\end{definition}

We stress that $\Sigma \in \itp$ (for an interpolant $\itp$) is in general not a tree sequent: $\Sigma$ contains no relational atoms and its labeled formulae may have distinct labels. Our interpolation algorithm is designed to ensure that all labeled formulae in an interpolant share the same label in the conclusion of an interpolation proof, so that the interpolant of the conclusion can be read as a formula, namely, a strong Lyndon interpolant (see \dfn~\ref{def:itp-formula}). 

As discussed in~\cite{LyoTiuGorClo20}, since interpolants are no longer formulae, we need a generalized notion of negation for them. The following definition suffices and defines the \emph{orthogonal} (i.e., negation) of an interpolant (cf.~\cite{LyoTiuGorClo20,Lyo21thesis,LyoKar24}).

\begin{definition}[Orthogonal~\cite{LyoTiuGorClo20}]\label{def:orthogonal} The \emph{orthogonal} of an interpolant $\itp = \{ \Sigma_{1}, \ldots, \Sigma_{n}\}$ is defined as follows:
$$
\orth{\itp} := \{w_{1} : \negnnf{A}_{1}, \ldots, w_{n} : \negnnf{A}_{n} \mid w_{i} : A_{i} \in \Sigma_{i} \mbox{ for each } i \in [n]\}.
$$
Intuitively, the orthogonal selects one labeled formula from each $\Sigma_i$, negates it, and collects the negations into a set, ranging over all possible such selections.
\end{definition}

\begin{example} If $\itp = \{(w : A), (w : B, u : C)\}$ is an interpolant, then its orthogonal is:
$$
\orth{\itp} = \{(w : \negnnf{A}, w : \negnnf{B}), (w : \negnnf{A}, u : \negnnf{C})\}.
$$
\end{example}

Note that the empty set $\emptyset$ is an interpolant as well, whose orthogonal is $\orth{\emptyset} = \set{\emptyset}$, and that $\orth{\set{\emptyset}} = \emptyset$. When interpolants are read as formulae in conjunctive normal form (\dfn~\ref{def:itp-formula}), the interpolant $\emptyset$ corresponds to $\top$ (the empty conjunction) and $\set{\emptyset}$ to $\bot$ (a conjunction whose only conjunct is the empty disjunction).

By \lem~\ref{lem:signature-negation}, the orthogonal swaps the positive and negative signature of an interpolant, i.e., $\sigp{\orth{\itp}} \subseteq \sign{\itp}$, $\sign{\orth{\itp}} \subseteq \sigp{\itp}$, and $\sigm{\orth{\itp}} \subseteq \sigm{\itp}$; moreover, $\lab(\orth{\itp}) \subseteq \lab(\itp)$. We also define special interpolants that are used when modalities are introduced into interpolants. In the definition below, we use the fact that every set $\Sigma$ of labeled formulae can be uniquely written as $\Sigma = \Sigma', u : B_{1}, \ldots, u : B_{k}$ with $k \geq 0$ and $u \not\in \lab(\Sigma')$.

\begin{definition}[Box Interpolant]\label{def:box-interpolant} Let $\itp$ be an interpolant, $\charx \in \albet$, and $w, u \in \lab$ with $w \neq u$. We define the interpolant $\xbi$ as follows, where the empty disjunction is $\bot$:
$$
\xbi := \Big\{ \Sigma', w : \xbox \!\!\!\! \bigvee_{1 \leq i \leq k} \!\!\!\! B_{i} \ \Big| \ (\Sigma', u : B_{1}, \ldots, u : B_{k}) \in \itp \mbox{ with } u \not\in \lab(\Sigma') \Big\}.
$$
\end{definition}

Observe that $\xbi$ adds the labeled formula $w : \xbox \bot$ to every element of $\itp$ that contains no labeled formula with the label $u$. As illustrated in Example~\ref{ex:interp-proof-detour}, this is necessary for the correctness of our method.

For a set $\tel$ of relational atoms and a set $V$ of labels, $\tel \rstr V := \set{w R_{\charx} u \in \tel \mid u \in V}$ is the set of relational atoms of $\tel$ whose target belongs to $V$.

\begin{definition}[Interpolation Sequent]\label{def:interp-sequent} We define an \emph{interpolation sequent} to be an expression of the form
$$
\iseq = \lsxy \sep \tel \sar \Gamma_{1} \splt \Gamma_{2} \sep \itp
$$
such that $\tagx, \tagy \in \set{\tagl,\tagr}$ with $\tagx \neq \tagy$, and:
\begin{itemize}

\item[(1)] $\tel \sar \Gamma_{1}, \Gamma_{2}$ is a tree sequent, called the \emph{underlying tree sequent} of $\iseq$;

\item[(2)] $\cst$, $\pst{\tagx}$, and $\pst{\tagy}$ are pairwise disjoint sets of labels whose union is the set of labels of the underlying tree sequent, the root of the underlying tree sequent belongs to $\cst$, and for $\tagz \in \set{\tagx, \tagy}$, every child of a label in $\pst{\tagz}$ belongs to $\pst{\tagz}$;

\item[(3)] every label of $\Gamma_{1}$ belongs to $\cst \cup \pst{\tagx}$, and every label of $\Gamma_{2}$ belongs to $\cst \cup \pst{\tagy}$;

\item[(4)] $\itp$ is an interpolant with $\lab(\itp) \subseteq \cst$.

\end{itemize}
\end{definition}

\begin{figure}[pos=t]
\noindent

\begin{center}
\AxiomC{}
\RightLabel{$\idIi$}
\UnaryInfC{$\lsets{\cst \cup \set{w}}{\pst{\tagx}}{\pst{\tagy}} \sep \tel \sar \Gamma_{1}, w : p \splt w : \notp, \Gamma_{2} \sep \{(w : \notp)\}$}
\DisplayProof
\end{center}

\begin{center}
\AxiomC{}
\RightLabel{$\idIii$}
\UnaryInfC{$\lsxy \sep \tel \sar \Gamma_{1} \splt \Gamma_{2}, w : p, w : \notp \sep \emptyset$}
\DisplayProof
\end{center}

\begin{center}
\AxiomC{}
\RightLabel{$\toprI$}
\UnaryInfC{$\lsxy \sep \tel \sar \Gamma_{1} \splt \Gamma_{2}, w : \top \sep \emptyset$}
\DisplayProof
\end{center}

\begin{center}
\AxiomC{$\lsxy \sep \tel \sar \Gamma_{1} \splt \Gamma_{2}, w : A, w : B \sep \itp$}
\RightLabel{$\disrI$}
\UnaryInfC{$\lsxy \sep \tel \sar \Gamma_{1} \splt \Gamma_{2}, w : A \lor B \sep \itp$}
\DisplayProof
\end{center}

\begin{center}
\AxiomC{$\lsxy \sep \tel \sar \Gamma_{1} \splt \Gamma_{2}, w : A \sep \itp_{1}$}
\AxiomC{$\lsxy \sep \tel \sar \Gamma_{1} \splt \Gamma_{2}, w : B \sep \itp_{2}$}
\RightLabel{$\conrI$}
\BinaryInfC{$\lsxy \sep \tel \sar \Gamma_{1} \splt \Gamma_{2}, w : A \land B \sep \itp_{1} \cup \itp_{2}$}
\DisplayProof
\end{center}

\begin{center}
\AxiomC{$\lsyx \sep \tel \sar \Gamma_{2} \spltrev \Gamma_{1} \sep \itp$}
\RightLabel{$\orthru$}
\UnaryInfC{$\lsxy \sep \tel \sar \Gamma_{1} \splt \Gamma_{2} \sep \orth{\itp}$}
\DisplayProof
\end{center}

\caption{Initial and propositional rules, and the orthogonal rule of $\tkmci$.\label{fig:interpolation-calc}}
\end{figure}

We use $\iseq$ (potentially annotated) to denote interpolation sequents. The tagged bar `$\splt$' separates the two components of $\iseq$: we refer to $\Gamma_{1}$ as the \emph{$\tagx$-component} and to $\Gamma_{2}$ as the \emph{$\tagy$-component} of $\iseq$. If $\tagx = \tagl$ (resp.\ $\tagx = \tagr$), then we also refer to the $\tagx$-component as the \emph{left component} (resp.\ \emph{right component}), and similarly for the $\tagy$-component. The labels in $\cst$ are \emph{common}, and the labels in $\pst{\tagz}$ are \emph{private} to the $\tagz$-component. Intuitively, both components may use common labels, whereas only the $\tagz$-component may use the labels private to it. 

We list the two sets of private labels in the same order as the components, i.e., first the labels private to the $\tagx$-component and then those private to the $\tagy$-component; in particular, an interpolation sequent with the tagged bar `$\splti$' lists its labels as $\lsets{\cst}{\lst}{\rst}$. In rules, a set written as $\cst \cup \set{w}$ indicates that $w$ belongs to it. 

The \emph{root} of an interpolation sequent, denoted by the meta-variable $\rt$, is the root of its underlying tree sequent. For example, the root of
$$
\lsets{\set{w, u}}{\set{v}}{\set{v'}} \sep w R_{\charx} u, u R_{\chary} v, u R_{\charz} v' \sar \Gamma_{1} \splti \Gamma_{2} \sep \itp
$$
is $w$, where $v$ is private to the left component and $v'$ is private to the right component. Condition (4) states that interpolants only refer to common labels; it is preserved by all rules of our interpolation calculus (\lem~\ref{lem:preservation} below).

Each component of an interpolation sequent can only `see' the common labels and its own private labels, which gives rise to the following notion.

\begin{definition}[Projections]\label{def:projections} For an interpolation sequent $\iseq$, the \emph{$\tagx$-projection} and the \emph{$\tagy$-projection} of $\iseq$ are defined as follows:
$$
\iseq|_{\tagx} := \tel \rstr (\cst \cup \pst{\tagx}) \sar \Gamma_{1}
\text{ and }
\iseq|_{\tagy} := \tel \rstr (\cst \cup \pst{\tagy}) \sar \Gamma_{2}.
$$
\end{definition}

That is, the $\tagz$-projection consists of the labeled formulae of the $\tagz$-component together with all relational atoms that do not point to a label private to the other component.

\begin{lemma}\label{lem:projections}
Let $\iseq$ be an interpolation sequent with root $\rt$. Then, $\iseq|_{\tagx}$ and $\iseq|_{\tagy}$ are tree sequents with root $\rt$, and $\cst$ is the set of labels occurring in both of them.
\end{lemma}

\begin{proof} Let $\tagz \in \set{\tagx, \tagy}$. By condition (2) of \dfn~\ref{def:interp-sequent}, the parent of a label in $\cst \cup \pst{\tagz}$ belongs to $\cst \cup \pst{\tagz}$ as well, and $\rt \in \cst$. Hence, $\tel \rstr (\cst \cup \pst{\tagz})$ is the subtree of the tree formed by $\tel$ whose labels are precisely those in $\cst \cup \pst{\tagz}$, which contains $\rt$. By condition (3), every labeled formula of the $\tagz$-component has a label in this subtree, so $\iseq|_{\tagz}$ is a tree sequent with root $\rt$. Last, $(\cst \cup \pst{\tagx}) \cap (\cst \cup \pst{\tagy}) = \cst$.
\end{proof}

The intended meaning of an interpolation sequent is that the interpolant $\itp$ is implied by the $\tagx$-projection and implies the $\tagy$-projection; more precisely, for the interpolation proofs constructed by our method, we will show the following in \sect~\ref{sec:main-theorem}:
\begin{itemize}

\item[(1)] for each $\Sigma \in \itp$, the tree sequent $\tel \rstr (\cst \cup \pst{\tagx}) \sar \Gamma_{1}, \Sigma$ is provable in $\tkmc$, and

\item[(2)] for each $\Sigma \in \orth{\itp}$, the tree sequent $\tel \rstr (\cst \cup \pst{\tagy}) \sar \Gamma_{2}, \Sigma$ is provable in $\tkmc$.

\end{itemize}
These are indeed tree sequents by \lem~\ref{lem:projections}, as $\lab(\Sigma) \subseteq \cst$ by condition (4) of \dfn~\ref{def:interp-sequent} and since $\lab(\orth{\itp}) \subseteq \lab(\itp)$.

\subsection{The Interpolation Calculus}\label{subsec:interp-calculus}

\begin{figure}[pos=t]
\noindent

\begin{center}
\AxiomC{$\lsxy \sep \tel \sar \Gamma_{1} \splt \Gamma_{2}, w : \xdia A, u : A \sep \itp$}
\RightLabel{${\xdiarI}^{\ddag(\fcons)}$}
\UnaryInfC{$\lsxy \sep \tel \sar \Gamma_{1} \splt \Gamma_{2}, w : \xdia A \sep \itp$}
\DisplayProof
\end{center}

\begin{center}
\AxiomC{$\lsets{\cst \cup \set{w, u}}{\pst{\tagx}}{\pst{\tagy}} \sep \tel, w R_{\charx} u \sar \Gamma_{1} \splt \Gamma_{2}, u : A \sep \itp$}
\RightLabel{${\xboxrIc}^{\dag}$}
\UnaryInfC{$\lsets{\cst \cup \set{w}}{\pst{\tagx}}{\pst{\tagy}} \sep \tel \sar \Gamma_{1} \splt \Gamma_{2}, w : \xbox A \sep \xbi$}
\DisplayProof
\end{center}

\begin{center}
\AxiomC{$\lsets{\cst}{\pst{\tagx}}{\pst{\tagy} \cup \set{u}} \sep \tel, w R_{\charx} u \sar \Gamma_{1} \splt \Gamma_{2}, u : A \sep \itp$}
\RightLabel{${\xboxrIf}^{\dag}$}
\UnaryInfC{$\lsxy \sep \tel \sar \Gamma_{1} \splt \Gamma_{2}, w : \xbox A \sep \itp$}
\DisplayProof
\end{center}

\begin{center}
\AxiomC{$\lsets{\cst \cup \set{w, u}}{\pst{\tagx}}{\pst{\tagy}} \sep \tel, w R_{\charx} u \sar \Gamma_{1} \splt \Gamma_{2} \sep \itp$}
\RightLabel{${\ddrIc}^{\dag}$}
\UnaryInfC{$\lsets{\cst \cup \set{w}}{\pst{\tagx}}{\pst{\tagy}} \sep \tel \sar \Gamma_{1} \splt \Gamma_{2} \sep \xbi$}
\DisplayProof
\end{center}

\begin{center}
\AxiomC{$\lsets{\cst}{\pst{\tagx}}{\pst{\tagy} \cup \set{u}} \sep \tel, w R_{\charx} u \sar \Gamma_{1} \splt \Gamma_{2} \sep \itp$}
\RightLabel{${\ddrIf}^{\dag}$}
\UnaryInfC{$\lsxy \sep \tel \sar \Gamma_{1} \splt \Gamma_{2} \sep \itp$}
\DisplayProof
\end{center}

\begin{flushleft}
\textbf{Side Conditions:}\\
$\ddag(\fcons) := \prgr{\iseq|_{\tagy}} \models w \prpath{L_{\charx}} u$ with $L_{\charx} = \glang(\charx)$ and $\iseq$ the conclusion.\\
$\dag :=$ $u$ is fresh, i.e., $u$ does not occur in the conclusion.
\end{flushleft}

\caption{Modal and seriality rules of $\tkmci$. The rules $\ddrIc$ and $\ddrIf$ are included \iffi $\D \in \fcons$.\label{fig:interpolation-calc-modal}}
\end{figure}

Our interpolation calculus $\tkmci$ consists of the rules shown in Figures~\ref{fig:interpolation-calc} and~\ref{fig:interpolation-calc-modal}, which include a rule $\xdiarI$, $\xboxrIc$, and $\xboxrIf$ for each $\charx \in \albet$, and the rules $\ddrIc$ and $\ddrIf$ for each $\charx \in \albet$ with $\D \in \fcons$. We refer to a proof in $\tkmci$ as an \emph{interpolation proof}, and use $\intprf$ and annotated versions thereof to denote such proofs. 

Observe that every rule of $\tkmci$ other than $\orthru$ acts on the $\tagy$-component of an interpolation sequent, leaving the $\tagx$-component untouched; the $\orthru$ rule lets us exchange the two components (together with their private labels), so that every rule is applicable to either side. This roughly halves the number of rules needed in the calculus. Moreover, disregarding the label sets, the partition, and all interpolants, every rule other than $\orthru$ is an instance of a rule of $\tkmc$.

The rules $\idIi$, $\idIii$, and $\toprI$ serve as the initial rules of the calculus. The $\idIi$ rule applies when the two complementary literals occur in distinct components, in which case their label $w$ is common and the literal of the $\tagy$-component serves as the interpolant. The $\idIii$ and $\toprI$ rules apply when the initial sequent is `internal' to the $\tagy$-component; in this case, the interpolant is the empty interpolant $\emptyset$, which corresponds to $\top$. 

The $\disrI$ and $\conrI$ rules are similar to their corresponding rules in $\tkmc$, but recast with interpolation sequents, and return either the interpolant of the premise, or the union of the interpolants of the premises, as the interpolant of the conclusion. The \emph{orthogonal rule} $\orthru$ is a special rule introduced in~\cite{LyoTiuGorClo20} and utilized in various interpolation systems for multi-modal and description logics~\cite{Lyo21thesis,LyoKar24}.

The label sets record which labels each component may use, and every rule respects them. In particular, interpolants only ever refer to common labels:

\begin{lemma}[Preservation]\label{lem:preservation}
If the premises of an instance of a rule of $\tkmci$ are interpolation sequents, then so is its conclusion. In particular, every rule of $\tkmci$ preserves the condition $\lab(\itp) \subseteq \cst$.
\end{lemma}

\begin{proof} Conditions (1)--(3) of \dfn~\ref{def:interp-sequent} are readily verified for each rule. For condition (4), the interpolant $\{(w : \notp)\}$ of $\idIi$ only contains the label $w$, which is common, and the interpolant $\emptyset$ of $\idIii$ and $\toprI$ contains no labels. The rules $\disrI$, $\conrI$, $\xdiarI$, $\xboxrIf$, and $\ddrIf$ neither change the set of common labels nor the labels of interpolants, and $\lab(\orth{\itp}) \subseteq \lab(\itp)$ for $\orthru$. Last, $\xbi$ replaces the label $u$ by $w$, so the labels of the interpolant of the conclusion of $\xboxrIc$ and $\ddrIc$ belong to $\cst \cup \set{w}$.
\end{proof}

Whereas the propositional rules have a one-to-one correspondence with rules in $\tkmc$, the modal and seriality rules do not; these rules are new and essential to our strengthened interpolation method. Their correctness relies on the following property of bridge-free grammar logics. We say that a propagation path \emph{traverses} a relational atom $u R_{\charx} v$ \iffi it contains the edge $(u, \charx, v)$ or the edge $(v, \conv{\charx}, u)$ (see \dfn~\ref{def:propagation-graph}). Since the relational atoms of a tree sequent form a tree, each edge of a propagation path stems from a unique relational atom.

\begin{lemma}\label{lem:bridge-free-lang}
Let $\kmc$ be bridge-free. If a propagation path whose string belongs to $\glang(\charx)$ traverses a relational atom $u R_{\chary} v$, then $\chary \in \set{\charx, \conv{\charx}}$.
\end{lemma}

\begin{proof} As $\kmc$ is bridge-free, the string of the propagation path is in $\set{\charx, \conv{\charx}}^{*}$ by \dfn~\ref{def:bridge-free}. As the path traverses $u R_{\chary} v$, its string contains $\chary$ or $\conv{\chary}$, and so, $\chary \in \set{\charx, \conv{\charx}}$.
\end{proof}

The \emph{diamond rule} $\xdiarI$ operates like the $\xdia$ rule of $\tkmc$: it (bottom-up) propagates the formula $A$ from $w$ to $u$ along a propagation path whose string belongs to $L_{\charx}$. The path must exist in the propagation graph of the $\tagy$-projection of the conclusion, that is, it may pass through common labels and labels private to the $\tagy$-component, but not through labels private to the $\tagx$-component. Hence, a formula of the $\tagy$-component is only propagated to labels that the $\tagy$-component may use.

The \emph{box rules} determine whether the label $u$ introduced (bottom-up) by a boxed formula $w : \xbox A$ of the $\tagy$-component becomes common or private to the $\tagy$-component. The \emph{common box rule} $\xboxrIc$ requires $w$ to be common and makes $u$ common, and is applied when the relational atom $w R_{\charx} u$ is also used by the $\tagx$-component, i.e., when the propagation path of some $\diap{\chary}^{\ilet}$ inference whose principal formula lies in the $\tagx$-component traverses $w R_{\charx} u$. By \lem~\ref{lem:bridge-free-lang}, we then have $\charx \in \set{\chary, \conv{\chary}}$, that is, the index of $w R_{\charx} u$ occurs in the modal signature of both components. It is therefore `safe' for $\xboxrIc$ to introduce the modality $\xbox$ into the interpolant by means of the box interpolant $\xbi$. This is precisely where bridge-freeness is needed: if, for instance, $(\R_{\charb} \subseteq \R_{\chara}) \in \fcons$, then an $\adia$-formula of one component could propagate along relational atoms indexed with $\charb$ that were introduced by the other component, so that modalities indexed with $\charb$ would enter the interpolant even though $\charb$ need not occur in both components (cf. \sect~\ref{subsec:strong-lip}).

The \emph{private box rule} $\xboxrIf$ makes $u$ private to the $\tagy$-component, and is applied when $w R_{\charx} u$ is only used by the $\tagy$-component. Observe that the interpolant $\itp$ of the $\xboxrIf$ rule is the same in the premise and the conclusion; this ensures that problematic modalities are not introduced into interpolants as we saw in the example at the beginning of the section. Since $u$ is not common, the interpolant of the premise of $\xboxrIf$ contains no formula labeled with $u$ by condition (4) of \dfn~\ref{def:interp-sequent}.

The \emph{seriality rules} $\ddrIc$ and $\ddrIf$ are interpolation variants of the $\ddr$ rule, and are included in $\tkmci$ \iffi $\D \in \fcons$. Unlike relational atoms introduced by box rules, a relational atom $w R_{\charx} u$ introduced by $\ddr$ does not stem from a formula of either component, so either component may use it. We nevertheless treat it exactly like a relational atom introduced by a box rule. 

The \emph{common seriality rule} $\ddrIc$ requires $w$ to be common and makes $u$ common, and is applied when both components use $w R_{\charx} u$, i.e., when the propagation paths of $\diap{\chary}^{\ilet}$ inferences of both components traverse it. As for the common box rule, the index of $w R_{\charx} u$ then occurs in the modal signature of both components by \lem~\ref{lem:bridge-free-lang}, so it is safe to introduce the modality $\xbox$ into the interpolant and use the box interpolant $\xbi$.

The \emph{private seriality rule} $\ddrIf$ makes $u$ private to the $\tagy$-component and leaves the interpolant unchanged, and is applied when at most one component uses $w R_{\charx} u$. Note that $w$ may be common or private to the $\tagy$-component in an instance of $\ddrIf$, but never private to the $\tagx$-component, as its premise satisfies condition~(2) of \dfn~\ref{def:interp-sequent}. 

The decision of which box or seriality rule to apply depends on how the introduced relational atom is used further up in the proof, and is determined by the partition algorithm described in the next section.

As an example, the problematic interpolation proof shown at the beginning of the section can be recast in $\tkmci$ as shown below, where $u$ is private to the right component. The constructed interpolant $\emptyset$ is no longer problematic and yields the strong Lyndon interpolant $\top$ of the conclusion.
\begin{center}
\small
\AxiomC{}
\RightLabel{$\idIii$}
\UnaryInfC{$\lsets{\set{w}}{\emptyset}{\set{u}} \sep w R_{\chary} u \sar w : \xdia \notp \splti w : \xbox p, u : q, u : \inot q \sep \emptyset$}
\RightLabel{$\disrI$}
\UnaryInfC{$\lsets{\set{w}}{\emptyset}{\set{u}} \sep w R_{\chary} u \sar w : \xdia \notp \splti w : \xbox p, u : q \lor \inot q \sep \emptyset$}
\RightLabel{$\ybox^{\ilet}_{p}$}
\UnaryInfC{$\lsets{\set{w}}{\emptyset}{\emptyset} \sep \emptyset \sar w : \xdia \notp \splti w : \xbox p, w : \ybox(q \lor \inot q) \sep \emptyset$}
\RightLabel{$\disrI$}
\UnaryInfC{$\lsets{\set{w}}{\emptyset}{\emptyset} \sep \emptyset \sar w : \xdia \notp \splti w : \xbox p \lor \ybox(q \lor \inot q) \sep \emptyset$}
\DisplayProof
\end{center}

\begin{figure}[pos=t]
\centering
\small
\AxiomC{}
\RightLabel{$\idIi$}
\UnaryInfC{$\lsets{\set{w, u}}{\emptyset}{\emptyset} \sep w R_{\charx} u \sar w : \xdia r, w : r \splti u : q, w : \inot r \sep \{(w : \inot r)\}$}
\RightLabel{$\orthru$}
\UnaryInfC{$\lsets{\set{w, u}}{\emptyset}{\emptyset} \sep w R_{\charx} u \sar u : q, w : \inot r \spltii w : \xdia r, w : r \sep \{(w : r)\}$}
\RightLabel{$\xdiarI$}
\UnaryInfC{$\lsets{\set{w, u}}{\emptyset}{\emptyset} \sep w R_{\charx} u \sar u : q, w : \inot r \spltii w : \xdia r \sep \{(w : r)\}$}
\RightLabel{$\orthru$}
\UnaryInfC{$\lsets{\set{w, u}}{\emptyset}{\emptyset} \sep w R_{\charx} u \sar w : \xdia r \splti u : q, w : \inot r \sep \{(w : \inot r)\}$}
\RightLabel{$\xboxrIc$}
\UnaryInfC{$\lsets{\set{w}}{\emptyset}{\emptyset} \sep \emptyset \sar w : \xdia r \splti w : \xbox q, w : \inot r \sep \{(w : \inot r, w : \xbox \bot)\}$}
\RightLabel{$\disrI$}
\UnaryInfC{$\lsets{\set{w}}{\emptyset}{\emptyset} \sep \emptyset \sar w : \xdia r \splti w : \xbox q \lor \inot r \sep \{(w : \inot r, w : \xbox \bot)\}$}
\DisplayProof
\caption{Example interpolation proof in $\tkmci$ with $\fcons = \{\R_{\charx\conv{\charx}} \subseteq \R_{\charx}\}$ (see Example~\ref{ex:interp-proof-detour}).\label{fig:example-proof}}
\end{figure}

\begin{example}\label{ex:interp-proof-detour} Let $\fcons := \{\R_{\charx\conv{\charx}} \subseteq \R_{\charx}\}$, so that $\g{\fcons} = \{\charx \pto \charx\conv{\charx}, \conv{\charx} \pto \charx\conv{\charx}\}$ and $\charx\conv{\charx} \in \glang(\charx)$. Note that $\kmc$ is bridge-free and $\xbox \inot r \imp \xbox q \lor \inot r \in \kmc$. \fig~\ref{fig:example-proof} shows an interpolation proof of $\sar w : \xdia r, w : \xbox q \lor \inot r$, where $w : \xdia r = w : \negnnf{\xbox \inot r}$ is placed in the left component.

The relational atom $w R_{\charx} u$ is introduced by the boxed formula $w : \xbox q$ of the right component, but is also used by the left component. In particular, the $\xdiarI$ inference propagates $r$ from $w$ to $w$ along the path $w \prpath{\charx} u \prpath{\conv{\charx}} w$ with string $\charx\conv{\charx} \in \glang(\charx)$. Hence, the label $u$ is common, and $w R_{\charx} u$ is introduced by $\xboxrIc$. No formula is ever placed at $u$ by the left component, and so, the interpolant of the premise of $\xboxrIc$ contains no labeled formula with the label $u$. The box interpolant nevertheless adds $w : \xbox \bot$, which yields the strong Lyndon interpolant $\inot r \lor \xbox \bot$ of the conclusion. Without $w : \xbox \bot$, we would obtain $\inot r$, which is not an interpolant, as $\xbox \inot r \imp \inot r \not\in \kmc$.
\end{example}

Let us now describe our interpolation algorithm for bridge-free grammar logics, which consists of a two-step process. (1) We take a proof $\prf$ in $\tkmc$ of a tree sequent $\sar w : \negnnf{A}, w : B$ as input. We first analyze $\prf$ to determine, for each relational atom, whether it is used to propagate formulae stemming from $\negnnf{A}$, from $B$, or from both. Based on this analysis, a \emph{tripartite partition algorithm} $\partalg$ bottom-up partitions $\prf$, starting from the partitioned conclusion $\lsets{\set{w}}{\emptyset}{\emptyset} \sep \emptyset \sar w : \negnnf{A} \splti w : B$, and declares each label common or private to the component that uses it. This yields a \emph{partitioned proof} $\partprf$, which consists of \emph{partitioned sequents}, that is, interpolation sequents without interpolants (\sect~\ref{sec:part-bias}). (2) We then utilize the rules of $\tkmci$ to assign interpolants to the partitioned sequents in $\partprf$, which yields an interpolation proof $\intprf$ whose conclusion $\lsets{\set{w}}{\emptyset}{\emptyset} \sep \emptyset \sar w : \negnnf{A} \splti w : B \sep \itp$ provides a strong Lyndon interpolant of $\negnnf{A} \lor B = A \imp B$ (\sect~\ref{sec:main-theorem}).

\section{Tripartite Partitioning}\label{sec:part-bias}

In this section, we describe the first step of our interpolation method, which transforms a proof $\prf$ into a \emph{partitioned proof} $\partprf$. A partitioned proof is a proof whose sequents split their labeled formulae into a left and a right component and their labels into common labels and labels private to either component. It has the shape of an interpolation proof, but interpolants have not yet been assigned to it.

Throughout the section, we fix a bridge-free grammar logic $\kmc$ and a proof $\prf$ in $\tkmc$ of $\sar \rta : \negnnf{A}, \rta : B$. Note that $\rta$ is the root of every tree sequent in $\prf$, since every rule of $\tkmc$ either bottom-up preserves the relational atoms of its conclusion or, in the case of the $\xboxr$ and $\ddr$ rules, adjoins a single edge to a fresh label, thus extending the tree by a new leaf. We first analyze $\prf$ to determine, for each relational atom, whether it is used to propagate formulae stemming from $\negnnf{A}$, from $B$, or from both (\sect~\ref{subsec:sides}). Based on this analysis, the \emph{partition algorithm} $\partalg$ translates $\prf$ bottom-up, rule by rule, into $\partprf$ (\sect~\ref{subsec:partition}). In \sect~\ref{subsec:part-properties}, we show that assigning interpolants to $\partprf$ according to the rules of $\tkmci$ yields an interpolation proof, permitting us to compute a strong Lyndon interpolant of the conclusion.

We assume w.l.o.g. that distinct $\xboxr$ and $\ddr$ inferences in $\prf$ introduce distinct fresh labels; this assumption is innocuous, as such a proof can always be obtained by a suitable renaming of fresh labels. Consequently, each non-root label $u$ of $\prf$ is introduced together with a unique relational atom $w R_{\charx} u$, and we call $u$ a \emph{child} of $w$. For a label $u \in \lab(\prf)$, we let $\desc(u)$ be the set containing $u$ and its \emph{descendants}, i.e., all labels obtained from $u$ by repeatedly taking children. For each $\xdiar$ inference of $\prf$ with a principal formula $w : \xdia A$ and auxiliary formula $u : A$, we fix a single propagation path $w \prpath{\stra} u$ witnessing its side condition $\dag_{2}(\fcons)$, and we say that the inference \emph{traverses} each relational atom of this path; thus, only $\xdiar$ inferences traverse relational atoms.

A \emph{partitioned sequent} is an expression $\lsxy \sep \tel \sar \Gamma_{1} \splt \Gamma_{2}$ satisfying conditions (1)--(3) of \dfn~\ref{def:interp-sequent}, i.e., an interpolation sequent without an interpolant. For a tag $\tagz \in \set{\tagl, \tagr}$, we define the \emph{opposite tag} $\optag{\tagz}$ as follows:
$$
\optag{\tagz} :=
\begin{cases}
\tagl & \text{if } \tagz = \tagr,\\
\tagr & \text{if } \tagz = \tagl.
\end{cases}
$$

\subsection{Sides, Usage, and Private Labels}\label{subsec:sides}

Since the labeled formulae of a sequent form a multiset, we fix for each inference of $\prf$ which labeled formula occurrences of its conclusion are principal; the remaining labeled formula occurrences are \emph{context occurrences}, and these correspond one-to-one to the context occurrences of each premise. 

We assign a \emph{side} $\tagz \in \set{\tagl, \tagr}$ to every labeled formula occurrence of $\prf$, proceeding bottom-up. In the conclusion of $\prf$, the occurrence $\rta : \negnnf{A}$ has side $\tagl$ and the occurrence $\rta : B$ has side $\tagr$. In each premise, every context occurrence inherits the side of the corresponding occurrence of the conclusion, and every auxiliary labeled formula, including the retained principal formula of a $\xdiar$ inference, receives the side of the principal formula. 

We say that an inference \emph{has side} $\tagz$ if its principal formula has side $\tagz$. Note that $\ddr$ inferences have no principal formula, and that the two principal formulae of an $\id$ inference may have distinct sides. As auxiliary labeled formulae are subformulae of their principal formula, we obtain the following.

\begin{lemma}\label{lem:sides}
Let $\tagz \in \set{\tagl, \tagr}$. If a sequent of $\prf$ contains a formula occurrence $u : C$ of side $\tagz$, then $C$ is a subformula of a formula occurrence of side $\tagz$ in every sequent below it on its branch. In particular, $C$ is a subformula of $\negnnf{A}$ if $\tagz = \tagl$, and of $B$ if $\tagz = \tagr$.
\end{lemma}


We now classify the relational atoms and labels of $\prf$ according to the sides that use them. We say that a relational atom $w R_{\charx} u$ of $\prf$ is \emph{used} by a side $\tagz \in \set{\tagl, \tagr}$ \iffi it is introduced by a $\xboxr$ inference with side $\tagz$, or some inference of $\prf$ with side $\tagz$ traverses it. As each relational atom of $\prf$ is bottom-up introduced by a $\xboxr$ or $\ddr$ rule and occurs only above that inference in $\prf$, every inference traversing it lies above that inference as well.

\begin{definition}[Common and Private Labels]\label{def:label-status} We assign to each label of $\prf$ a \emph{status}, which is either \emph{common} or \emph{$\tagz$-private} for some $\tagz \in \set{\tagl, \tagr}$, by recursion along the tree, starting from the root: the root $\rta$ is common, and if $u$ is a child of $w$, introduced together with the relational atom $w R_{\charx} u$, then:
\begin{itemize}

\item[(1)] if $w$ is $\tagz$-private, then $u$ is $\tagz$-private;

\item[(2)] if $w$ is common and $w R_{\charx} u$ is used by both sides, then $u$ is common;

\item[(3)] if $w$ is common and $w R_{\charx} u$ is used by exactly one side $\tagz$, then $u$ is $\tagz$-private;

\item[(4)] if $w$ is common and $w R_{\charx} u$ is used by neither side, then $u$ is $\tagr$-private.

\end{itemize}
A relational atom is \emph{common} if its target is common, and \emph{$\tagz$-private} if its target is $\tagz$-private.
\end{definition}

Intuitively, walking down the tree from the root, a label remains common as long as both sides use the relational atom leading to it, and the first relational atom that is not used by both sides starts a subtree of private labels (see \fig~\ref{fig:tagged-subtrees} for an example). Case (4) only applies to relational atoms introduced by $\ddr$ inferences, as a relational atom introduced by a $\xboxr$ inference is used by the side of that inference; the choice of $\tagr$ in this case is arbitrary, as neither side uses the relational atom leading to the resulting subtree. 

Note that, by definition, every common relational atom has a common source and is used by both sides. This terminology matches that of \sect~\ref{sec:inter-calc}: the partition algorithm places the common labels of each sequent in the set $\cst$ and its $\tagz$-private labels in the set $\pst{\tagz}$ (\lem~\ref{lem:part} below). Each subtree of $\tagz$-private labels is \emph{one-sided}: the other side never places formulae into it and never propagates formulae through it (\lem~\ref{lem:one-sided} below). For the reader's convenience, these notions are summarized in Table~\ref{tab:terminology} of Appendix~\ref{app:terminology}, together with their effect on the partition algorithm described in \sect~\ref{subsec:partition}.

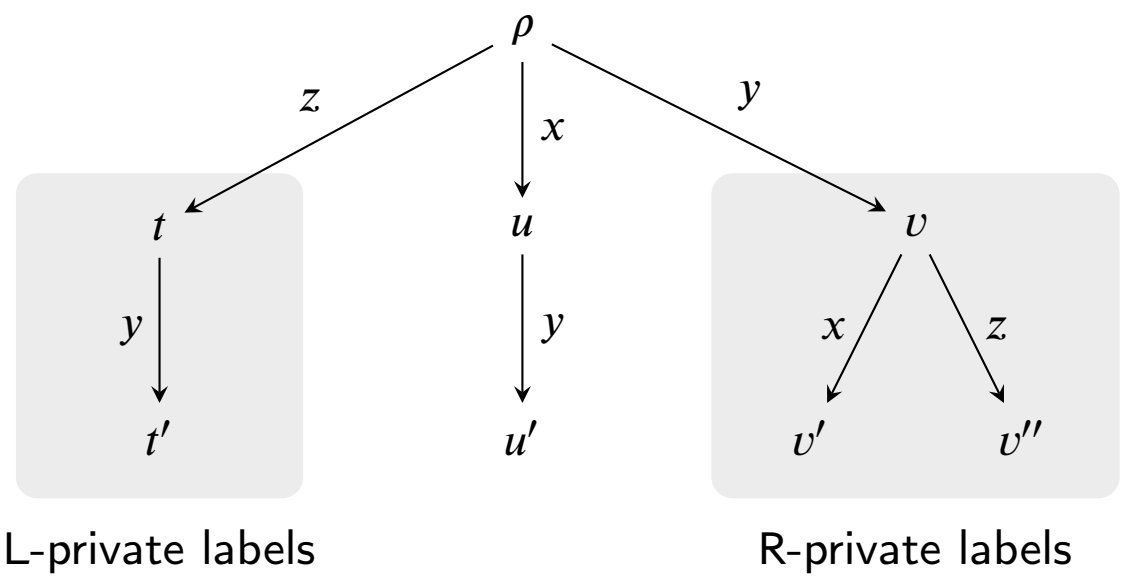
\begin{figure}[pos=t]
\centering
\begin{tikzpicture}[>=stealth]
\fill[gray!15, rounded corners] (-3.35,-0.95) rectangle (-1.45,-3.1);
\fill[gray!15, rounded corners] (1.25,-0.95) rectangle (3.95,-3.1);
\node (w) at (0,0) {$\rta$};
\node (t) at (-2.4,-1.3) {$t$};
\node (t1) at (-2.4,-2.7) {$t'$};
\node (u) at (0,-1.3) {$u$};
\node (u1) at (0,-2.7) {$u'$};
\node (v) at (2.6,-1.3) {$v$};
\node (v1) at (1.9,-2.7) {$v'$};
\node (v2) at (3.3,-2.7) {$v''$};
\draw[->] (w) -- node[above left] {$\charz$} (t);
\draw[->] (w) -- node[right] {$\charx$} (u);
\draw[->] (w) -- node[above right] {$\chary$} (v);
\draw[->] (t) -- node[left] {$\chary$} (t1);
\draw[->] (u) -- node[right] {$\chary$} (u1);
\draw[->] (v) -- node[left] {$\charx$} (v1);
\draw[->] (v) -- node[right] {$\charz$} (v2);
\node at (-2.4,-3.45) {$\tagl$-private labels};
\node at (2.6,-3.45) {$\tagr$-private labels};
\end{tikzpicture}
\caption{An example of how the labels of a tree sequent of $\prf$ are sorted by the partition algorithm. Each edge of the tree represents a relational atom and is labeled with its character. Suppose that $\rta R_{\charz} t$ is only used by the left side, $\rta R_{\chary} v$ is only used by the right side, and $\rta R_{\charx} u$ and $u R_{\chary} u'$ are used by both sides. Then the labels in $\desc(t)$ and $\desc(v)$ (gray regions) are $\tagl$-private and $\tagr$-private, respectively, and all other labels are common. Accordingly, the partition algorithm places the labels of the two gray regions in $\lst$ and $\rst$, respectively, and all other labels in $\cst$. Removing the right (left) gray region thus yields the tree of the $\tagl$-projection ($\tagr$-projection) of the corresponding partitioned sequent.\label{fig:tagged-subtrees}}
\end{figure}

\begin{lemma}[One-Sidedness]\label{lem:one-sided}
Let $\tagz \in \set{\tagl, \tagr}$. Then:
\begin{itemize}

\item[(1)] no formula occurrence of side $\optag{\tagz}$ has a $\tagz$-private label;

\item[(2)] no inference with side $\optag{\tagz}$ traverses a $\tagz$-private relational atom.

\end{itemize}
\end{lemma}

\begin{proof} By \dfn~\ref{def:label-status}, every $\tagz$-private label lies in a set $U := \desc(u)$, where $u$ is the topmost $\tagz$-private label on its branch. The parent $w$ of $u$ is common, all labels in $U$ are $\tagz$-private, and the relational atom $w R_{\charx} u$ is not used by $\optag{\tagz}$, by cases (3) and (4) of \dfn~\ref{def:label-status}. We fix such a label $u$ and let $\iota$ be the $\xboxr$ or $\ddr$ inference introducing $w R_{\charx} u$. It suffices to prove both claims for the labels in $U$, which only occur in the premise of $\iota$ and above it, as $u$ is fresh for $\iota$; in particular, the claims hold trivially for all other sequents, such as the conclusion of $\iota$. Our argument rests on the following observation:
\begin{flushleft}
$(\ast)$ \emph{A path of an inference with side $\optag{\tagz}$ starting outside $U$ never reaches a label in $U$.}
\end{flushleft}
Indeed, $w R_{\charx} u$ is the only relational atom connecting $U$ to the remaining labels of a tree sequent, so such a path could only reach $U$ by traversing $w R_{\charx} u$, which no inference with side $\optag{\tagz}$ does, as $\optag{\tagz}$ does not use $w R_{\charx} u$.

For claim (1), we proceed upward from the premise of $\iota$. If $\iota$ is a $\xboxr$ inference, then the only formula occurrence with a label in $U$ in its premise is the auxiliary formula $u : A$, which has the side of $\iota$, i.e., $\tagz$, as $\optag{\tagz}$ does not use $w R_{\charx} u$; if $\iota$ is a $\ddr$ inference, then its premise contains no formula occurrence with a label in $U$. Moving upward, context occurrences keep their side and label, so a new formula occurrence of side $\optag{\tagz}$ only arises as an auxiliary formula of an inference with side $\optag{\tagz}$, whose principal formula has a label outside of $U$ by induction. Such an auxiliary formula is labeled with the label of the principal formula (for $\disr$, $\conr$, and the retained principal formula of $\diap{\chary}$), with a fresh child of it, which differs from $u$ (for $\ybox$), or with the endpoint of its path (for $\diap{\chary}$). None of these labels lies in $U$, where we use $(\ast)$ in the last case.

For claim (2), observe that a relational atom with its target in $U$ is either $w R_{\charx} u$ or has both of its labels in $U$, so an inference traversing it reaches a label in $U$ along its path. By claim (1), the path of an inference with side $\optag{\tagz}$ starts outside of $U$, and so, it never does so by $(\ast)$.
\end{proof}

The following lemma establishes that common relational atoms indeed have their index in the common modal signature, which is the key ingredient for showing that interpolants lie in the common modal signature in \sect~\ref{sec:main-theorem}. It is the only place in this section where bridge-freeness is used.

\begin{lemma}[Common Relational Atoms]\label{lem:common-atoms} If $w R_{\charx} u$ is a common relational atom, then $\set{\charx, \conv{\charx}} \subseteq \sigm{A} \cap \sigm{B}$.
\end{lemma}

\begin{proof} Let $A_{\tagl} := \negnnf{A}$ and $A_{\tagr} := B$. By \dfn~\ref{def:label-status}, $w R_{\charx} u$ is used by both sides. Let $\tagz \in \set{\tagl, \tagr}$. If $w R_{\charx} u$ is introduced by a $\xboxr$ inference with side $\tagz$, then its principal formula $w : \xbox C$ has side $\tagz$, so $\xbox C$ is a subformula of $A_{\tagz}$ by \lem~\ref{lem:sides}, and $\set{\charx, \conv{\charx}} \subseteq \sigm{\xbox C}$. Otherwise, $w R_{\charx} u$ is traversed by a $\diap{\chary}$ inference with side $\tagz$ and principal formula $w' : \diap{\chary} D$. By \lem~\ref{lem:bridge-free-lang}, $\charx \in \set{\chary, \conv{\chary}}$, so $\set{\charx, \conv{\charx}} = \set{\chary, \conv{\chary}} \subseteq \sigm{\diap{\chary} D}$, and $\diap{\chary} D$ is a subformula of $A_{\tagz}$ by \lem~\ref{lem:sides}. Hence, $\set{\charx, \conv{\charx}} \subseteq \sigm{A_{\tagl}} \cap \sigm{A_{\tagr}}$, and the claim follows from $\sigm{\negnnf{A}} = \sigm{A}$ (\lem~\ref{lem:signature-negation}).
\end{proof}

\subsection{Tripartite Partition Algorithm}\label{subsec:partition}

The partition algorithm $\partalg$ processes $\prf$ bottom-up, rule by rule, replacing each inference of $\prf$ by an inference of the corresponding rule of $\tkmci$ (disregarding interpolants), possibly preceded by $\orthru$. Each formula occurrence is placed in the component of its side, and each label is placed, once and for all, in $\cst$ if it is common, and in $\pst{\tagz}$ if it is $\tagz$-private; the relational atoms of each partitioned sequent are simply those of the corresponding sequent of $\prf$. Consequently, a $\xboxr$ or $\ddr$ inference introducing a common label is translated into an instance of the common rule $\xboxrIc$ or $\ddrIc$, respectively, and every other $\xboxr$ or $\ddr$ inference into an instance of the private rule $\xboxrIf$ or $\ddrIf$.

Formally, $\partalg$ takes as input a subproof $\prf'$ of $\prf$ together with a partitioned sequent $\iseq$. The underlying tree sequent of $\iseq$ is the conclusion of $\prf'$, its $\tagl$- and $\tagr$-components contain the formula occurrences of the respective sides, and its labels are placed as described above. The algorithm returns a partitioned proof $\partalg(\prf', \iseq)$ of $\iseq$. It is defined by the clauses (P1)--(P8) below, in which $\prf_{1}$ and $\prf_{2}$ denote the immediate subproofs of $\prf'$, and $\iseq_{1}$ and $\iseq_{2}$ the partitioned sequents supplied to the corresponding recursive calls.

\begin{itemize}

\item[(P1)] \emph{Initialization.} The \emph{partitioned proof} of $\prf$ is
$$
\partprf := \partalg(\prf, \lsets{\set{\rta}}{\emptyset}{\emptyset} \sep \emptyset \sar \rta : \negnnf{A} \splti \rta : B).
$$
In other words, $\rta : \negnnf{A}$ is placed in the left component, $\rta : B$ in the right component, and the root $\rta$ is common.

\item[(P2)] \emph{Initial sequents.} If $\prf'$ consists of an $\id$ or $\topr$ inference, then $\partalg(\prf', \iseq)$ consists of $\iseq$ alone, viewed as an instance of
\begin{itemize}
\item[$\bullet$] $\idIii$ if $\iseq = \lsxy \sep \tel \sar \Gamma_{1} \splt \Gamma_{2}, w : p, w : \inot p$;
\item[$\bullet$] $\idIi$ if $\iseq = \lsxy \sep \tel \sar \Gamma_{1}, w : p \splt w : \inot p, \Gamma_{2}$;
\item[$\bullet$] $\toprI$ if $\iseq = \lsxy \sep \tel \sar \Gamma_{1} \splt \Gamma_{2}, w : \top$.
\end{itemize}

\item[(P3)] \emph{Disjunction.} Suppose the $\disr$ inference has principal formula $w : A \lor B$, so that $\prf'$ is as shown below left, $\iseq = \lsxy \sep \tel \sar \Gamma_{1} \splt \Gamma_{2}, w : A \lor B$, and $\iseq_{1} := \lsxy \sep \tel \sar \Gamma_{1} \splt \Gamma_{2}, w : A, w : B$. Then $\partalg(\prf',\iseq)$ is as shown below right.
\begin{center}
\begin{tabular}{c @{\hspace{2em}} c}
\AxiomC{$\prf_{1}$}
\noLine
\UnaryInfC{$\tel \sar \Gamma_{1}, \Gamma_{2}, w : A, w : B$}
\RightLabel{$\disr$}
\UnaryInfC{$\tel \sar \Gamma_{1}, \Gamma_{2}, w : A \lor B$}
\DisplayProof

&

\AxiomC{$\partalg(\prf_{1},\iseq_{1})$}
\noLine
\UnaryInfC{$\iseq_{1}$}
\RightLabel{$\disrI$}
\UnaryInfC{$\iseq$}
\DisplayProof
\end{tabular}
\end{center}

\item[(P4)] \emph{Conjunction.} Suppose the $\conr$ inference has principal formula $w : A \land B$, so that $\prf'$ is as shown below left, and let $\iseq = \lsxy \sep \tel \sar \Gamma_{1} \splt \Gamma_{2}, w : A \land B$, $\iseq_{1} := \lsxy \sep \tel \sar \Gamma_{1} \splt \Gamma_{2}, w : A$, and $\iseq_{2} := \lsxy \sep \tel \sar \Gamma_{1} \splt \Gamma_{2}, w : B$. Then $\partalg(\prf',\iseq)$ is as shown below right.
\begin{center}
\begin{tabular}{c @{\hspace{2em}} c}
\AxiomC{$\prf_{1}$}
\noLine
\UnaryInfC{$\tel \sar \Gamma_{1}, \Gamma_{2}, w : A$}
\AxiomC{$\prf_{2}$}
\noLine
\UnaryInfC{$\tel \sar \Gamma_{1}, \Gamma_{2}, w : B$}
\RightLabel{$\conr$}
\BinaryInfC{$\tel \sar \Gamma_{1}, \Gamma_{2}, w : A \land B$}
\DisplayProof

&

\AxiomC{$\partalg(\prf_{1},\iseq_{1})$}
\noLine
\UnaryInfC{$\iseq_{1}$}
\AxiomC{$\partalg(\prf_{2},\iseq_{2})$}
\noLine
\UnaryInfC{$\iseq_{2}$}
\RightLabel{$\conrI$}
\BinaryInfC{$\iseq$}
\DisplayProof
\end{tabular}
\end{center}

\item[(P5)] \emph{Box rule.} Suppose the $\xboxr$ inference has principal formula $w : \xbox A$ and introduces the relational atom $w R_{\charx} u$, so that $\prf'$ is as shown below left and $\iseq = \lsxy \sep \tel \sar \Gamma_{1} \splt \Gamma_{2}, w : \xbox A$. As the principal formula $w : \xbox A$ has side $\tagy$, the relational atom $w R_{\charx} u$ is used by $\tagy$, and $w$ is not $\tagx$-private by \lem~\ref{lem:one-sided}-(1); hence, $u$ is either common or $\tagy$-private by \dfn~\ref{def:label-status}. If $u$ is common, then so is $w$, and $u$ is placed in $\cst$, i.e., $\iseq_{1} := \lsets{\cst \cup \set{u}}{\pst{\tagx}}{\pst{\tagy}} \sep \tel, w R_{\charx} u \sar \Gamma_{1} \splt \Gamma_{2}, u : A$, and $\partalg(\prf',\iseq)$ is as shown below middle. If $u$ is $\tagy$-private, then $u$ is placed in $\pst{\tagy}$, i.e., $\iseq_{1} := \lsets{\cst}{\pst{\tagx}}{\pst{\tagy} \cup \set{u}} \sep \tel, w R_{\charx} u \sar \Gamma_{1} \splt \Gamma_{2}, u : A$, and $\partalg(\prf',\iseq)$ is as shown below right.
\begin{center}
\begin{tabular}{c @{\hspace{2em}} c @{\hspace{2em}} c}
\AxiomC{$\prf_{1}$}
\noLine
\UnaryInfC{$\tel, w R_{\charx} u \sar \Gamma_{1}, \Gamma_{2}, u : A$}
\RightLabel{$\xboxr$}
\UnaryInfC{$\tel \sar \Gamma_{1}, \Gamma_{2}, w : \xbox A$}
\DisplayProof

&

\AxiomC{$\partalg(\prf_{1},\iseq_{1})$}
\noLine
\UnaryInfC{$\iseq_{1}$}
\RightLabel{$\xboxrIc$}
\UnaryInfC{$\iseq$}
\DisplayProof

&

\AxiomC{$\partalg(\prf_{1},\iseq_{1})$}
\noLine
\UnaryInfC{$\iseq_{1}$}
\RightLabel{$\xboxrIf$}
\UnaryInfC{$\iseq$}
\DisplayProof
\end{tabular}
\end{center}

\item[(P6)] \emph{Diamond rule.} Suppose the $\xdiar$ inference has principal formula $w : \xdia A$ with $w \prpath{\stra} u$ its fixed propagation path, so that $\prf'$ is as shown below left, $\iseq = \lsxy \sep \tel \sar \Gamma_{1} \splt \Gamma_{2}, w : \xdia A$, and $\iseq_{1} := \lsxy \sep \tel \sar \Gamma_{1} \splt \Gamma_{2}, w : \xdia A, u : A$. Then $\partalg(\prf',\iseq)$ is as shown below right.
\begin{center}
\begin{tabular}{c @{\hspace{2em}} c}
\AxiomC{$\prf_{1}$}
\noLine
\UnaryInfC{$\tel \sar \Gamma_{1}, \Gamma_{2}, w : \xdia A, u : A$}
\RightLabel{$\xdiar$}
\UnaryInfC{$\tel \sar \Gamma_{1}, \Gamma_{2}, w : \xdia A$}
\DisplayProof

&

\AxiomC{$\partalg(\prf_{1},\iseq_{1})$}
\noLine
\UnaryInfC{$\iseq_{1}$}
\RightLabel{$\xdiarI$}
\UnaryInfC{$\iseq$}
\DisplayProof
\end{tabular}
\end{center}

\item[(P7)] \emph{Seriality rule.} Suppose the $\ddr$ inference introduces the relational atom $w R_{\charx} u$, so that $\prf'$ is as shown below left and $\iseq = \lsxy \sep \tel \sar \Gamma_{1} \splt \Gamma_{2}$. By \dfn~\ref{def:label-status}, $u$ is common \iffi $w$ is common and both sides use $w R_{\charx} u$. If $u$ is common, then $\iseq_{1} := \lsets{\cst \cup \set{u}}{\pst{\tagx}}{\pst{\tagy}} \sep \tel, w R_{\charx} u \sar \Gamma_{1} \splt \Gamma_{2}$, and $\partalg(\prf',\iseq)$ is as shown below middle. If $u$ is $\tagy$-private, then $\iseq_{1} := \lsets{\cst}{\pst{\tagx}}{\pst{\tagy} \cup \set{u}} \sep \tel, w R_{\charx} u \sar \Gamma_{1} \splt \Gamma_{2}$, and $\partalg(\prf',\iseq)$ is as shown below right.
\begin{center}
\begin{tabular}{c @{\hspace{2em}} c @{\hspace{2em}} c}
\AxiomC{$\prf_{1}$}
\noLine
\UnaryInfC{$\tel, w R_{\charx} u \sar \Gamma_{1}, \Gamma_{2}$}
\RightLabel{$\ddr$}
\UnaryInfC{$\tel \sar \Gamma_{1}, \Gamma_{2}$}
\DisplayProof

&

\AxiomC{$\partalg(\prf_{1},\iseq_{1})$}
\noLine
\UnaryInfC{$\iseq_{1}$}
\RightLabel{$\ddrIc$}
\UnaryInfC{$\iseq$}
\DisplayProof

&

\AxiomC{$\partalg(\prf_{1},\iseq_{1})$}
\noLine
\UnaryInfC{$\iseq_{1}$}
\RightLabel{$\ddrIf$}
\UnaryInfC{$\iseq$}
\DisplayProof
\end{tabular}
\end{center}

\item[(P8)] \emph{Swapping components.} If none of (P2)--(P7) applies, then the principal formulae of the last inference of $\prf'$ are positioned as required by one of (P2)--(P6) after swapping the two components, or the last inference of $\prf'$ is a $\ddr$ inference introducing an $\tagx$-private label. Hence, $\partalg$ first applies $\orthru$, and then proceeds with the respective clause. Letting $\iseq = \lsxy \sep \tel \sar \Gamma_{1} \splt \Gamma_{2}$ and $\iseq_{1} := \lsyx \sep \tel \sar \Gamma_{2} \spltrev \Gamma_{1}$, we define:
\begin{center}
$\partalg(\prf',\iseq) :=$
\AxiomC{$\partalg(\prf', \iseq_{1})$}
\RightLabel{$\orthru$}
\UnaryInfC{$\iseq$}
\DisplayProof
\end{center}

\end{itemize}

As $\partalg$ only partitions the sequents of $\prf$ and inserts instances of $\orthru$, $\partprf$ is a proof of partitioned sequents whose underlying tree sequents are those of $\prf$. Labeled formulae stay in the component of their side, and labels never change their set once introduced. Note that (P5) and (P7) are precisely the rules of \fig~\ref{fig:interpolation-calc-modal} read bottom-up. The common rules $\xboxrIc$ and $\ddrIc$ are used precisely for common relational atoms, whose indices occur in the modal signatures of both $A$ and $B$ (\lem~\ref{lem:common-atoms}), whereas private relational atoms are introduced by the private rules $\xboxrIf$ and $\ddrIf$, and never contribute modalities to interpolants.

\begin{figure}[pos=t]
\centering
\small
\AxiomC{}
\RightLabel{$\idIi$}
\UnaryInfC{$\lsets{\set{\rta, u}}{\emptyset}{\set{v}} \sep \rta R_{\chary} v, \rta R_{\charx} u \sar \rta : \xdia r, \rta : r \splti u : q, \rta : \inot r, v : s$}
\RightLabel{$\orthru$}
\UnaryInfC{$\lsets{\set{\rta, u}}{\set{v}}{\emptyset} \sep \rta R_{\chary} v, \rta R_{\charx} u \sar u : q, \rta : \inot r, v : s \spltii \rta : \xdia r, \rta : r$}
\RightLabel{$\xdiarI$}
\UnaryInfC{$\lsets{\set{\rta, u}}{\set{v}}{\emptyset} \sep \rta R_{\chary} v, \rta R_{\charx} u \sar u : q, \rta : \inot r, v : s \spltii \rta : \xdia r$}
\RightLabel{$\orthru$}
\UnaryInfC{$\lsets{\set{\rta, u}}{\emptyset}{\set{v}} \sep \rta R_{\chary} v, \rta R_{\charx} u \sar \rta : \xdia r \splti u : q, \rta : \inot r, v : s$}
\RightLabel{$\xboxrIc$}
\UnaryInfC{$\lsets{\set{\rta}}{\emptyset}{\set{v}} \sep \rta R_{\chary} v \sar \rta : \xdia r \splti \rta : \xbox q, \rta : \inot r, v : s$}
\RightLabel{$\ybox^{\ilet}_{p}$}
\UnaryInfC{$\lsets{\set{\rta}}{\emptyset}{\emptyset} \sep \emptyset \sar \rta : \xdia r \splti \rta : \xbox q, \rta : \inot r, \rta : \ybox s$}
\RightLabel{$\disrI \times 2$}
\UnaryInfC{$\lsets{\set{\rta}}{\emptyset}{\emptyset} \sep \emptyset \sar \rta : \xdia r \splti \rta : (\xbox q \lor \inot r) \lor \ybox s$}
\DisplayProof
\caption{The partitioned proof $\partprf$ for the proof $\prf$ of Example~\ref{ex:part-bias}. In the sequents with the tagged bar `$\spltii$', the labels are listed as $\lsets{\cst}{\rst}{\lst}$.\label{fig:example-partitioned-proof}}
\end{figure}

\begin{example}\label{ex:part-bias} Let $\fcons := \{\R_{\charx\conv{\charx}} \subseteq \R_{\charx}\}$ and consider the following proof $\prf$, which proves the implication $\xbox \inot r \imp (\xbox q \lor \inot r) \lor \ybox s$.
\begin{center}
\AxiomC{}
\RightLabel{$\id$}
\UnaryInfC{$\rta R_{\chary} v, \rta R_{\charx} u \sar \rta : \xdia r, \rta : r, u : q, \rta : \inot r, v : s$}
\RightLabel{$\xdiar$}
\UnaryInfC{$\rta R_{\chary} v, \rta R_{\charx} u \sar \rta : \xdia r, u : q, \rta : \inot r, v : s$}
\RightLabel{$\xbox$}
\UnaryInfC{$\rta R_{\chary} v \sar \rta : \xdia r, \rta : \xbox q, \rta : \inot r, v : s$}
\RightLabel{$\ybox$}
\UnaryInfC{$\sar \rta : \xdia r, \rta : \xbox q, \rta : \inot r, \rta : \ybox s$}
\RightLabel{$\disr \times 2$}
\UnaryInfC{$\sar \rta : \xdia r, \rta : (\xbox q \lor \inot r) \lor \ybox s$}
\DisplayProof
\end{center}
The $\xdiar$ inference has side $\tagl$ and propagates $r$ from $\rta$ back to $\rta$ along the path $\rta \prpath{\charx} u \prpath{\conv{\charx}} \rta$, whose string $\charx\conv{\charx}$ belongs to $\glang(\charx)$, whereas the $\xbox$ and $\ybox$ inferences have side $\tagr$. Hence, $\rta R_{\charx} u$ is used by both sides, as it is introduced by a boxed formula of side $\tagr$ and traversed by the $\xdiar$ inference, which has side $\tagl$, whereas $\rta R_{\chary} v$ is only used by $\tagr$. Consequently, $u$ is common and $v$ is $\tagr$-private.

The partitioned proof $\partprf$ is shown in \fig~\ref{fig:example-partitioned-proof}, where these two facts can be read off the label sets directly. Reading bottom-up, the $\disr$ inferences are translated by (P3), and the $\ybox$ and $\xbox$ inferences by (P5), which places $v$ in $\rst$ and $u$ in $\cst$. The $\xdiar$ inference has side $\tagl$, so (P8) applies $\orthru$, after which (P6) translates it into a $\xdiarI$ inference, whose path only passes through the common labels $\rta$ and $u$. Last, (P8) and (P2) translate the $\id$ inference into $\orthru$ followed by $\idIi$. Assigning interpolants according to the rules of $\tkmci$ yields the interpolant $\{(\rta : \inot r, \rta : \xbox \bot)\}$ of the conclusion, i.e., the strong Lyndon interpolant $\inot r \lor \xbox \bot$ (cf.\ Example~\ref{ex:interp-proof-detour}). Had we translated the $\ybox$ inference into a $\ybox^{\ilet}_{c}$ inference, making $v$ common, we would have obtained the Lyndon interpolant $\inot r \lor \xbox \bot \lor \ybox \bot$, which contains the character $\chary$, even though $\chary$ does not occur in $\xbox \inot r$.
\end{example}

\subsection{Properties of Partitioned Proofs}\label{subsec:part-properties}

We now show that the partitioned proof $\partprf$ has the shape of an interpolation proof, so that the rules of $\tkmci$ can be used to assign interpolants to it.

\begin{lemma}\label{lem:part}
The partitioned proof $\partprf$ is a finite derivation with conclusion $\lsets{\set{\rta}}{\emptyset}{\emptyset} \sep \emptyset \sar \rta : \negnnf{A} \splti \rta : B$ such that:
\begin{itemize}

\item[(1)] each sequent of $\partprf$ is a partitioned sequent $\lsxy \sep \tel \sar \Gamma_{1} \splt \Gamma_{2}$, where $\tel \sar \Gamma_{1}, \Gamma_{2}$ is the corresponding sequent of $\prf$, and $\Gamma_{1}$ and $\Gamma_{2}$ consist of its formula occurrences of side $\tagx$ and $\tagy$, respectively;

\item[(2)] in each sequent $\lsxy \sep \tel \sar \Gamma_{1} \splt \Gamma_{2}$ of $\partprf$, the set $\cst$ consists of the common labels of the sequent, and $\pst{\tagz}$ consists of its $\tagz$-private labels, for $\tagz \in \set{\tagx, \tagy}$;

\item[(3)] each inference of $\partprf$ is an instance of a rule of $\tkmci$ (disregarding interpolants), and each initial sequent of $\partprf$ is an instance of $\idIi$, $\idIii$, or $\toprI$.

\end{itemize}
\end{lemma}

\begin{proof} Claims (1) and (2) hold by construction: $\partalg$ places every formula occurrence in the component of its side, and every label in $\cst$ or $\pst{\tagz}$ according to whether it is common or $\tagz$-private, where (P5) and (P7) place the new label $u$ according to \dfn~\ref{def:label-status}. It remains to show that each sequent satisfies conditions (1)--(3) of \dfn~\ref{def:interp-sequent}. Condition (1) holds as the underlying tree sequent is a sequent of $\prf$. For condition (2), the sets $\cst$, $\pst{\tagx}$, and $\pst{\tagy}$ are pairwise disjoint, as each label has exactly one status, the root $\rta$ is common, and every child of a $\tagz$-private label is $\tagz$-private by \dfn~\ref{def:label-status}. Condition (3) holds by \lem~\ref{lem:one-sided}(1), as formula occurrences of side $\tagz$ have no $\optag{\tagz}$-private labels.

For claim (3), we inspect the clauses of $\partalg$. In (P2), the principal formulae of an $\idIi$ inference share a label that labels formula occurrences of both sides, which is common by \lem~\ref{lem:one-sided}(1). In (P5) and (P7), the label $w$ is common if $u$ is common by \dfn~\ref{def:label-status}, as required by $\xboxrIc$ and $\ddrIc$. For (P5) and (P7), the label $u$ is fresh, as it is fresh in $\prf$. For (P6), the path of the inference traverses no relational atom with an $\tagx$-private target by \lem~\ref{lem:one-sided}(2), so it only passes through labels in $\cst \cup \pst{\tagy}$, i.e., it is a propagation path in $\prgr{\iseq|_{\tagy}}$.
\end{proof}

Combining the above lemmas, we obtain the main result of this section.

\begin{theorem}\label{thm:interp-proof}
Let $\prf$ be a proof in $\tkmc$ of $\sar \rta : \negnnf{A}, \rta : B$, where $\kmc$ is bridge-free. Then, assigning interpolants to the partitioned proof $\partprf$ according to the rules of $\tkmci$ yields an interpolation proof $\intprf$ of $\lsets{\set{\rta}}{\emptyset}{\emptyset} \sep \emptyset \sar \rta : \negnnf{A} \splti \rta : B \sep \itp$ for some interpolant $\itp$.
\end{theorem}

\begin{proof}
By \lem~\ref{lem:part}(3), each inference of $\partprf$ is an instance of a rule of $\tkmci$ (disregarding interpolants), and each initial sequent is an instance of an initial rule of $\tkmci$. We assign interpolants to the initial sequents and propagate them downward according to the rules of $\tkmci$, which yields interpolation sequents by \lem~\ref{lem:preservation}. The freshness conditions of the rules only concern the underlying tree sequents, which are those of $\prf$, and are thus satisfied.
\end{proof}

We note that by condition (4) of \dfn~\ref{def:interp-sequent}, the interpolants of $\intprf$ only contain common labels, and so, private relational atoms never contribute modalities to interpolants. Our method is effective: the sides of formula occurrences and the classification of relational atoms and labels are computed from $\prf$ directly, and $\partalg$ adds at most one instance of $\orthru$ per inference of $\prf$, so the number of inferences in the resulting interpolation proof is linear in the size of $\prf$. However, the size of the interpolants themselves may grow exponentially in the height of $\prf$, since $\orthru$ can be repeatedly applied along a proof branch. Thus, while the inference count is linear, the total size of an interpolation proof (i.e., the number of symbols it consists of) may be exponential in the worst case.

\section{Strong Lyndon Interpolation}\label{sec:main-theorem}

We have now laid enough groundwork to prove our main result, namely, that every bridge-free grammar logic has the strong LIP. Throughout this section, we fix a bridge-free grammar logic $\kmc$ and a proof $\prf$ in $\tkmc$ of $\sar \rta : \negnnf{A}, \rta : B$, where we assume the conventions of \sect~\ref{sec:part-bias}. We let $\intprf$ be the interpolation proof obtained by assigning interpolants to the partitioned proof $\partprf$ according to the rules of $\tkmci$ (\thm~\ref{thm:interp-proof}), and we let $\lsets{\set{\rta}}{\emptyset}{\emptyset} \sep \emptyset \sar \rta : \negnnf{A} \splti \rta : B \sep \itp$ be its conclusion. For a tree sequent $\tseq = \tel \sar \Gamma$ and a set $\Sigma$ of labeled formulae, we write $\tseq, \Sigma$ to denote $\tel \sar \Gamma, \Sigma$.

Our proof proceeds in three steps. First, we show that interpolants lie in the common signature (\sect~\ref{subsec:itp-labels-sig}). Second, we show that the interpolant of each interpolation sequent is implied by its $\tagx$-projection and implies its $\tagy$-projection (\sect~\ref{subsec:itp-lemma}). Third, we interpret the interpolant of the conclusion as a formula, which yields a strong Lyndon interpolant (\sect~\ref{subsec:strong-lip-thm}).

\subsection{Signatures of Interpolants}\label{subsec:itp-labels-sig}

By condition (4) of \dfn~\ref{def:interp-sequent}, the labels occurring in interpolants are common. In particular, private relational atoms never contribute modalities to interpolants, which lets us circumvent the issues discussed at the beginning of \sect~\ref{sec:inter-calc}. We now show that interpolants respect the common signature. The interpolant of an interpolation sequent is implied by its $\tagx$-component, which plays the role of a negated antecedent, and implies its $\tagy$-component. Hence, atoms occurring positively (negatively) in an interpolant must occur negatively (positively) in the $\tagx$-component, whereas polarities agree with those in the $\tagy$-component.


\begin{lemma}[Signatures of Interpolants]\label{lem:itp-signature}
Let $\lsxy \sep \tel \sar \Gamma_{1} \splt \Gamma_{2} \sep \itp$ be an interpolation sequent of $\intprf$. Then:
\begin{itemize}

\item[(1)] $\sigp{\itp} \subseteq \sign{\Gamma_{1}} \cap \sigp{\Gamma_{2}}$ and $\sign{\itp} \subseteq \sigp{\Gamma_{1}} \cap \sign{\Gamma_{2}}$;

\item[(2)] $\sigm{\itp} \subseteq \sigm{A} \cap \sigm{B}$.

\end{itemize}
\end{lemma}

\begin{proof} By induction on the height of the subproof ending with the interpolation sequent.

\textit{Base case.} For an instance of $\idIi$ with principal formulae $w : p$ in $\Gamma_{1}$ and $w : \notp$ in $\Gamma_{2}$, we have $\itp = \{(w : \notp)\}$, so $\sigp{\itp} = \emptyset$, $\sign{\itp} = \set{p} \subseteq \sigp{\Gamma_{1}} \cap \sign{\Gamma_{2}}$, and $\sigm{\itp} = \emptyset$. For instances of $\idIii$ and $\toprI$, we have $\itp = \emptyset$, whose signatures are empty.

\textit{Inductive step.} We first observe two facts about the rules of $\tkmci$ other than $\orthru$. First, for each $\isym \in \set{+,-}$, the $\isym$-signature of each component of a premise is contained in the $\isym$-signature of the corresponding component of the conclusion. This holds because auxiliary formulae are subformulae of the principal formula in the same component, which preserves polarity in negation normal form, while the rest of each component remains unchanged. Second, the box interpolant $\xbi$ has the same positive and negative signature as $\itp$, since it only prefixes disjunctions with modalities and adds the formula $\xbox \bot$, whose positive and negative signatures are empty.

(1) For all rules other than $\orthru$, claim (1) follows from \ih\ and the two facts above, as the interpolant of the conclusion is either a premise interpolant, the union of the premise interpolants, or a box interpolant of the premise interpolant. For $\orthru$, let the premise be $\lsyx \sep \tel \sar \Gamma_{2} \spltrev \Gamma_{1} \sep \itp'$ with $\itp = \orth{\itp'}$. By \ih, $\sigp{\itp'} \subseteq \sign{\Gamma_{2}} \cap \sigp{\Gamma_{1}}$ and $\sign{\itp'} \subseteq \sigp{\Gamma_{2}} \cap \sign{\Gamma_{1}}$. Since the orthogonal swaps the positive and negative signature (\sect~\ref{subsec:interp-sequents}), we obtain $\sigp{\itp} \subseteq \sign{\itp'} \subseteq \sign{\Gamma_{1}} \cap \sigp{\Gamma_{2}}$ and $\sign{\itp} \subseteq \sigp{\itp'} \subseteq \sigp{\Gamma_{1}} \cap \sign{\Gamma_{2}}$.

(2) The modal signature of the interpolant of the conclusion is contained in that of the premise interpolants, except possibly in the cases of $\xboxrIc$ and $\ddrIc$. In both cases, $\sigm{\xbi} \subseteq \sigm{\itp} \cup \set{\charx, \conv{\charx}}$, and $\set{\charx, \conv{\charx}} \subseteq \sigm{A} \cap \sigm{B}$ by \lem~\ref{lem:common-atoms}, as the relational atom $w R_{\charx} u$ introduced by these rules has a common target, and is thus common by \lem~\ref{lem:part}(2).
\end{proof}

\subsection{The Interpolation Lemma}\label{subsec:itp-lemma}

We now show that every interpolation sequent of $\intprf$ satisfies the intended meaning of interpolation sequents stated in \sect~\ref{subsec:interp-sequents}. The following lemma was proven in~\cite[Lemma~8]{LyoTiuGorClo20} and is used to handle the $\orthru$ rule.

\begin{lemma}[Persistence~\cite{LyoTiuGorClo20}]\label{lem:orth-lemma}
If $\Delta \in \orth{\orth{\itp}}$, then there exists a $\Gamma \in \itp$ such that $\Gamma \subseteq \Delta$.
\end{lemma}

The modal cases of the Interpolation Lemma below rely on the following lemma, which shows how box interpolants transfer provability from a premise to a conclusion. Recall that $\xdiar$ can propagate a formula from $w$ to $u$ along a relational atom $w R_{\charx} u$, as the string $\charx$ belongs to $L_{\charx}$.

\begin{lemma}[Box Interpolants]\label{lem:box-interpolants}
Let $\itp$ be an interpolant, let $w$ and $u$ be distinct labels, and let $\tel$ be a set of relational atoms in which $u$ does not occur.
\begin{itemize}

\item[(1)] Let $\tel \sar \Gamma$ be a tree sequent in which $w$ occurs and $u$ does not occur, such that $\lab(\itp) \subseteq \lab(\tel \sar \Gamma) \cup \set{u}$. If $\tkmc \proves (\tel, w R_{\charx} u \sar \Gamma, \Sigma)$ for each $\Sigma \in \itp$, then $\tkmc \proves (\tel \sar \Gamma, \Sigma)$ for each $\Sigma \in \xbi$.

\item[(2)] Let $\tel, w R_{\charx} u \sar \Gamma$ be a tree sequent with $\lab(\itp) \subseteq \lab(\tel, w R_{\charx} u \sar \Gamma)$. If $\tkmc \proves (\tel, w R_{\charx} u \sar \Gamma, \Sigma)$ for each $\Sigma \in \orth{\itp}$, then $\tkmc \proves (\tel, w R_{\charx} u \sar \Gamma, \Sigma)$ for each $\Sigma \in \orth{\xbi}$.

\end{itemize}
\end{lemma}

\begin{proof} We prove both claims in turn.

(1) Let $(\Sigma', w : \xbox \bigvee_{i \in [k]} B_{i}) \in \xbi$ stem from $(\Sigma', u : B_{1}, \ldots, u : B_{k}) \in \itp$ with $u \not\in \lab(\Sigma')$. By assumption, $\tkmc \proves (\tel, w R_{\charx} u \sar \Gamma, \Sigma', u : B_{1}, \ldots, u : B_{k})$. If $k \geq 1$, then $k{-}1$ applications of $\disr$ yield $\tel, w R_{\charx} u \sar \Gamma, \Sigma', u : \bigvee_{i \in [k]} B_{i}$, and if $k = 0$, then $\wk$ yields $\tel, w R_{\charx} u \sar \Gamma, \Sigma', u : \bot$. In both cases, an application of $\xboxr$, whose side condition holds as $u$ does not occur in $\tel \sar \Gamma, \Sigma'$, yields $\tel \sar \Gamma, \Sigma', w : \xbox \bigvee_{i \in [k]} B_{i}$.

(2) Let $\tel' := \tel, w R_{\charx} u$ and let $\itp = \{(\Sigma_{j}, u : B_{j,1}, \ldots, u : B_{j,k_{j}}) \mid j \in [n]\}$ with $u \not\in \lab(\Sigma_{j})$ for each $j \in [n]$, so that $\xbi = \{(\Sigma_{j}, w : \xbox \bigvee_{i \in [k_{j}]} B_{j,i}) \mid j \in [n]\}$. Let $C_{j} := \bigwedge_{i \in [k_{j}]} \negnnf{B_{j,i}}$, so that $\negnnf{\xbox \bigvee_{i \in [k_{j}]} B_{j,i}} = \xdia C_{j}$. Each element of $\orth{\xbi}$ is of the form $(\Theta, \Delta)$ for some $J \subseteq [n]$, where $\Delta := \{w : \xdia C_{j} \mid j \in J\}$ and $\Theta$ contains, for each $j \in [n] \setminus J$, the negation of a labeled formula from $\Sigma_{j}$. If $k_{j} = 0$ for some $j \in J$, then $C_{j} = \top$, and $\tel' \sar \Gamma, \Theta, \Delta, u : \top$ is an instance of $\top$. Otherwise, for every function $f$ mapping each $j \in J$ to an element of $[k_{j}]$, we have $(\Theta, \{u : \negnnf{B_{j,f(j)}} \mid j \in J\}) \in \orth{\itp}$, so $\tel' \sar \Gamma, \Theta, \Delta, \{u : \negnnf{B_{j,f(j)}} \mid j \in J\}$ is provable by assumption and $\wk$, and applying $\conr$ to these sequents yields $\tel' \sar \Gamma, \Theta, \Delta, \{u : C_{j} \mid j \in J\}$. In either case, applications of $\xdiar$, which propagate along the path $w \prpath{\charx} u$, yield $\tel' \sar \Gamma, \Theta, \Delta$.
\end{proof}

In the lemma below, note that $\iseq|_{\tagx}, \Sigma$ and $\iseq|_{\tagy}, \Sigma$ are tree sequents by \lem~\ref{lem:projections}, as the labels of $\Sigma$ are common by condition (4) of \dfn~\ref{def:interp-sequent} and $\lab(\orth{\itp}) \subseteq \lab(\itp)$.

\begin{lemma}[Interpolation Lemma]\label{lem:interpolation-lemma}
For every interpolation sequent $\iseq = \lsxy \sep \tel \sar \Gamma_{1} \splt \Gamma_{2} \sep \itp$ of $\intprf$, the following hold:
\begin{itemize}

\item[(1)] for each $\Sigma \in \itp$, we have $\tkmc \proves \iseq|_{\tagx}, \Sigma$, i.e., $\tkmc \proves (\tel \rstr (\cst \cup \pst{\tagx}) \sar \Gamma_{1}, \Sigma)$;

\item[(2)] for each $\Sigma \in \orth{\itp}$, we have $\tkmc \proves \iseq|_{\tagy}, \Sigma$, i.e., $\tkmc \proves (\tel \rstr (\cst \cup \pst{\tagy}) \sar \Gamma_{2}, \Sigma)$.

\end{itemize}
\end{lemma}

\begin{proof} By induction on the height of the subproof ending with $\iseq$. In each case, we let $\iseq'$ (or $\iseq_{1}$ and $\iseq_{2}$) denote the premise(s) of the last inference. We freely use the height-preserving admissibility of $\wk$ (see \thm~\ref{thm:tree-properties}(2)).

\textit{Base case.} For $\idIi$, we have $\itp = \{(w : \notp)\}$ and $\orth{\itp} = \{(w : p)\}$, where $w : p \in \Gamma_{1}$ and $w : \notp \in \Gamma_{2}$. Hence, $\iseq|_{\tagx}, w : \notp$ and $\iseq|_{\tagy}, w : p$ are instances of $\id$. For $\idIii$ and $\toprI$, we have $\itp = \emptyset$ and $\orth{\itp} = \set{\emptyset}$, so claim (1) holds vacuously, and claim (2) states that $\iseq|_{\tagy}$ is provable, which is an instance of $\id$ or $\top$, as $w : p, w : \notp \in \Gamma_{2}$ or $w : \top \in \Gamma_{2}$, respectively.

\textit{Inductive step.} We consider each rule of $\tkmci$ in turn.

\textit{$\orthru$.} The premise is $\iseq' = \lsyx \sep \tel \sar \Gamma_{2} \spltrev \Gamma_{1} \sep \itp'$ with $\itp = \orth{\itp'}$. Observe that $\iseq'|_{\tagy} = \iseq|_{\tagy}$ and $\iseq'|_{\tagx} = \iseq|_{\tagx}$, where the $\tagy$-component is the first component of $\iseq'$. (1) Let $\Sigma \in \orth{\itp'}$. Then, $\tkmc \proves \iseq|_{\tagx}, \Sigma$ by claim (2) of \ih. (2) Let $\Sigma \in \orth{\orth{\itp'}}$. By \lem~\ref{lem:orth-lemma}, there exists a $\Sigma' \in \itp'$ with $\Sigma' \subseteq \Sigma$. By claim (1) of \ih, $\tkmc \proves \iseq|_{\tagy}, \Sigma'$, and so, $\tkmc \proves \iseq|_{\tagy}, \Sigma$ by $\wk$.

\textit{$\disrI$.} The premise $\iseq'$ has the same $\tagx$-projection and interpolant as $\iseq$, which proves claim (1). For claim (2), let $\Sigma \in \orth{\itp}$. By \ih, $\tkmc \proves \iseq'|_{\tagy}, \Sigma$, where $\iseq'|_{\tagy}$ is obtained from $\iseq|_{\tagy}$ by replacing $w : A \lor B$ with $w : A$ and $w : B$. Applying $\disr$ yields $\iseq|_{\tagy}, \Sigma$.

\textit{$\conrI$.} The premises $\iseq_{1}$ and $\iseq_{2}$ have the same $\tagx$-projection as $\iseq$, and $\itp = \itp_{1} \cup \itp_{2}$, which implies claim (1). For claim (2), observe that every element of $\orth{\itp_{1} \cup \itp_{2}}$ is of the form $(\Sigma_{1}, \Sigma_{2})$ with $\Sigma_{i} \in \orth{\itp_{i}}$ for $i \in \set{1,2}$. By \ih\ and $\wk$, we have $\tkmc \proves \iseq_{1}|_{\tagy}, \Sigma_{1}, \Sigma_{2}$ and $\tkmc \proves \iseq_{2}|_{\tagy}, \Sigma_{1}, \Sigma_{2}$, which contain $w : A$ and $w : B$, respectively. Applying $\conr$ yields $\iseq|_{\tagy}, \Sigma_{1}, \Sigma_{2}$.

\textit{$\xdiarI$.} The premise $\iseq'$ has the same label sets and relational atoms as $\iseq$, so it has the same projections as $\iseq$, except that $u : A$ is added to its $\tagy$-projection, and it has the same interpolant $\itp$. Hence, claim (1) follows from \ih. For claim (2), let $\Sigma \in \orth{\itp}$. By \ih, $\tkmc \proves \iseq|_{\tagy}, \Sigma, u : A$. The side condition of $\xdiarI$ provides a propagation path from $w$ to $u$ in $\prgr{\iseq|_{\tagy}}$ whose string belongs to $L_{\charx}$, which is also a propagation path in the propagation graph of $\iseq|_{\tagy}, \Sigma$. Hence, applying $\xdiar$ to remove $u : A$ yields $\iseq|_{\tagy}, \Sigma$, as $w : \xdia A \in \Gamma_{2}$.

\textit{$\xboxrIc$.} In this case, $\Gamma_{2} = \Gamma_{2}', w : \xbox A$ with $w \in \cst$, and we let $\itp$ denote the interpolant of the premise $\iseq' = \lsets{\cst \cup \set{u}}{\pst{\tagx}}{\pst{\tagy}} \sep \tel, w R_{\charx} u \sar \Gamma_{1} \splt \Gamma_{2}', u : A \sep \itp$, so that the interpolant of the conclusion is $\xbi$. As $u$ is common in $\iseq'$, the $\tagx$-projection of $\iseq'$ is obtained from $\iseq|_{\tagx}$ by adding $w R_{\charx} u$, and its $\tagy$-projection is obtained from $\iseq|_{\tagy}$ by adding $w R_{\charx} u$ and replacing $w : \xbox A$ with $u : A$. Moreover, $w$ occurs in both projections of $\iseq$, as it is common (\lem~\ref{lem:projections}), whereas $u$ does not occur in $\iseq$, and $\lab(\itp) \subseteq \cst \cup \set{u}$ by condition (4) of \dfn~\ref{def:interp-sequent}. Hence, claim (1) follows from \ih\ by \lem~\ref{lem:box-interpolants}(1). For claim (2), \ih\ and \lem~\ref{lem:box-interpolants}(2) yield $\tkmc \proves \iseq'|_{\tagy}, \Sigma$ for each $\Sigma \in \orth{\xbi}$. As $u$ occurs neither in $\iseq|_{\tagy}$ nor in $\Sigma$ (since $\xbi$, and hence its orthogonal, contains no labeled formula with the label $u$), applying $\xboxr$ with principal formula $w : \xbox A$ to $\iseq'|_{\tagy}, \Sigma$ yields $\iseq|_{\tagy}, \Sigma$.

\textit{$\xboxrIf$.} In this case, $\Gamma_{2} = \Gamma_{2}', w : \xbox A$, and the premise is $\iseq' = \lsets{\cst}{\pst{\tagx}}{\pst{\tagy} \cup \set{u}} \sep \tel, w R_{\charx} u \sar \Gamma_{1} \splt \Gamma_{2}', u : A \sep \itp$. As $u$ is private to the $\tagy$-component, we have $\iseq'|_{\tagx} = \iseq|_{\tagx}$, so claim (1) follows from \ih. For claim (2), let $\Sigma \in \orth{\itp}$. By \ih, $\tkmc \proves \iseq'|_{\tagy}, \Sigma$, where $\iseq'|_{\tagy}$ is obtained from $\iseq|_{\tagy}$ by adding $w R_{\charx} u$ and replacing $w : \xbox A$ with $u : A$. As $u$ is not common, we have $u \not\in \lab(\itp)$ by condition (4) of \dfn~\ref{def:interp-sequent}, and so, $u$ occurs neither in $\Sigma$, as $\lab(\orth{\itp}) \subseteq \lab(\itp)$, nor in $\iseq|_{\tagy}$, as $u$ is fresh. Hence, applying $\xboxr$ with principal formula $w : \xbox A$ to $\iseq'|_{\tagy}, \Sigma$ yields $\iseq|_{\tagy}, \Sigma$.

\textit{$\ddrIc$.} In this case, we let $\itp$ denote the interpolant of the premise $\iseq' = \lsets{\cst \cup \set{u}}{\pst{\tagx}}{\pst{\tagy}} \sep \tel, w R_{\charx} u \sar \Gamma_{1} \splt \Gamma_{2} \sep \itp$, where $w \in \cst$, so that the interpolant of the conclusion is $\xbi$. As $u$ is common in $\iseq'$, both projections of $\iseq'$ are obtained from those of $\iseq$ by adding $w R_{\charx} u$. As in the $\xboxrIc$ case, claim (1) follows from \ih\ by \lem~\ref{lem:box-interpolants}(1), and \ih\ together with \lem~\ref{lem:box-interpolants}(2) yields $\tkmc \proves \iseq'|_{\tagy}, \Sigma$ for each $\Sigma \in \orth{\xbi}$. As $u$ does not occur in $\iseq|_{\tagy}, \Sigma$ and $w$ occurs in $\iseq|_{\tagy}$, an application of $\ddr$, which is a rule of $\tkmc$ as $\D \in \fcons$, yields $\iseq|_{\tagy}, \Sigma$, which proves claim (2).

\textit{$\ddrIf$.} The premise is $\iseq' = \lsets{\cst}{\pst{\tagx}}{\pst{\tagy} \cup \set{u}} \sep \tel, w R_{\charx} u \sar \Gamma_{1} \splt \Gamma_{2} \sep \itp$, where $w \in \cst \cup \pst{\tagy}$ by condition (2) of \dfn~\ref{def:interp-sequent}. As in the $\xboxrIf$ case, $\iseq'|_{\tagx} = \iseq|_{\tagx}$, which proves claim (1). For claim (2), let $\Sigma \in \orth{\itp}$. By \ih, $\tkmc \proves \iseq'|_{\tagy}, \Sigma$, where $\iseq'|_{\tagy}$ is obtained from $\iseq|_{\tagy}$ by adding $w R_{\charx} u$. As $u \not\in \lab(\itp)$ by condition (4) of \dfn~\ref{def:interp-sequent} and $w$ occurs in $\iseq|_{\tagy}$, applying $\ddr$ yields $\iseq|_{\tagy}, \Sigma$.
\end{proof}

\subsection{The Strong Lyndon Interpolation Theorem}\label{subsec:strong-lip-thm}

By condition (4) of \dfn~\ref{def:interp-sequent}, the interpolant $\itp$ of the conclusion of $\intprf$ contains no label other than $\rta$, which is the only common label of the conclusion. Such an interpolant can be interpreted as a formula in conjunctive normal form (cf.~\cite{Lyo21thesis,LyoTiuGorClo20}).

\begin{definition}[Formula Interpretation of Interpolants]\label{def:itp-formula}
Let $\itp = \{(w : C_{i,1}, \ldots, w : C_{i,k_{i}}) \mid i \in [n]\}$ be an interpolant with $\lab(\itp) \subseteq \set{w}$. We define the \emph{formula interpretation} of $\itp$ to be the formula
$$
\textstyle\bigwedge \bigvee \itp := \displaystyle\bigwedge_{i \in [n]} \bigvee_{j \in [k_{i}]} C_{i,j},
$$
where we fix an arbitrary order and bracketing of the conjuncts and disjuncts, and where the empty disjunction and empty conjunction are taken to be $\bot$ and $\top$, respectively.
\end{definition}

Observe that $\sigb{\bigwedge \bigvee \itp} = \sigb{\itp}$ for each $\isym \in \set{+,-,\albet}$, as $\top$ and $\bot$ have empty signatures. Moreover, by \dfn~\ref{def:negation}, $\negnnf{\bigwedge \bigvee \itp} = \bigvee_{i \in [n]} \bigwedge_{j \in [k_{i}]} \negnnf{C_{i,j}}$, where the empty conjunction and disjunction are $\top$ and $\bot$, respectively.

\begin{lemma}\label{lem:sub-interpolation-lemma-1}
Let $\tel \sar \Gamma$ be a tree sequent with $w \in \lab(\tel \sar \Gamma)$ and let $\itp$ be an interpolant with $\lab(\itp) \subseteq \set{w}$. If $\tkmc \proves (\tel \sar \Gamma, \Sigma)$ for each $\Sigma \in \itp$, then $\tkmc \proves (\tel \sar \Gamma, w : \bigwedge \bigvee \itp)$.
\end{lemma}

\begin{proof} Let $\itp = \{(w : C_{i,1}, \ldots, w : C_{i,k_{i}}) \mid i \in [n]\}$. For each $i \in [n]$, we obtain a proof of $\tel \sar \Gamma, w : \bigvee_{j \in [k_{i}]} C_{i,j}$ from the proof of $\tel \sar \Gamma, w : C_{i,1}, \ldots, w : C_{i,k_{i}}$ by $k_{i} {-} 1$ applications of $\disr$ if $k_{i} \geq 1$, and by an application of $\wk$ introducing $w : \bot$ if $k_{i} = 0$. If $n \geq 1$, then $n{-}1$ applications of $\conr$ yield the desired conclusion. If $n = 0$, then $\bigwedge \bigvee \itp = \top$, and the desired conclusion is an instance of $\top$.
\end{proof}

\begin{lemma}\label{lem:sub-interpolation-lemma-2}
Let $\tel \sar \Gamma$ be a tree sequent with $w \in \lab(\tel \sar \Gamma)$ and let $\itp$ be an interpolant with $\lab(\itp) \subseteq \set{w}$. If $\tkmc \proves (\tel \sar \Gamma, \Sigma)$ for each $\Sigma \in \orth{\itp}$, then $\tkmc \proves (\tel \sar \Gamma, w : \negnnf{\bigwedge \bigvee \itp})$.
\end{lemma}

\begin{proof} Let $\itp = \{(w : C_{i,1}, \ldots, w : C_{i,k_{i}}) \mid i \in [n]\}$, so that $\negnnf{\bigwedge \bigvee \itp} = \bigvee_{i \in [n]} D_{i}$ with $D_{i} := \bigwedge_{j \in [k_{i}]} \negnnf{C_{i,j}}$. If $n = 0$, then $\orth{\itp} = \{\emptyset\}$, so $\tel \sar \Gamma$ is provable by assumption, and $\tel \sar \Gamma, w : \bot$ is provable by $\wk$. Suppose that $n \geq 1$. By $n{-}1$ applications of $\disr$, it suffices to prove $\tel \sar \Gamma, w : D_{1}, \ldots, w : D_{n}$. If $k_{i} = 0$ for some $i \in [n]$, then $D_{i} = \top$, and this sequent is an instance of $\top$. Otherwise, for every function $f$ mapping each $i \in [n]$ to an element of $[k_{i}]$, we have $(w : \negnnf{C_{1,f(1)}}, \ldots, w : \negnnf{C_{n,f(n)}}) \in \orth{\itp}$ by \dfn~\ref{def:orthogonal}, and so, $\tel \sar \Gamma, w : \negnnf{C_{1,f(1)}}, \ldots, w : \negnnf{C_{n,f(n)}}$ is provable by assumption. Applying $\conr$ to these sequents for each conjunction $D_{i}$ yields the desired sequent.
\end{proof}

We are now in a position to prove our main theorem.

\begin{theorem}[Strong Lyndon Interpolation]\label{thm:strong-lip}
Every bridge-free grammar logic $\kmc$ has the strong LIP.
\end{theorem}

\begin{proof} Suppose that $A \imp B \in \kmc$, where $A \imp B = \negnnf{A} \lor B$. By \thm~\ref{thm:tkma-sound-compl}, $\tkmc \proves \ \sar \rta : \negnnf{A} \lor B$, and so, by the invertibility of $\disr$ (\thm~\ref{thm:tree-properties}(3)), there exists a proof $\prf$ of $\sar \rta : \negnnf{A}, \rta : B$ in $\tkmc$. Let $\intprf$ be the interpolation proof of $\lsets{\set{\rta}}{\emptyset}{\emptyset} \sep \emptyset \sar \rta : \negnnf{A} \splti \rta : B \sep \itp$ obtained from $\prf$ by \thm~\ref{thm:interp-proof}, and let $I := \bigwedge \bigvee \itp$, which is well-defined as $\lab(\itp) \subseteq \set{\rta}$ by condition (4) of \dfn~\ref{def:interp-sequent}. We show that $I$ is a strong Lyndon interpolant of $A \imp B$.

\textit{Condition (3).} The $\tagl$-projection of the conclusion of $\intprf$ is $\sar \rta : \negnnf{A}$ and its $\tagr$-projection is $\sar \rta : B$. By \lem~\ref{lem:interpolation-lemma}, $\tkmc \proves  (\sar \rta : \negnnf{A}, \Sigma)$ for each $\Sigma \in \itp$, and $\tkmc \proves (\sar \rta : B, \Sigma)$ for each $\Sigma \in \orth{\itp}$. By Lemmas~\ref{lem:sub-interpolation-lemma-1} and~\ref{lem:sub-interpolation-lemma-2}, we obtain $\tkmc \proves ( \sar \rta : \negnnf{A}, \rta : I)$ and $\tkmc \proves ( \sar \rta : \negnnf{I}, \rta : B)$, and so, $\tkmc \proves ( \sar \rta : \negnnf{A} \lor I)$ and $\tkmc \proves ( \sar \rta : \negnnf{I} \lor B)$ by $\disr$. By \thm~\ref{thm:tkma-sound-compl}, $A \imp I \in \kmc$ and $I \imp B \in \kmc$.

\textit{Conditions (1) and (2).} By \lem~\ref{lem:itp-signature}(1) applied to the conclusion of $\intprf$, together with \lem~\ref{lem:signature-negation}, we have
$$
\begin{array}{l}
\sigp{I} = \sigp{\itp} \subseteq \sign{\negnnf{A}} \cap \sigp{B} = \sigp{A} \cap \sigp{B}, \\[1ex]
\sign{I} = \sign{\itp} \subseteq \sigp{\negnnf{A}} \cap \sign{B} = \sign{A} \cap \sign{B}.
\end{array}
$$
Last, $\sigm{I} = \sigm{\itp} \subseteq \sigm{A} \cap \sigm{B}$ by \lem~\ref{lem:itp-signature}(2).
\end{proof}

We note that our proof of \thm~\ref{thm:strong-lip} is constructive: given a proof of $\sar \rta : \negnnf{A}, \rta : B$ in $\tkmc$, the algorithm $\partalg$, together with the rules of $\tkmci$, computes a strong Lyndon interpolant, and the proofs of $A \imp I$ and $I \imp B$ in $\tkmc$ can be extracted from the proofs of Lemmas~\ref{lem:interpolation-lemma}, \ref{lem:sub-interpolation-lemma-1}, and~\ref{lem:sub-interpolation-lemma-2}. As each of the path conditions $\T$, $\four$, $\B$, and $\five$ is of the form $\R_{\stra} \subseteq \R_{\charx}$ with $\stra \in \set{\charx, \conv{\charx}}^{*}$, and seriality conditions do not affect bridge-freeness, we obtain the following corollary by \lem~\ref{lem:bridge-free-char}.

\begin{corollary}\label{cor:strong-lip-examples}
Let $\fcons$ be a set of path and seriality conditions such that each condition in $\fcons$ is of the form $\D$, $\T$, $\four$, $\B$, or $\five$ for some $\charx \in \albet$. Then, $\kmc$ has the strong LIP. In particular, this holds for the multi-modal tense logics obtained by imposing any combination of seriality, reflexivity, transitivity, symmetry, and Euclideanity on the accessibility relations $\R_{\charx}$ with $\charx \in \albet$.
\end{corollary}


\iffalse
\begin{corollary}\label{cor:strong-lip-examples}
\begin{itemize}

\item[(1)] Let $\fcons$ be a set of path and seriality conditions such that each condition in $\fcons$ is of the form $\D$, $\T$, $\four$, $\B$, or $\five$ for some $\charx \in \albet$. Then, $\kmc$ has the strong LIP. In particular, this holds for the multi-modal tense logics obtained by imposing any combination of seriality, reflexivity, transitivity, symmetry, and Euclideanity on the accessibility relations $\R_{\charx}$ with $\charx \in \albet$.

\item[(2)] Let $\mathcal{O}$ be an ontology consisting of RIAs of the form $r_{1} \circ \cdots \circ r_{n} \sqsubseteq s$ with $r_{1}, \ldots, r_{n} \in \set{s, s^{-}}$ and of seriality axioms of the form $\top \sqsubseteq \exists s.\top$. If $C \sqsubseteq D$ holds in every model of $\mathcal{O}$, where $C$ and $D$ are $\mathcal{ALCI}$ concepts, then there exists an $\mathcal{ALCI}$ concept $I$ such that $C \sqsubseteq I$ and $I \sqsubseteq D$ hold in every model of $\mathcal{O}$, every role name occurring in $I$ occurs in both $C$ and $D$, and every concept name occurring positively (negatively) in $I$ occurs positively (negatively) in both $C$ and $D$.

\end{itemize}
\end{corollary}

\begin{proof} (1) Each of the path conditions $\T$, $\four$, $\B$, and $\five$ is of the form $\R_{\stra} \subseteq \R_{\charx}$ with $\stra \in \set{\charx, \conv{\charx}}^{*}$, and seriality conditions do not affect bridge-freeness. Hence, $\kmc$ is bridge-free by \lem~\ref{lem:bridge-free-char}, and the claim follows from \thm~\ref{thm:strong-lip}.

(2) Under the standard correspondence between modal and description logics, concept names correspond to propositional atoms, role names and their inverses to characters and their converses, and the concept constructors $\forall r$ and $\exists r$ to the modalities $\xbox$ and $\xdia$, where $\charx$ is the character corresponding to $r$. Models of $\mathcal{ALCI}$ correspond to the models of \dfn~\ref{def:model}, where the converse condition of \dfn~\ref{def:frame} ensures that inverse roles are interpreted correctly. Moreover, an RIA $r_{1} \circ \cdots \circ r_{n} \sqsubseteq s$ corresponds to the path condition $\R_{\charx_{1} \cdots \charx_{n}} \subseteq \R_{\charx}$, where $\charx_{1}, \ldots, \charx_{n}, \charx$ are the characters corresponding to $r_{1}, \ldots, r_{n}, s$, and a seriality axiom $\top \sqsubseteq \exists s.\top$ to the seriality condition $\D$. Let $\fcons$ be the set of path and seriality conditions corresponding to $\mathcal{O}$, and let $A$ and $B$ be the formulae corresponding to the negation normal forms of $C$ and $D$, respectively. Then, $C \sqsubseteq D$ holds in every model of $\mathcal{O}$ \iffi $A \imp B$ is true in every $\fcons$-model, i.e., \iffi $A \imp B \in \kmc$ by \dfn~\ref{def:fcons-valid}. As each RIA of $\mathcal{O}$ only involves a single role name and its inverse, $\kmc$ is bridge-free by \lem~\ref{lem:bridge-free-char}, so \thm~\ref{thm:strong-lip} yields a strong Lyndon interpolant of $A \imp B$. The concept $I$ corresponding to this interpolant has the desired properties: $C \sqsubseteq I$ and $I \sqsubseteq D$ hold in every model of $\mathcal{O}$ by condition (3) of \dfn~\ref{def:strong-lip}, the restriction on role names is condition (2), and the restriction on concept names is condition (1).
\end{proof}
\fi

\section{Concluding Remarks}\label{sec:conclusion}

In this paper, we established the strong Lyndon interpolation property for bridge-free grammar logics with converses and seriality, which requires interpolants to respect both the polarities of propositional atoms and the indices of modalities, in line with signature-based interpolation in description logics. This class of logics includes the multi-modal tense logics obtained by extending $\km$ with any combination of seriality, reflexivity, transitivity, symmetry, and Euclideanity for each pair of converse modalities, and thus subsumes the indexed tense logics of Sano and Yamasaki~\cite{SanYam26} as well as (since converse modalities can be eliminated in the presence of symmetry) the multi-agent logic $\mathsf{S5}_{n}$ considered by B{\'\i}lkov{\'a} et al.~\cite{BilFusKuz25}.

We also showed that the strong LIP fails for certain grammar logics with bridge axioms, which makes the strong LIP a proper strengthening of the LIP over the class of grammar logics. To obtain our main result, we introduced a tripartite partition method for tree sequent proofs, which sorts the labels of each sequent into common labels and labels private to the left or right component, and which only permits relational atoms pointing to common labels to contribute modalities to interpolants. This method was realized by means of a partition algorithm and an interpolation calculus, which together compute a strong Lyndon interpolant, along with proofs of the corresponding implications, from any tree sequent proof of an implication. Our approach is therefore constructive and purely syntactic, and requires no post-processing of the computed interpolants. 

Our results also clarify the status of Craig interpolation in description logics (DLs). Under the standard correspondence between modal and description logics, propositional atoms correspond to concept names, modal indices to role names, converse modalities to inverse roles, path conditions to complex role inclusion axioms (RIAs) $r_{1} \circ \cdots \circ r_{n} \sqsubseteq s$, and seriality conditions to seriality axioms $\top \sqsubseteq \exists s.\top$. In DLs, interpolation is formulated relative to ontologies: a DL has the CIP if, whenever $\mathcal{O}_{1} \cup \mathcal{O}_{2} \models C_{1} \sqsubseteq C_{2}$, there exists a concept $I$ with $\mathcal{O}_{1} \cup \mathcal{O}_{2} \models C_{1} \sqsubseteq I$ and $\mathcal{O}_{1} \cup \mathcal{O}_{2} \models I \sqsubseteq C_{2}$ that only uses concept \emph{and} role names occurring both in $\mathcal{O}_{1}$ or $C_{1}$ and in $\mathcal{O}_{2}$ or $C_{2}$~\cite{ArtJunMazOzaWol23,TenFraSey13}.

It is known that the CIP fails in the presence of role inclusions: already $\mathcal{ALCH}$, the extension of $\mathcal{ALC}$ with role inclusions, lacks the CIP, and so do standard DLs containing it, such as $\mathcal{RIQ}$~\cite{ArtJunMazOzaWol23}. Consequently, deciding the existence of interpolants in such DLs becomes a non-trivial problem, even if the ontologies only contain role inclusions~\cite{ArtJunMazOzaWol23,JunMazWol22}. Positive results therefore either weaken the notion of interpolant, as in the case of the \emph{concept} interpolation property for $\mathcal{RIQ}$, which only requires common concept names in an interpolant~\cite{LyoKar24}, or restrict the admissible ontologies, as we explain below.

The present work shows that complex RIAs are not necessarily an obstacle, at least for ontologies without general concept inclusions. By \lem~\ref{lem:bridge-free-char}, a set of RIAs is bridge-free \iffi each of its RIAs only involves a single role name and its inverse, i.e., is of the form $r_{1} \circ \cdots \circ r_{n} \sqsubseteq s$ with $r_{1}, \ldots, r_{n} \in \set{s, s^{-}}$, as in the case of transitivity $s \circ s \sqsubseteq s$ or symmetry $s^{-} \sqsubseteq s$. Let $\mathcal{O}$ be an ontology consisting of bridge-free RIAs and possibly seriality axioms, and let $\fcons$ be the corresponding set of path and seriality conditions. As the models of $\mathcal{O}$ correspond to $\fcons$-models, $\mathcal{O} \models C \sqsubseteq D$ holds for $\mathcal{ALCI}$ concepts $C$ and $D$ \iffi the corresponding implication belongs to the bridge-free grammar logic $\kmc$ (cf.\ \dfn~\ref{def:fcons-valid}). Hence, \thm~\ref{thm:strong-lip} yields an $\mathcal{ALCI}$ concept $I$ with $\mathcal{O} \models C \sqsubseteq I$ and $\mathcal{O} \models I \sqsubseteq D$ that only uses concept and role names occurring in both $C$ and $D$, and that additionally respects the polarities of concept names. 
As the union of two such ontologies is again of this form, this yields the CIP in the above sense for $\mathcal{ALCI}$ with ontologies consisting of bridge-free RIAs and seriality axioms, and even its Lyndon variant, in which every concept name occurring positively (negatively) in the interpolant additionally occurs positively (negatively) in both $C_{1}$ and $C_{2}$. By contrast, RIAs that relate distinct role names, such as $r \sqsubseteq s$, are bridges, and the strong LIP may fail in their presence (see the counter-example in \sect~\ref{subsec:strong-lip}).

There are several directions for future research. First, it would be interesting to investigate the strong LIP for broader classes of multi-modal logics. While bridge-freeness is sufficient for the strong LIP, our counter-example (\sect~\ref{subsec:strong-lip}) only shows that the property fails in the presence of certain bridge axioms. This raises the question of whether grammar logics with converses that enjoy the strong LIP can be characterized precisely, for instance, in terms of the production rules of their grammars. Second, it would be natural to adapt the tripartite partition method to multi-modal logics beyond grammar logics, whose frame conditions cannot be expressed as path conditions, such as extensions with certain Scott--Lemmon axioms; for such logics, labeled sequent calculi~\cite{Kuz16,Sim94,Vig00} appear better suited than tree sequents. 
Third, it would be natural to extend our results to DL ontologies with general concept inclusions, which correspond to the global consequence relation and relative to which signature-based interpolation and Beth definability are usually formulated in DLs~\cite{TenFraSey13}; this does not follow from our results by the usual argument, and it would be interesting to determine for which bridge-free grammar logics the strong LIP extends to this setting, and whether explicit definitions can then be computed in the style of~\cite{LyoKar24}.

\appendix

\section{Supplementary Material}

\subsection{Summary of Terminology}\label{app:terminology}

Table~\ref{tab:terminology} summarizes the notions used in \sect~\ref{subsec:sides} to classify the formula occurrences, relational atoms, and labels of the input proof $\prf$, together with their effect on the partitioned proof $\partprf$ computed by the partition algorithm of \sect~\ref{subsec:partition}. In $\partprf$, sides determine the component of each formula occurrence, and the status of each label determines whether it belongs to the set $\cst$, $\lst$, or $\rst$ of each partitioned sequent (\lem~\ref{lem:part}); the relational atoms of each partitioned sequent are simply those of the corresponding sequent of $\prf$.

\begin{table}[pos=H]
\caption{The notions of \sect~\ref{subsec:sides} for objects of $\prf$, and their effect on the partitioned proof $\partprf$ (\sect~\ref{subsec:partition}), where $\tagz \in \set{\tagl, \tagr}$.\label{tab:terminology}}
\centering
\small
\renewcommand{\arraystretch}{1.2}
\begin{tabular}{@{}>{\raggedright\arraybackslash}p{0.20\textwidth} >{\raggedright\arraybackslash}p{0.14\textwidth} >{\raggedright\arraybackslash}p{0.31\textwidth} >{\raggedright\arraybackslash}p{0.25\textwidth}@{}}
\toprule
Object & Notion & Meaning in $\prf$ & Effect in $\partprf$ \\
\midrule
formula occurrence & side $\tagz$ & stems from $\rta : \negnnf{A}$ if $\tagz = \tagl$, and from $\rta : B$ if $\tagz = \tagr$ & lies in the $\tagz$-component \\
\midrule
relational atom & used by $\tagz$ & introduced by a $\xboxr$ inference with side $\tagz$, or traversed by an inference with side $\tagz$ & --- \\
\midrule
label $v$ & common & the root, or a child of a common label via a relational atom used by both sides & lies in $\cst$; occurs in both projections \\
 & $\tagz$-private & a child of a $\tagz$-private label, or of a common label via a relational atom used only by $\tagz$ (or, if $\tagz = \tagr$, by neither side) & lies in $\pst{\tagz}$; never occurs in an interpolant \\
\midrule
relational atom & common & its target is common & introduced by $\xboxrIc$ or $\ddrIc$ \\
 & $\tagz$-private & its target is $\tagz$-private & introduced by $\xboxrIf$ or $\ddrIf$ \\
\bottomrule
\end{tabular}
\end{table}

The following facts are useful to keep in mind when reading \sect~\ref{sec:part-bias}: a common relational atom has a common source and is used by both sides (\dfn~\ref{def:label-status}); no formula occurrence of side $\optag{\tagz}$ has a $\tagz$-private label; and no inference with side $\optag{\tagz}$ traverses a $\tagz$-private relational atom (\lem~\ref{lem:one-sided}).

\bibliographystyle{elsarticle-num-names-alpha}
\bibliography{bibliography}

@article{ArtJunMazOzaWol23,
  author  = {Artale, Alessandro and Jung, Jean Christoph and Mazzullo, Andrea and Ozaki, Ana and Wolter, Frank},
  title   = {Living without {B}eth and {C}raig: Definitions and interpolants in description and modal logics with nominals and role inclusions},
  journal = {ACM Transactions on Computational Logic},
  volume  = {24},
  number  = {4},
  pages   = {1--51},
  year    = {2023},
  doi     = {10.1145/3597301}
}

@article{vBe97,
  author  = {van Benthem, Johan},
  title   = {Modal foundations for predicate logic},
  journal = {Logic Journal of the IGPL},
  volume  = {5},
  number  = {2},
  pages   = {259--286},
  year    = {1997},
  doi     = {10.1093/jigpal/5.2.259}
}

@inproceedings{JunMazWol22,
  author    = {Jung, Jean Christoph and Mazzullo, Andrea and Wolter, Frank},
  title     = {More on interpolants and explicit definitions for description logics with nominals and/or role inclusions},
  booktitle = {Proceedings of the 35th International Workshop on Description Logics ({DL} 2022)},
  editor    = {Arieli, Ofer and Homola, Martin and Jung, Jean Christoph and Mugnier, Marie-Laure},
  series    = {CEUR Workshop Proceedings},
  volume    = {3263},
  publisher = {CEUR-WS.org},
  year      = {2022}
}

@inproceedings{BorTomWed16,
  author    = {Borgida, Alexander and Toman, David and Weddell, Grant E.},
  title     = {On referring expressions in query answering over first order knowledge bases},
  booktitle = {Principles of Knowledge Representation and Reasoning: Proceedings of the Fifteenth International Conference ({KR} 2016)},
  editor    = {Baral, Chitta and Delgrande, James P. and Wolter, Frank},
  publisher = {AAAI Press},
  year      = {2016},
  pages     = {319--328}
}

@inproceedings{HudTomWed15,
  author    = {Hudek, Alexander and Toman, David and Weddell, Grant},
  title     = {On enumerating query plans using analytic tableau},
  booktitle = {Automated Reasoning with Analytic Tableaux and Related Methods},
  editor    = {de Nivelle, Hans},
  series    = {Lecture Notes in Computer Science},
  volume    = {9323},
  publisher = {Springer},
  year      = {2015},
  pages     = {339--354},
  doi       = {10.1007/978-3-319-24312-2_23}
}

@incollection{Aqv84,
  author    = {{\AA}qvist, Lennart},
  title     = {Deontic logic},
  booktitle = {Handbook of Philosophical Logic, Volume {II}: Extensions of Classical Logic},
  editor    = {Gabbay, Dov and Guenthner, Franz},
  publisher = {D. Reidel},
  address   = {Dordrecht},
  year      = {1984},
  pages     = {605--714},
  doi       = {10.1007/978-94-009-6259-0_11}
}

@book{Pri67,
  author    = {Prior, Arthur N.},
  title     = {Past, present and future},
  publisher = {Clarendon Press},
  address   = {Oxford},
  year      = {1967}
}

@book{Pri57,
  author    = {Arthur N. Prior},
  title     = {Time and modality},
  publisher = {Clarendon Press},
  address   = {Oxford},
  year      = {1957}
}

@inproceedings{BalGioMar98,
  author    = {Baldoni, Matteo and Giordano, Laura and Martelli, Alberto},
  title     = {A tableau calculus for multimodal logics and some (un)decidability results},
  booktitle = {Automated Reasoning with Analytic Tableaux and Related Methods},
  editor    = {de Swart, Harrie},
  series    = {Lecture Notes in Computer Science},
  volume    = {1397},
  publisher = {Springer},
  year      = {1998},
  pages     = {44--59},
  doi       = {10.1007/3-540-69778-0_13}
}

@inproceedings{BilFusKuz25,
  author    = {B{\'\i}lkov{\'a}, Marta and Fussner, Wesley and Kuznets, Roman},
  title     = {Agent interpolation in distributed systems},
  booktitle = {Relational and Algebraic Methods in Computer Science ({RAMiCS} 2026)},
  editor    = {Fahrenberg, Uli and Fussner, Wesley and Santocanale, Luigi},
  series    = {Lecture Notes in Computer Science},
  volume    = {16526},
  publisher = {Springer},
  year      = {2026},
  pages     = {95--113},
  doi       = {10.1007/978-3-032-22469-9_6}
}

@inproceedings{MurSan20,
  author    = {Murai, Ryo and Sano, Katsuhiko},
  title     = {Craig interpolation of epistemic logics with distributed knowledge},
  booktitle = {Foundations of Information and Knowledge Systems},
  editor    = {Herzig, Andreas and Kontinen, Juha},
  series    = {Lecture Notes in Computer Science},
  volume    = {12012},
  publisher = {Springer},
  year      = {2020},
  pages     = {211--221},
  doi       = {10.1007/978-3-030-39951-1_13}
}

@inproceedings{Mar98,
  author    = {Marx, Maarten},
  title     = {Interpolation in modal logic},
  booktitle = {Algebraic Methodology and Software Technology},
  series    = {Lecture Notes in Computer Science},
  volume    = {1548},
  publisher = {Springer},
  year      = {1998},
  pages     = {154--163},
  doi       = {10.1007/3-540-49253-4_13}
}

@incollection{Kur07,
  author    = {Kurucz, Agi},
  title     = {Combining modal logics},
  booktitle = {Handbook of Modal Logic},
  editor    = {Blackburn, Patrick and van Benthem, Johan and Wolter, Frank},
  series    = {Studies in Logic and Practical Reasoning},
  volume    = {3},
  publisher = {Elsevier},
  address   = {Amsterdam},
  year      = {2007},
  pages     = {869--924},
  doi       = {10.1016/S1570-2464(07)80018-8}
}

@misc{Lyo26,
  author        = {Lyon, Tim S.},
  title         = {Nested sequents for intuitionistic multi-modal logics: Modularity, cut-elimination, and undecidability},
  year          = {2026},
  note          = {Preprint, arXiv:2511.22174}
}

@article{Bru09,
  author  = {Br{\"u}nnler, Kai},
  title   = {Deep sequent systems for modal logic},
  journal = {Archive for Mathematical Logic},
  volume  = {48},
  number  = {6},
  pages   = {551--577},
  year    = {2009},
  doi     = {10.1007/s00153-009-0137-3}
}

@incollection{CerHer95,
  author    = {Fari{\~n}as del Cerro, Luis and Herzig, Andreas},
  title     = {Modal deduction with applications in epistemic and temporal logics},
  booktitle = {Handbook of Logic in Artificial Intelligence and Logic Programming, Volume 4: Epistemic and Temporal Reasoning},
  editor    = {Gabbay, Dov M. and Hogger, C. J. and Robinson, J. A.},
  publisher = {Oxford University Press},
  address   = {Oxford},
  year      = {1995},
  pages     = {499--594}
}

@article{CerPen88,
  author  = {Fari{\~n}as del Cerro, Luis and Penttonen, Martti},
  title   = {Grammar logics},
  journal = {Logique et Analyse},
  volume  = {31},
  number  = {121/122},
  pages   = {123--134},
  year    = {1988}
}

@article{CiaLyoRamTiu21,
  author  = {Ciabattoni, Agata and Lyon, Tim S. and Ramanayake, Revantha and Tiu, Alwen},
  title   = {Display to labelled proofs and back again for tense logics},
  journal = {ACM Transactions on Computational Logic},
  volume  = {22},
  number  = {3},
  pages   = {1--31},
  year    = {2021}
}

@article{Cra57,
  author  = {Craig, William},
  title   = {Three uses of the {H}erbrand--{G}entzen theorem in relating model theory and proof theory},
  journal = {The Journal of Symbolic Logic},
  volume  = {22},
  number  = {3},
  pages   = {269--285},
  year    = {1957},
  doi     = {10.2307/2963594}
}

@article{DemNiv05,
  author  = {Demri, St{\'e}phane and de Nivelle, Hans},
  title   = {Deciding regular grammar logics with converse through first-order logic},
  journal = {Journal of Logic, Language and Information},
  volume  = {14},
  number  = {3},
  pages   = {289--329},
  year    = {2005}
}

@article{Dem01,
  author  = {Demri, St{\'e}phane},
  title   = {The complexity of regularity in grammar logics and related modal logics},
  journal = {Journal of Logic and Computation},
  volume  = {11},
  number  = {6},
  pages   = {933--960},
  year    = {2001}
}

@article{Ewa86,
  author  = {Ewald, W. B.},
  title   = {Intuitionistic tense and modal logic},
  journal = {The Journal of Symbolic Logic},
  volume  = {51},
  number  = {1},
  pages   = {166--179},
  year    = {1986}
}

@book{FagMosHalVar95,
  author    = {Fagin, Ronald and Halpern, Joseph Y. and Moses, Yoram and Vardi, Moshe Y.},
  title     = {Reasoning about knowledge},
  publisher = {MIT Press},
  address   = {Cambridge, MA},
  year      = {1995}
}

@article{FitKuz15,
  author  = {Fitting, Melvin and Kuznets, Roman},
  title   = {Modal interpolation via nested sequents},
  journal = {Annals of Pure and Applied Logic},
  volume  = {166},
  number  = {3},
  pages   = {274--305},
  year    = {2015}
}

@inproceedings{GorRam12,
  author    = {Gor{\'e}, Rajeev and Ramanayake, Revantha},
  title     = {Labelled tree sequents, tree hypersequents and nested (deep) sequents},
  booktitle = {Advances in Modal Logic, Volume 9},
  editor    = {Bolander, Thomas and Bra{\"u}ner, Torben and Ghilardi, Silvio and Moss, Lawrence S.},
  publisher = {College Publications},
  year      = {2012},
  pages     = {279--299}
}

@article{GorPosTiu11,
  author  = {Gor{\'e}, Rajeev and Postniece, Linda and Tiu, Alwen},
  title   = {On the correspondence between display postulates and deep inference in nested sequent calculi for tense logics},
  journal = {Logical Methods in Computer Science},
  volume  = {7},
  number  = {2:8},
  pages   = {1--38},
  year    = {2011},
  doi     = {10.2168/LMCS-7(2:8)2011}
}

@inproceedings{GorNgu05,
  author    = {Gor{\'e}, Rajeev and Nguyen, Linh Anh},
  title     = {A tableau calculus with automaton-labelled formulae for regular grammar logics},
  booktitle = {Automated Reasoning with Analytic Tableaux and Related Methods},
  editor    = {Beckert, Bernhard},
  series    = {Lecture Notes in Computer Science},
  volume    = {3702},
  publisher = {Springer},
  year      = {2005},
  pages     = {138--152}
}

@article{HorSat04,
  author  = {Horrocks, Ian and Sattler, Ulrike},
  title   = {Decidability of {SHIQ} with complex role inclusion axioms},
  journal = {Artificial Intelligence},
  volume  = {160},
  number  = {1--2},
  pages   = {79--104},
  year    = {2004}
}

@inproceedings{IshKik07,
  author    = {Ishigaki, Ryo and Kikuchi, Kentaro},
  title     = {Tree-sequent methods for subintuitionistic predicate logics},
  booktitle = {Automated Reasoning with Analytic Tableaux and Related Methods},
  editor    = {Olivetti, Nicola},
  series    = {Lecture Notes in Computer Science},
  volume    = {4548},
  publisher = {Springer},
  year      = {2007},
  pages     = {149--164}
}

@article{Kas94,
  author  = {Kashima, Ryo},
  title   = {Cut-free sequent calculi for some tense logics},
  journal = {Studia Logica},
  volume  = {53},
  number  = {1},
  pages   = {119--135},
  year    = {1994}
}

@article{KihOno10,
  author  = {Kihara, Hitoshi and Ono, Hiroakira},
  title   = {Interpolation properties, {B}eth definability properties and amalgamation properties for substructural logics},
  journal = {Journal of Logic and Computation},
  volume  = {20},
  number  = {4},
  pages   = {823--875},
  year    = {2010}
}

@inproceedings{KuzLel18,
  author    = {Kuznets, Roman and Lellmann, Bj{\"o}rn},
  title     = {Interpolation for intermediate logics via hyper- and linear nested sequents},
  booktitle = {Advances in Modal Logic, Volume 12},
  editor    = {Bezhanishvili, Guram and D'Agostino, Giovanna and Metcalfe, George and Studer, Thomas},
  publisher = {College Publications},
  year      = {2018},
  pages     = {473--492}
}

@article{Kuz18,
  author  = {Kuznets, Roman},
  title   = {Multicomponent proof-theoretic method for proving interpolation properties},
  journal = {Annals of Pure and Applied Logic},
  volume  = {169},
  number  = {12},
  pages   = {1369--1418},
  year    = {2018}
}

@incollection{Kuz16b,
  author    = {Kuznets, Roman},
  title     = {Craig interpolation via hypersequents},
  booktitle = {Concepts of Proof in Mathematics, Philosophy, and Computer Science},
  editor    = {Probst, Dieter and Schuster, Peter},
  series    = {Ontos Mathematical Logic},
  volume    = {6},
  publisher = {De Gruyter},
  address   = {Berlin},
  year      = {2016},
  pages     = {193--214},
  doi       = {10.1515/9781501502620-012}
}

@inproceedings{Kuz16,
  author    = {Kuznets, Roman},
  title     = {Proving {C}raig and {L}yndon interpolation using labelled sequent calculi},
  booktitle = {Logics in Artificial Intelligence},
  editor    = {Michael, Loizos and Kakas, Antonis},
  series    = {Lecture Notes in Computer Science},
  volume    = {10021},
  publisher = {Springer},
  year      = {2016},
  pages     = {320--335}
}

@article{Bet56,
  author  = {Beth, Evert Willem},
  title   = {On {P}adoa's method in the theory of definition},
  journal = {Indagationes Mathematicae},
  volume  = {15},
  year    = {1953},
  pages   = {330--339},
  doi     = {10.1016/S1385-7258(53)50042-3}
}

@article{Lyn59,
  author  = {Lyndon, Roger C.},
  title   = {An interpolation theorem in the predicate calculus},
  journal = {Pacific Journal of Mathematics},
  volume  = {9},
  number  = {1},
  pages   = {129--142},
  year    = {1959}
}

@inproceedings{Lyo21b,
  author    = {Lyon, Tim S.},
  title     = {A framework for intuitionistic grammar logics},
  booktitle = {Logic and Argumentation},
  editor    = {Baroni, Pietro and Benzm{\"u}ller, Christoph and W{\'a}ng, Y{\`\i} N.},
  series    = {Lecture Notes in Computer Science},
  volume    = {13040},
  publisher = {Springer},
  year      = {2021},
  pages     = {495--503},
  doi       = {10.1007/978-3-030-89391-0_29}
}

@phdthesis{Lyo21thesis,
  author = {Lyon, Tim S.},
  title  = {Refining labelled systems for modal and constructive logics with applications},
  school = {Technische Universit{\"a}t Wien},
  year   = {2021}
}

@inproceedings{LutWol11,
  author    = {Lutz, Carsten and Wolter, Frank},
  title     = {Foundations for uniform interpolation and forgetting in expressive description logics},
  booktitle = {Proceedings of the Twenty-Second International Joint Conference on Artificial Intelligence ({IJCAI} 2011)},
  editor    = {Walsh, Toby},
  publisher = {IJCAI/AAAI},
  year      = {2011},
  pages     = {989--995},
  doi       = {10.5591/978-1-57735-516-8/IJCAI11-170}
}

@inproceedings{LyoTiuGorClo20,
  author    = {Lyon, Tim S. and Tiu, Alwen and Gor{\'e}, Rajeev and Clouston, Ranald},
  title     = {Syntactic interpolation for tense logics and bi-intuitionistic logic via nested sequents},
  booktitle = {28th {EACSL} Annual Conference on Computer Science Logic ({CSL} 2020)},
  editor    = {Fern{\'a}ndez, Maribel and Muscholl, Anca},
  series    = {Leibniz International Proceedings in Informatics ({LIPIcs})},
  volume    = {152},
  publisher = {Schloss Dagstuhl -- Leibniz-Zentrum f{\"u}r Informatik},
  year      = {2020},
  pages     = {28:1--28:16},
  doi       = {10.4230/LIPIcs.CSL.2020.28}
}

@article{Mae60,
  author  = {Maehara, Shoji},
  title   = {On the interpolation theorem of {C}raig},
  journal = {S{\^u}gaku},
  volume  = {12},
  number  = {4},
  pages   = {235--237},
  year    = {1960}
}

@incollection{McM18,
  author    = {McMillan, Kenneth L.},
  title     = {Interpolation and model checking},
  booktitle = {Handbook of Model Checking},
  editor    = {Clarke, Edmund M. and Henzinger, Thomas A. and Veith, Helmut and Bloem, Roderick},
  publisher = {Springer},
  address   = {Cham},
  year      = {2018},
  pages     = {421--446},
  doi       = {10.1007/978-3-319-10575-8_14}
}

@article{TenFraSey13,
  author  = {ten Cate, Balder and Franconi, Enrico and Seylan, {\.I}nan{\c{c}}},
  title   = {Beth definability in expressive description logics},
  journal = {Journal of Artificial Intelligence Research},
  volume  = {48},
  pages   = {347--414},
  year    = {2013}
}

@article{SanYam26,
  author  = {Sano, Katsuhiko and Yamasaki, Sakiko},
  title   = {Craig interpolation of indexed tense logics via analytic sequent calculi},
  journal = {Studia Logica},
  year    = {2026},
  note    = {Online first},
  doi     = {10.1007/s11225-026-10263-8}
}

@article{NguSza11,
  author  = {Nguyen, Linh Anh and Sza{\l}as, Andrzej},
  title   = {{ExpTime} tableau decision procedures for regular grammar logics with converse},
  journal = {Studia Logica},
  volume  = {98},
  number  = {3},
  pages   = {387--428},
  year    = {2011}
}

@inproceedings{PloSti86,
  author    = {Plotkin, Gordon and Stirling, Colin},
  title     = {A framework for intuitionistic modal logics: Extended abstract},
  booktitle = {Proceedings of the 1st Conference on Theoretical Aspects of Reasoning about Knowledge ({TARK} 1986)},
  editor    = {Halpern, Joseph Y.},
  publisher = {Morgan Kaufmann},
  year      = {1986},
  pages     = {399--406}
}

@incollection{Pog09,
  author    = {Poggiolesi, Francesca},
  title     = {The method of tree-hypersequents for modal propositional logic},
  booktitle = {Towards Mathematical Philosophy},
  editor    = {Makinson, David and Malinowski, Jacek and Wansing, Heinrich},
  series    = {Trends in Logic},
  volume    = {28},
  publisher = {Springer},
  address   = {Dordrecht},
  year      = {2009},
  pages     = {31--51},
  doi       = {10.1007/978-1-4020-9084-4_3}
}

@article{Pos47,
  author  = {Post, Emil L.},
  title   = {Recursive unsolvability of a problem of {T}hue},
  journal = {The Journal of Symbolic Logic},
  volume  = {12},
  number  = {1},
  pages   = {1--11},
  year    = {1947}
}

@article{Fis84,
  author  = {Fischer Servi, Gis{\`e}le},
  title   = {Axiomatizations for some intuitionistic modal logics},
  journal = {Rendiconti del Seminario Matematico Universit{\`a} e Politecnico di Torino},
  volume  = {42},
  number  = {3},
  pages   = {179--194},
  year    = {1984}
}

@phdthesis{Sim94,
  author = {Simpson, Alex K.},
  title  = {The proof theory and semantics of intuitionistic modal logic},
  school = {University of Edinburgh},
  year   = {1994}
}

@incollection{LelPog24,
  author    = {Lellmann, Bj{\"o}rn and Poggiolesi, Francesca},
  title     = {Nested sequents or tree-hypersequents---a survey},
  booktitle = {Saul {K}ripke on Modal Logic},
  editor    = {Weiss, Yale and Birman, Romina},
  publisher = {Springer},
  address   = {Cham},
  year      = {2024},
  pages     = {243--301},
  doi       = {10.1007/978-3-031-57635-5_11}
}

@inproceedings{LyoKar24,
  author    = {Lyon, Tim S. and Karge, Jonas},
  title     = {Constructive interpolation and concept-based {B}eth definability for description logics via sequents},
  booktitle = {Proceedings of the Thirty-Third International Joint Conference on Artificial Intelligence ({IJCAI} 2024)},
  publisher = {ijcai.org},
  year      = {2024},
  pages     = {3484--3492}
}

@inproceedings{TiuIanGor12,
  author    = {Tiu, Alwen and Ianovski, Egor and Gor{\'e}, Rajeev},
  title     = {Grammar logics in nested sequent calculus: Proof theory and decision procedures},
  booktitle = {Advances in Modal Logic, Volume 9},
  editor    = {Bolander, Thomas and Bra{\"u}ner, Torben and Ghilardi, Silvio and Moss, Lawrence S.},
  publisher = {College Publications},
  year      = {2012},
  pages     = {516--537}
}

@inproceedings{Vak86,
  author    = {Vakarelov, Dimiter},
  title     = {Abstract characterization of some knowledge representation systems and the logic {NIL} of nondeterministic information},
  booktitle = {Artificial Intelligence {II}: Methodology, Systems, Applications},
  editor    = {Jorrand, Ph. and Sgurev, V.},
  publisher = {North-Holland},
  year      = {1987},
  pages     = {255--260}
}

@book{Vig00,
  author    = {Vigan{\`o}, Luca},
  title     = {Labelled non-classical logics},
  publisher = {Kluwer Academic Publishers},
  address   = {Dordrecht},
  year      = {2000}
}

\end{document}